\documentclass[aps,pre,reprint]{revtex4-2}
\usepackage{dcolumn}
\usepackage{multirow}
\usepackage{bm}
\usepackage{enumitem}
\usepackage[percent]{overpic}
\usepackage{lipsum}
\usepackage{bbold}
\usepackage{mathtools}
\usepackage[normalem]{ulem}
\usepackage{tikz}
\usetikzlibrary{math}
\usetikzlibrary{positioning}
\usepackage{graphicx}
\usepackage{overpic}
\usepackage[caption=false]{subfig}
\usepackage{listings}
\usepackage{subfig}
\usepackage{amsthm}
\usepackage{amsmath,amssymb}
\usepackage[breaklinks,colorlinks = true,linkcolor = magenta,urlcolor=magenta,citecolor=red,hypertexnames=false]{hyperref}
\begin{document}
\title{Energy-Weighted Site Percolation in Two Dimensions}
\author{Sayan Sircar}
\email{sircars@tifrh.res.in}
\affiliation{Tata Institute of Fundamental Research, Hyderabad - 500046, India}
\author{Kabir Ramola}
\email{kramola@tifrh.res.in}
\affiliation{Tata Institute of Fundamental Research, Hyderabad - 500046, India}
\begin{abstract}
We study a generalization of two-dimensional site percolation by assigning an energy cost $\varepsilon$ to bonds between nearest-neighbor occupied sites. This leads to a competition between entropy-driven cluster growth and energetic suppression (or enhancement) of connectivity.
Varying $\varepsilon$ continuously interpolates between dense ferromagnetic-like clusters, ordinary classical percolation, and a dilute regime of minimally connected isolated clusters. Using Monte Carlo simulations and real-space renormalization-group (RG) methods, we show that bond energy shifts the percolation threshold smoothly. We define an energy-weighted correlation length that remains finite at the classical site occupation threshold ($p_c(\varepsilon=0)$) and shrinks with increasing $\varepsilon$, capturing the energetic suppression of large-scale connectivity. The cluster size distribution exhibits an energy-dependent cutoff that drives the transition from percolation-like clusters to isolated clusters. A real-space RG with Kadanoff block recursions reveals a systematic evolution of the correlation-length exponent $\nu$ from $\nu=1/2$ (dense clusters) to $\nu=4/3$ (classical percolation), approaching $\nu=1$ (minimally connected isolated clusters), in agreement with Coulomb-gas predictions for loop models where bond energy renormalizes loop fugacity. For large values of \(\varepsilon\) (isotropic case), the suppression of nearest-neighbor bonds results in the emergence of antiferromagnetic sub-lattice ordering at high densities. Additionally, anisotropic bond energies lead to directionally selective cluster growth. Finally, we also discuss a lattice gas RG approach and scenarios where bond energy is renormalized across different scales.
\end{abstract}
\maketitle
\section{Introduction}
Percolation theory provides a fundamental framework for understanding 
geometric phase transitions in disordered systems~\cite{StaufferAharony, BollobasRiordan, Kesten}. 
In its classical formulation, each site or bond of a lattice is independently occupied 
with probability \(p\), and large-scale connectivity emerges when \(p\) exceeds a critical 
threshold \(p_c\) \cite{StaufferAharony}. Near this threshold, the system exhibits universal scaling behavior 
governed by a diverging correlation length, fractal cluster geometry, and well-characterized 
critical exponents. These features place percolation within a broad universality class 
shared across problems ranging from transport in porous media to epidemic spreading, 
polymer networks, and the onset of rigidity in disordered solids~\cite{Sahimi, PhysRevB.66.075417,Kirkpatrick, PhysRevB.72.125121}.
Beyond the traditional connectivity-based framework, several generalized percolation models incorporate additional geometric, algebraic, and topological constraints. These include rigidity percolation \cite{Jacobs1995PebbleGame,Zhang2015Rigidity,Duxbury1999Rigidity,deSouza2009Rigidity}, bootstrap percolation \cite{JChalupa_1979,Holroyd2003Sharp,Balogh2012Bootstrap,Baxter2010Bootstrap}, quantum percolation \cite{Soukoulis1991Quantum,PhysRevB.45.7724,PhysRevB.90.174203}, among other generalizations \cite{PhysRevX.12.021058}, where global order does not depend only on local occupancy. Standard or classical percolation models treat all occupied nearest-neighbor connections as equivalent. 
However, many physical systems exhibit additional energetic constraints or interactions 
that bias cluster formation. Examples include polymer networks with bending or stretching 
energies, resistor networks with weighted conductances~\cite{Sahimi, StaufferAharony, PhysRevLett.95.178102, PhysRevE.76.031906, PhysRevLett.91.108102}, and random-cluster models where cluster weights depend on geometric or topological features~\cite{fortuin1972random,swendsen1987nonuniversal}. In such systems, connectivity is governed not only by quenched disorder but also by energetic weights that modify the probability of realizing particular cluster configurations. Understanding how such energetic factors modify percolation transitions and critical scaling remains an important question in statistical mechanics, network science, and the physics of disordered materials.

Percolation can also be defined on correlated, scale-invariant backgrounds, rather than solely on uncorrelated lattices. A notable example of this is found in critical configurations of the two-dimensional \( O(n) \) loop model \cite{PhysRevE.79.061118}, where fractal loops partition the lattice into extensive regions.
By introducing probabilistic connections between neighboring regions, a percolation transition can occur within this critical geometric framework. Despite the strong correlations created by the loop structure, this transition has been demonstrated to fall within the universality class of Kasteleyn–Fortuin (FK) random-cluster models, which are linked to the Potts universality class \cite{PhysRevE.79.061118}. This indicates that universal percolative behavior persists even in systems where connectivity arises from an interacting framework.

In this work, we introduce and analyze an energy-weighted site-percolation (EWSP) model, which is defined by the site occupation probability. In this model, each bond between nearest-neighbor occupied sites incurs an energy cost denoted by \(\varepsilon\). A cluster consisting of \(T\) such bonds then contributes a Boltzmann weight of \(e^{-\varepsilon T}\). This results in a grand canonical ensemble of clusters, where the combinatorial multiplicity of a cluster with a given number of occupied sites competes with an energetic penalty that is proportional to its connectivity. When \(\varepsilon = 0\), the model simplifies to classical site percolation. As \(\varepsilon\) approaches infinity, the model resembles an isolated cluster phase, where clusters with the fewest bonds between nearest-neighbor occupied sites dominate. Therefore, the tuning parameter \(\varepsilon\) allows for a continuous interpolation between classical percolation universality and a regime with strongly suppressed or enhanced clusters.
The introduction of the energy cost \(\varepsilon\) significantly alters both the cluster-size 
distribution and the correlation length. 
Consequently, the correlation length remains 
finite even at \(p=p_c(\varepsilon=0)\). This behavior parallels the suppression of FK clusters in the \(q\)-state Potts model \cite{FORTUIN1972536,PhysRevLett.58.86,RevModPhys.54.235} and admits a natural interpretation within the Coulomb-gas formulation, 
where tuning \(\varepsilon\) corresponds to modifying the loop fugacity and thereby the thermal scaling field~\cite{Nienhuis1984, DOTSENKO2001523,Kapec2021}. Energy-weighted site percolation thus provides a tunable extension of classical percolation in which energetic constraints reshape connectivity, critical scaling, and emergent spatial order. \par
To quantitatively characterize this model, we employ real-space RG 
methods based on Kadanoff block scaling \cite{PhysRevB.4.3174}. These allow the site occupation probability \(p\) to renormalize under coarse-graining. Using different spanning-rules, we construct explicit RG recursion relations and compute fixed points and correlation-length exponents. Our results demonstrate a smooth change of the exponent \(\nu\) with increasing \(\varepsilon\), consistent with Coulomb-gas predictions in which the thermal exponent interpolates between classical 
percolation (\(\nu = 4/3\)) and the dilute cluster limit (\(\nu = 1\)). Furthermore we formulate a lattice-gas representation defined by site fugacity \( e^{\mu} = \frac{p}{1-p} \) and bond fugacity \( e^{J=-\varepsilon} \). This framework produces exact recursion relations for block-renormalized parameters, allowing us to determine the RG flow. In our analysis using lattice gas RG, we explore two scenarios: one where the bond energy is tunable and does not depend on the length scale, and another where the bond energy cost is dependent on the length scale and undergoes renormalization. Overall, the EWSP model introduced here provides a tunable extension of classical percolation with rich crossover behavior and nontrivial modifications to critical scaling. 
Unlike traditional interacting percolation models \cite{PhysRevA.38.4198,sun2023dynamic}, which generate correlations through spin interactions or dynamic occupation rules, the EWSP model incorporates an explicit energetic cost for nearest-neighbor occupied bonds. Consequently, connectivity is influenced by this energetic factor, creating a tunable competition between entropy and bond energetics. \par
The paper is organized as follows: In Section \ref{sec:mo1}, we introduce the model of EWSP on a square lattice, the partition function, and we define the associated parameters that will be used in subsequent sections. Section \ref{Sec:moo9} discusses our Monte Carlo simulation of the lattice gas in the grand canonical ensemble. It examines the cluster configurations for both isotropic and anisotropic bond energy cases. In the isotropic case, we illustrate the emergence of antiferromagnetic sub-lattice ordering in site occupancy, while in the anisotropic case, we focus on strip-like ordering. In Section \ref{sec:RG1}, we compute real-space RG recursion relations and determine the critical site occupation probability, and the correlation length critical exponent (both isotropic and anisotropic). In Section \ref{sec:CG1}, we use results from the Coulomb gas to explain different limits of energy cost, denoted as \(\varepsilon\), and the corresponding changes in the critical exponent associated with these limits. In Section \ref{sec:mo2}, we derive the expression for the correlation length related to energy-weighted percolation in the canonical ensemble and discuss cluster size statistics. Finally, in Section \ref{sec:mo6}, we derive the RG recursion relation using the lattice gas analogy expressed in terms of site fugacity and bond fugacity.
\begin{figure}[t!]
\centering
\begin{tikzpicture}[scale=1.2, every node/.style={scale=0.9}]
\foreach \x in {0,1,2,3}{
    \foreach \y in {0,1,2,3}{
        \coordinate (n\x\y) at (\x,\y);
    }
}
\foreach \coord in {(0,0),(1,0),(2,0),(3,0),
                    (0,1),(1,1),(2,1),
                    (0,2),(1,2),(2,2),
                    (1,3),(2,3)}{
    \fill[blue] \coord circle (4pt);
}
\foreach \coord in {(3,1),(3,2),(0,3),(3,3)}{
    \fill[red] \coord circle (4pt);
}
\draw[thick] (0,0)--(1,0) node[midway, below, yshift=-2pt] {$\varepsilon_x$};
\draw[thick] (1,0)--(2,0) node[midway, below, yshift=-2pt] {$\varepsilon_x$};
\draw[thick] (2,0)--(3,0) node[midway, below, yshift=-2pt] {$\varepsilon_x$};
\draw[thick] (0,0)--(0,1) node[midway, left, xshift=-2pt] {$\varepsilon_y$};
\draw[thick] (1,0)--(1,1) node[midway, left, xshift=-2pt] {$\varepsilon_y$};
\draw[thick] (2,0)--(2,1) node[midway, right, xshift=2pt] {$\varepsilon_y$};
\draw[thick] (0,1)--(1,1) node[midway, below, yshift=-2pt] {$\varepsilon_x$};
\draw[thick] (1,1)--(2,1) node[midway, below, yshift=-2pt] {$\varepsilon_x$};
\draw[thick] (0,1)--(0,2) node[midway, left, xshift=-2pt] {$\varepsilon_y$};
\draw[thick] (1,1)--(1,2) node[midway, left, xshift=-2pt] {$\varepsilon_y$};
\draw[thick] (2,1)--(2,2) node[midway, right, xshift=2pt] {$\varepsilon_y$};
\draw[thick] (0,2)--(1,2) node[midway, below, yshift=-2pt] {$\varepsilon_x$};
\draw[thick] (1,2)--(2,2) node[midway, below, yshift=-2pt] {$\varepsilon_x$};
\draw[thick] (1,2)--(1,3) node[midway, left, xshift=-2pt] {$\varepsilon_y$};
\draw[thick] (2,2)--(2,3) node[midway, right, xshift=2pt] {$\varepsilon_y$};
\draw[thick] (1,3)--(2,3) node[midway, below, yshift=-2pt] {$\varepsilon_x$};
\draw[->, thick] (-0.5,0) -- (4.2,0) node[right] {$\mathbf{x}$};
\draw[->, thick] (0,-0.5) -- (0,4.2) node[above] {$\mathbf{y}$};
\end{tikzpicture}
\caption{ A $4\times4$ square lattice with 12 occupied (blue) and 4 unoccupied (red) sites. Bonds between nearest-neighbor occupied sites are shown in black and have an energy cost $\varepsilon_x$ for bonds along horizontal $\mathbf{x}$ direction, and energy cost $\varepsilon_y$ for bonds along vertical $\mathbf{y}$ direction. The connected occupied sites form a cluster. In this diagram, there is only one cluster containing $12$ occupied sites and $16$ bonds connecting nearest-neighbor occupied sites.}
\label{fig:mod1}
\end{figure}
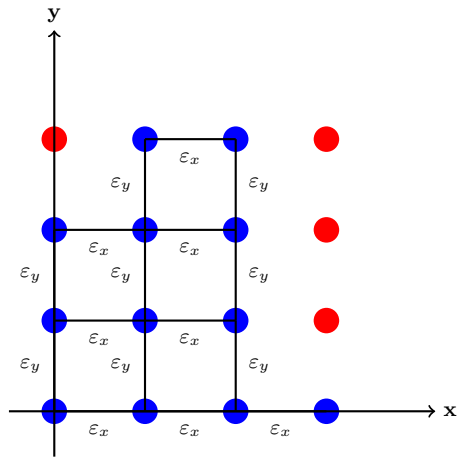
\begin{figure*}[ht!]
    \centering
    \hspace{-1cm}
    \includegraphics[keepaspectratio, width=1.0\textwidth]{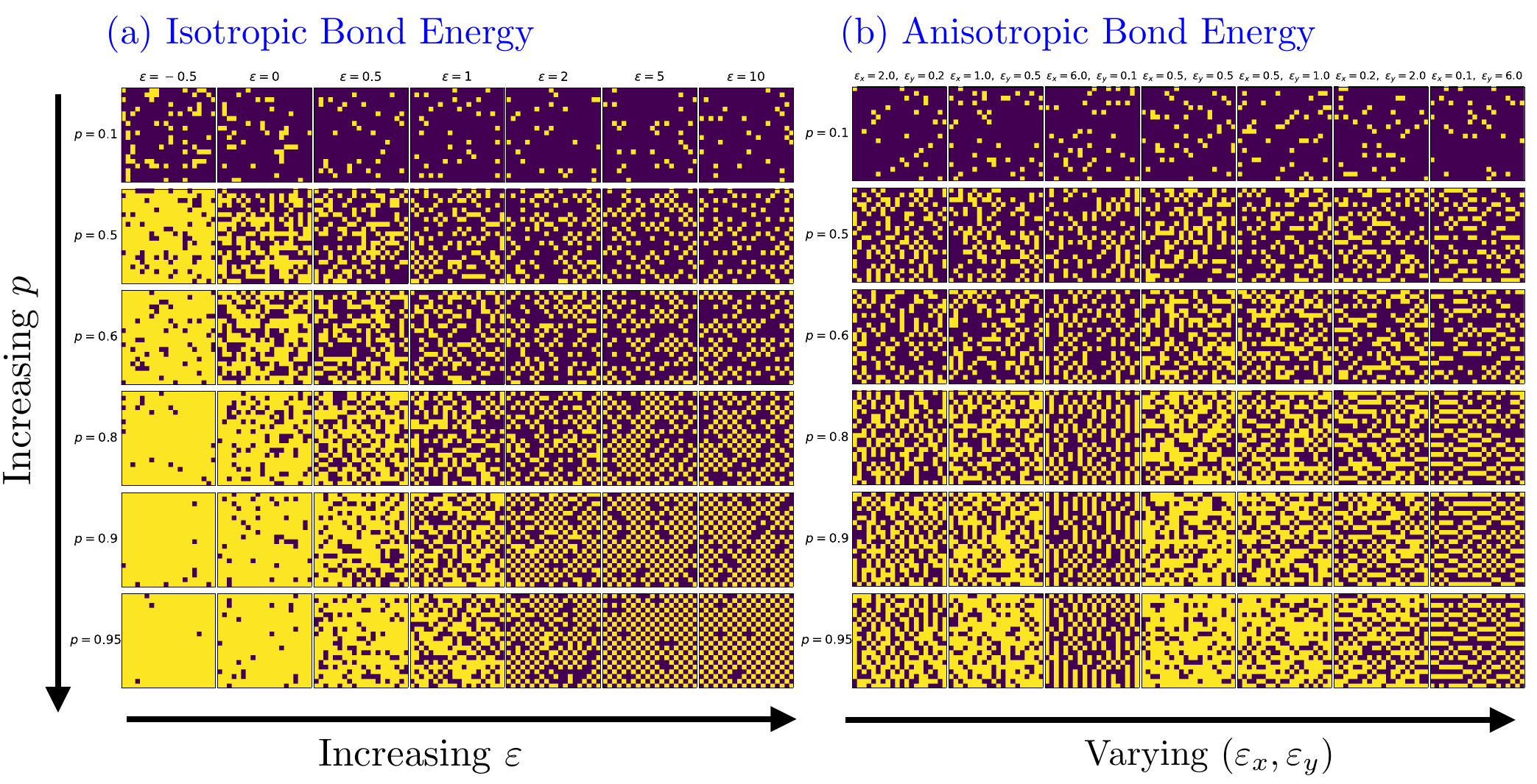}
    \caption{The left plot \textbf{(a)} illustrates the variation in cluster configurations as a function of site-occupation probability \( p \) and energy cost \( \varepsilon \), and the right plot \textbf{(b)} illustrates the variation in cluster configurations for different values of site occupation probability \( p \) and energy cost \( \varepsilon_x \) for bonds along the direction of $x$ axis and \( \varepsilon_y \) for bonds along the direction of $y$ axis. The computation is performed on a square grid with a total of \( N = 400 \) sites. The colors indicate the occupancy variable $n_{ij}$, with $n_{ij}=1$ indicating that a particular site is occupied (yellow), and $n_{ij}=0$ indicating that a particular site is not occupied (blue).}
    \label{fig:cs1}
\end{figure*}
\section{THE Model}\label{sec:mo1}
We propose an energy-weighted generalization of site percolation on a two-dimensional square lattice of linear dimension $L$, with $N=L^2$ total sites, as shown in Fig.~\ref{fig:mod1}. Each lattice site $i$ is associated with a binary occupation variable $n_i \in \{0,1\}$, where $n_i=1$ denotes an occupied site and $n_i=0$ an unoccupied site. In the absence of energetic constraints, sites are independently occupied with probability $p$, recovering ordinary site percolation. Undirected bonds connect occupied nearest-neighbor sites. A connected component of occupied sites defines a cluster $C$. For a given cluster $C$, we denote by $s(C)$ the number of occupied sites and by $T(C)$ the total number of nearest-neighbor bonds between occupied sites within the cluster, counting each bond once. In standard percolation, all configurations with the same number of occupied sites or bonds are treated equivalently. Here, we introduce an additional energetic constraint by assigning an energy cost $\varepsilon$ to each bond connecting nearest-neighbor occupied sites. The total energy of a cluster is therefore defined as
\begin{equation}
E(C) = \varepsilon\, T(C).
\end{equation}
A lattice configuration $\mathcal{C}$ consists of a collection of disjoint clusters $\{C_k\}$. The partition function of the EWSP model can be computed as:
\begin{equation}\label{eq:paf1}
    Z_{\mathbf{EWSP}}=\sum_{\mathcal{C}}p^{N_{\mathrm{occ}}}(1-p)^{N-N_{\mathrm{occ}}}
\prod_k e^{-\varepsilon T(C_k)}.
\end{equation}
Using the partition function defined in Eq.~\eqref{eq:paf1}, a probability measure can be assigned to each configuration as:
\begin{equation}\label{eq:ptf1}
W_{\mathbf{EWSP}}(\mathcal{C})= \frac{1}{Z_{\mathbf{EWSP}}}p^{N_{\mathrm{occ}}}(1-p)^{N-N_{\mathrm{occ}}}
\prod_k e^{-\varepsilon T(C_k)},
\end{equation}
where $N_{\mathrm{occ}}=\sum_i n_i$ is the total number of occupied sites in a particular configuration. This defines a grand-canonical ensemble in which both the number of occupied sites and the internal connectivity of a configuration fluctuate. The site occupation probability $p$ controls the density of occupied sites, while the parameter $\varepsilon$ penalizes highly connected clusters for $\varepsilon>0$. From a statistical-mechanical perspective, the model may be viewed as a lattice gas with site fugacity
\begin{equation}
e^{\mu} = \frac{p}{1-p},
\end{equation}
and bond fugacity $e^{-\varepsilon}$. The resulting ensemble reflects a competition between entropic combinatorial factors favoring large clusters and energetic suppression or enhancement of configurations with many internal bonds. The lattice gas partition function in terms of this site and bond fugacity can be written as:
\begin{equation}\label{eq:pflg1}
    Z_{\mathbf{LG}}=\sum_{\mathcal{C}}e^{\mu \sum_{i}n_i-\varepsilon\sum_{\langle ij\rangle}n_i n_j}.
\end{equation}
\begin{figure}[t!]
    \centering
    \includegraphics[keepaspectratio, width=1.02\linewidth]
    {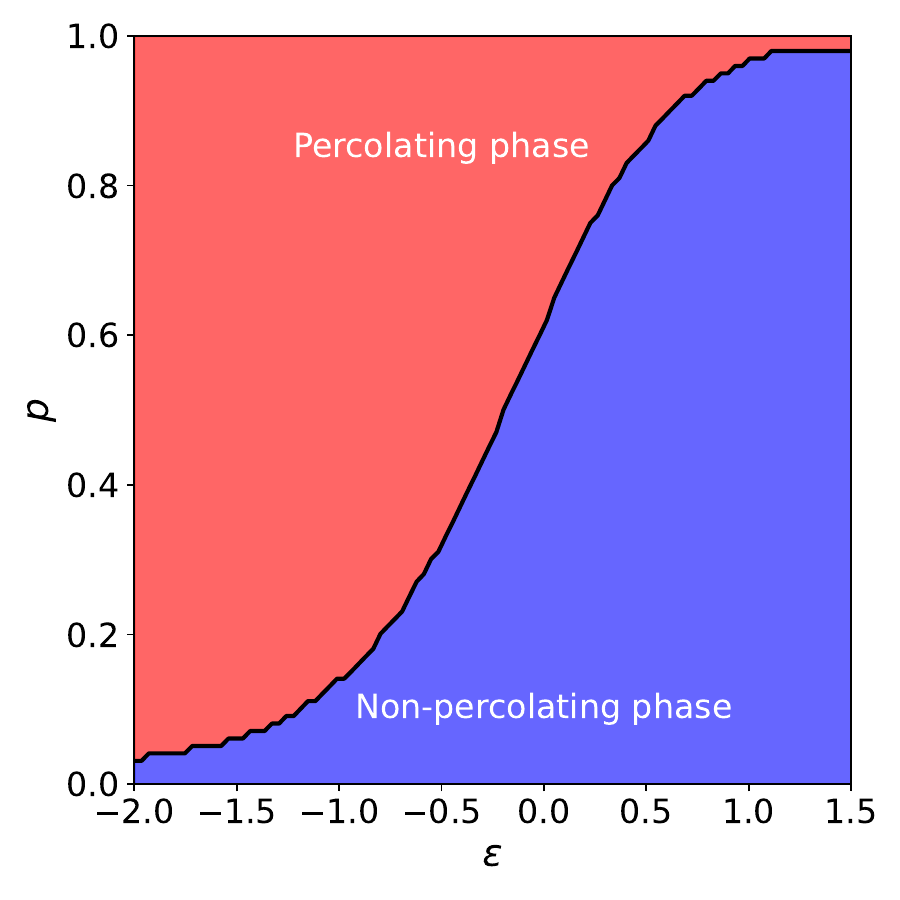}
     \caption{Phase diagram indicating percolating and non-percolating phase separated by critical curve $p_c(\varepsilon)$, in the $\varepsilon-p$ plane. The phase diagram is obtained using numerical Monte Carlo simulation on a square lattice with dimensions $N=40000$, as detailed in Sec.~\ref {Sec:moo9}. The blue colored region indicates the non-percolating phase, while the red colored region indicates the percolating phase.}
    \label{fig:PD1}
\end{figure}
In this context, $n_{i}$ represents the occupation variable at site $i$, taking values of either $(0,1)$. For $n_i=0$ the site is unoccupied, while $n_i=1$ indicates that the site is occupied, $\langle ij\rangle$ denotes the nearest neighbor sites $i$ and $j$. 
One can easily observe the connection between the lattice gas and site percolation in the hard-core on-site exclusion limit. The grand canonical partition function for the lattice gas, as shown in Eq.~\eqref{eq:pflg1}, can be computed in the hard-core on-site exclusion limit as:
\begin{equation}
    Z_{\mathbf{HLG}}=(1+e^{\mu})^{N},
\end{equation} where $N$ is the local number of sites on the lattice, and $\mathbf{HLG}$ stands for hard-core lattice gas. Using this partition function, the density of occupied sites can be computed as:
\begin{equation}
    \rho(\varepsilon=0)=\frac{1}{N}\frac{\partial \ln Z_{\mathbf{HLG}}}{\partial \mu}=\frac{e^{\mu}}{1+e^{\mu}}=p.
\end{equation}
Thus, in the case of a hard-core on-site exclusion limit, the density of occupied sites is precisely equal to the site occupation probability. However, in a finite $\varepsilon$ domain, the density of occupied sites is influenced by both the site occupation probability \( p \), through site fugacity, and the bond energy \( \varepsilon \), which reflects the interactions between nearest neighbor occupied sites. This mapping of the lattice gas model in the finite $\varepsilon$ domain corresponds directly to that of the Ising model in a finite magnetic field \cite{PhysRev.85.808, PhysRev.87.410}. Our study essentially seeks to understand the percolation phase transition in the Lee-Yang lattice gas \cite{PhysRev.87.410}, which is extensively studied in the literature regarding the order-disorder phase transition \cite{PhysRevE.90.012120,PhysRevE.100.022109}. \par
Increasing $p$ promotes connectivity, while increasing $\varepsilon>0$ suppresses it. Several limiting cases are of particular interest. When $\varepsilon = 0$, all clusters carry weight proportional to the number of occupied sites and the model reduces exactly to standard or classical site percolation, with critical behavior governed by the standard percolation universality class. For $\varepsilon>0$, clusters with many internal bonds are exponentially suppressed, leading to a reduced cutoff in the cluster-size distribution as shown in Fig.~\ref{fig:combined365iu1} and a diminished correlation length as shown in Fig.~\ref{fig:combined3658u1}.
In the limit $\varepsilon \to \infty$, only minimally connected clusters contribute significantly, and the system is dominated by isolated cluster-like structures reminiscent of dilute lattice walks.
The introduction of the energy cost $\varepsilon$ therefore provides a continuous tuning parameter that interpolates between classical percolation and an isolated-cluster regime with antiferromagnetic ordering in site occupancy, as shown in Fig.~\ref{fig:cs1}. In defining the partition function in Eq.~\eqref{eq:paf1}, we assume that the energy cost per bond, denoted as \(\varepsilon\), is a tunable parameter that remains constant across all length scales. This means we allow the bond variable to remain unchanged during the renormalization process in our real-space RG scheme presented in Sec.~\ref{sec:RG1}. However, in the lattice gas RG framework presented in Sec.~\ref{sec:mo6}, we will relax this assumption. We will consider \(\varepsilon\) not as a constant tunable parameter, but rather as a variable that depends on the length scale, similar to site occupation probability, and that gets renormalized. Unlike classical percolation, which is controlled by a single parameter $p$, the present model exhibits a two-parameter phase space $(p,\varepsilon)$.  
As we show below, increasing $\varepsilon$ shifts the effective percolation threshold $p_c(\varepsilon)$ and can render the correlation length finite even at the classical percolation point $p_c(\varepsilon=0)$. In what follows, clusters are identified as connected components of occupied sites, and all observables, including cluster-size distributions and correlation lengths, are computed as ensemble averages over the energy-weighted measure defined above. This formulation provides the foundation for the numerical simulations, correlation-length computation, and real-space RG calculations presented in the subsequent sections. Through the lattice-gas simulation discussed in Sec.~\ref {Sec:moo9}, we obtained a phase diagram that illustrates the percolating and non-percolating phases with varying bond energy costs \(\varepsilon\), as shown in Fig.~\ref{fig:PD1}. In this phase diagram, it is evident that as \(\varepsilon\) approaches \(-\infty\), the critical site occupation probability \(p_c(\varepsilon)\) decreases. This allows for a transition from the non-percolating phase to the percolating phase at very low occupation probabilities, as the bond energy in this scenario enhances connectivity. Conversely, as \(\varepsilon\) approaches \(+\infty\), the critical site occupation probability \(p_c(\varepsilon)\) increases, leading to a transition from the non-percolating phase to the percolating phase at very high occupation probabilities, since the bond energy in this case suppresses connectivity. Overall, this phase diagram effectively illustrates the competition between entropy and energy costs in determining global connectivity within a disordered network. \par
Interestingly, this model can be exactly mapped to the high-temperature expansion of the \( O(n) \) spin model \cite{PhysRevLett.58.86}, especially in the high and low-energy limit, where \( \varepsilon \to \pm\infty \). In this regime, we can also apply the results from the Coulomb gas, as demonstrated in Sec.~\ref {sec:CG1}.
In this representation, the partition function can be written as a sum over loop configurations
\(\mathcal{G}\),
\begin{equation}\label{eq:Zon}
    Z_{\mathbf{O(n)}} = \sum_{\mathcal{G}} n^{C(\mathcal{G})} K^{|\mathcal{G}|},
\end{equation}
where \(C(\mathcal{G})\) is the number of connected components (closed loops),
\(|\mathcal{G}|\) is the total number of occupied bonds, and \(K\) is the bond fugacity \cite{PhysRev.158.546}.
This formulation emphasizes that the statistical weight depends explicitly on the connectivity
of the underlying graph, a feature shared with percolation-type models. In the limit \(n \to 0\), configurations containing closed loops are suppressed, and only graphs with a single open trajectory contribute.
A closely related structure appears in generalized percolation models, particularly in bond percolation \cite{PhysRevLett.58.86}, most notably in the random-cluster (FK) formulation. The partition function for this is given by:
\begin{equation}\label{eq:rc1}
Z_{\mathbf{RC}} =
\sum_{\mathcal{G}}
p^{|\mathcal{G}|} (1-p)^{|E|-|\mathcal{G}|} q^{C(\mathcal{G})},
\end{equation}
where \(p\) represents the bond occupation probability, \(|E|\) is the total number of edges in the lattice, and \(q\) is a fugacity that controls the number of connected clusters. In ordinary percolation, this corresponds to the limit \(q \to 1\) \cite{fortuin1972random}. \par
The explicit expression for the partition function of our model is derived in Sec.~\ref {sec:mo1}. Its connection to the FK cluster partition function is established in Sec.~\ref{sec:CG1}. By comparing Eq.~\eqref{eq:Zon} and Eq.~\eqref{eq:rc1}, it becomes evident that the loop fugacity \(n\) in the \(O(n)\) model plays a role analogous to the cluster fugacity \(q\) in the random-cluster model. The limit of single open trajectories as \(n \to 0\) corresponds to a generalized percolation regime where closed clusters are suppressed, allowing a single non-self-intersecting path to dominate connectivity. 
\section{Simulation details}\label{Sec:moo9}
\begin{figure}[t!]
    \centering
    \hspace{-1cm}
    \includegraphics[keepaspectratio, width=1.02\linewidth]
    {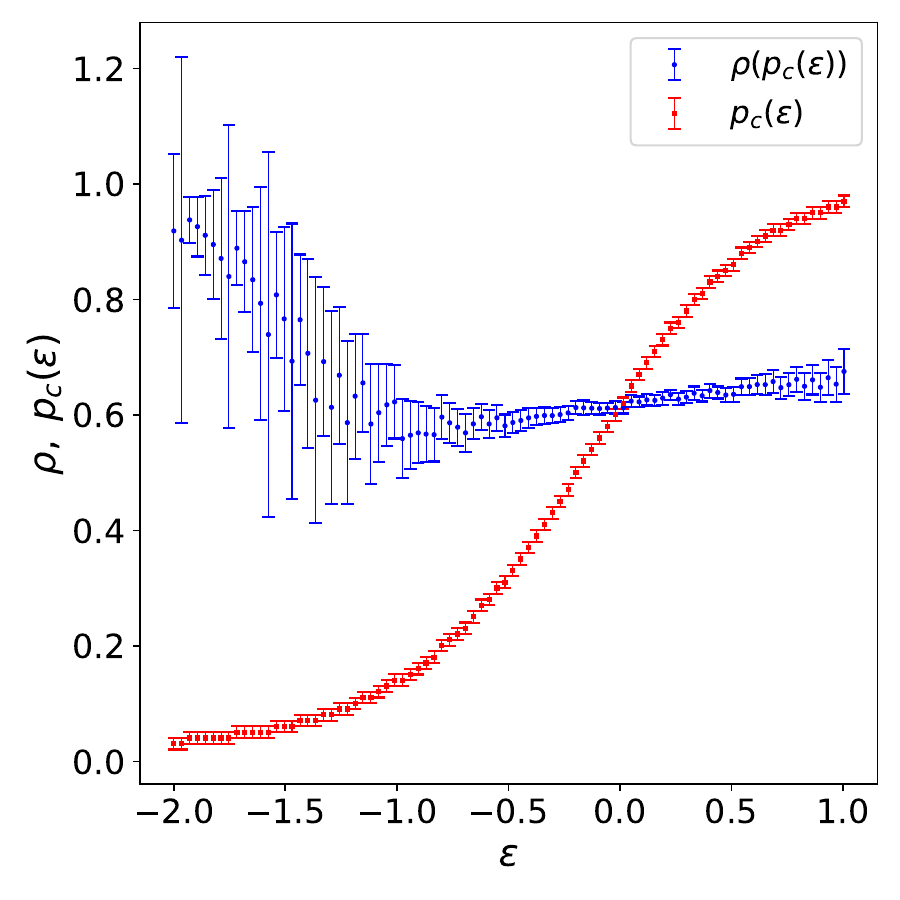}
     \caption{The plot illustrates the density of occupied sites, denoted as \(\rho(p_c(\varepsilon))\), which is evaluated at the critical site occupation probability \(p_c(\varepsilon)\) for various values of the bond energy cost \(\varepsilon\). The data for \(p_c(\varepsilon)\) is  also presented in Fig.~\ref{fig:PD1}. Additionally, the corresponding mean density of occupied sites, \(\rho(p_c(\varepsilon), \varepsilon)\), is computed for a system size of \(N = 400\). 
     }
    \label{fig:PD2aa1}
\end{figure}

\begin{figure*}[ht!]
\centering
\setlength{\tabcolsep}{4pt}
\begin{tabular}{ccc}
    \includegraphics[width=0.33\textwidth]{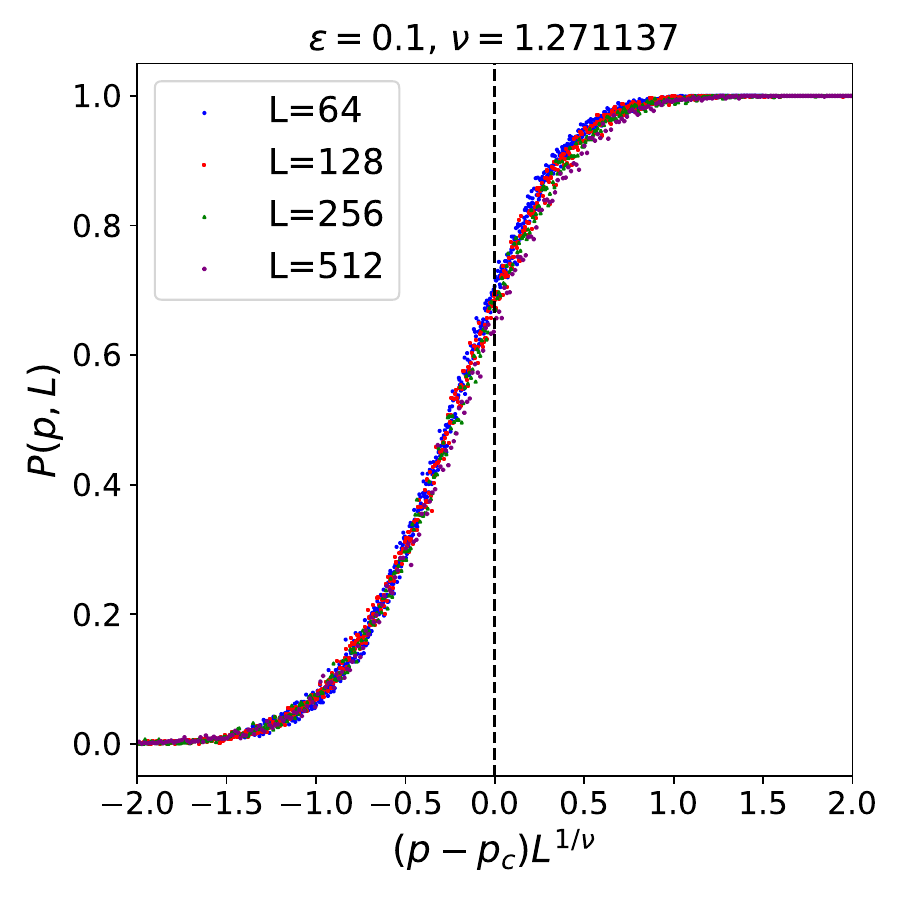} &
    \includegraphics[width=0.33\textwidth]{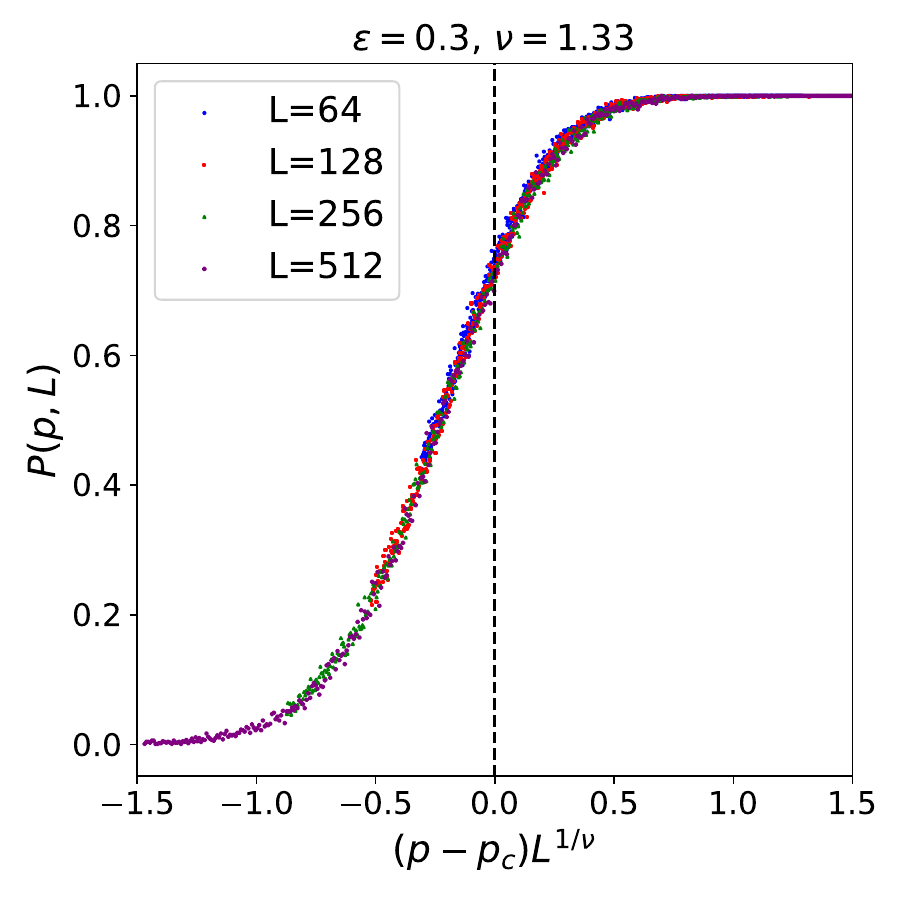} &
    \includegraphics[width=0.33\textwidth]{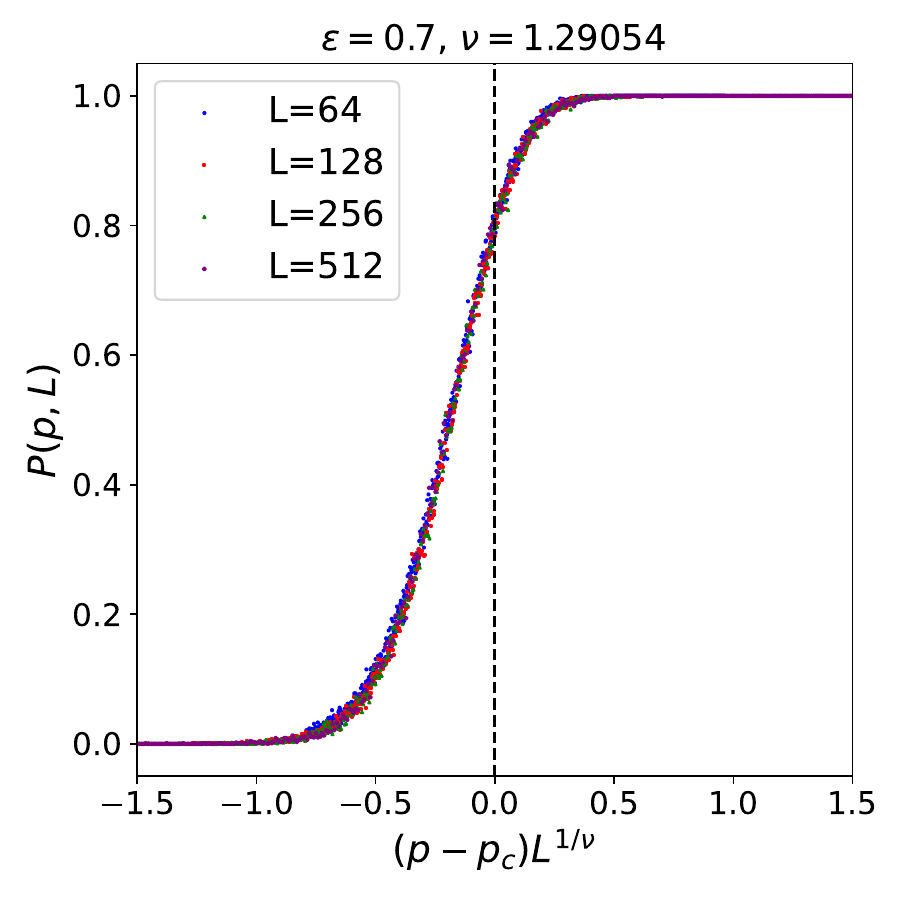} \\[-0.2cm]
    \textbf{(a) $\varepsilon = 0.1$ } & \textbf{(b) $\varepsilon = 0.3$} & \textbf{(c) $\varepsilon = 0.7$} \\[0.25cm]
    \includegraphics[width=0.33\textwidth]{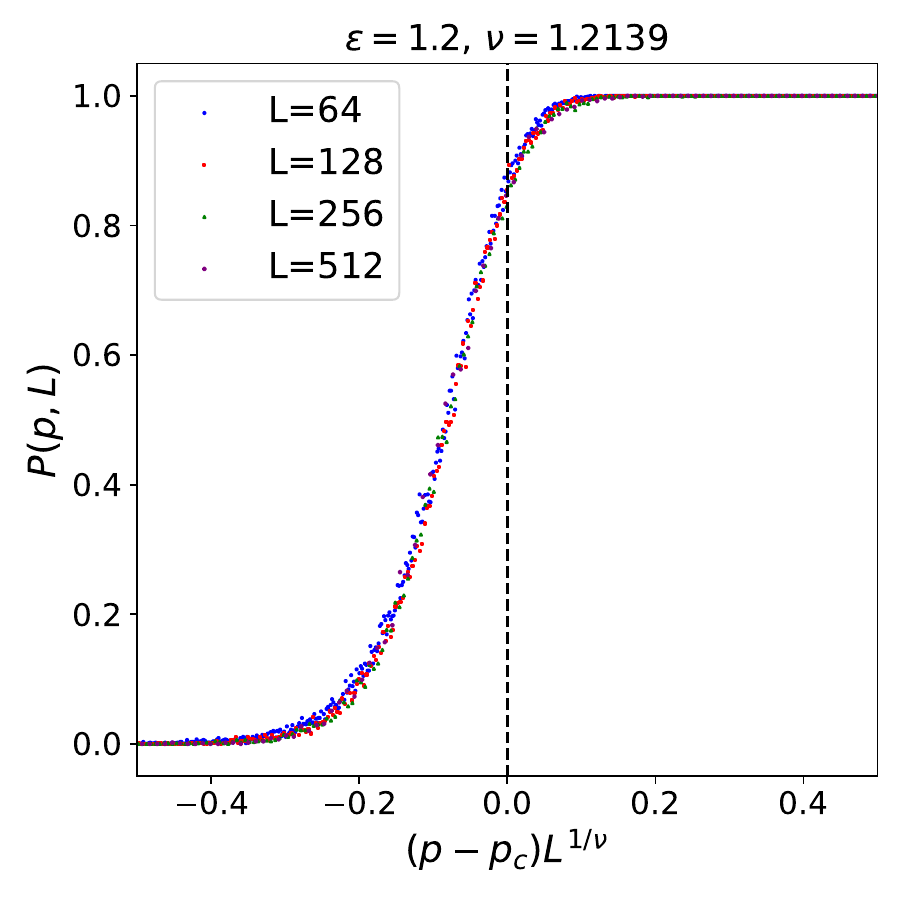} &
    \includegraphics[width=0.33\textwidth]{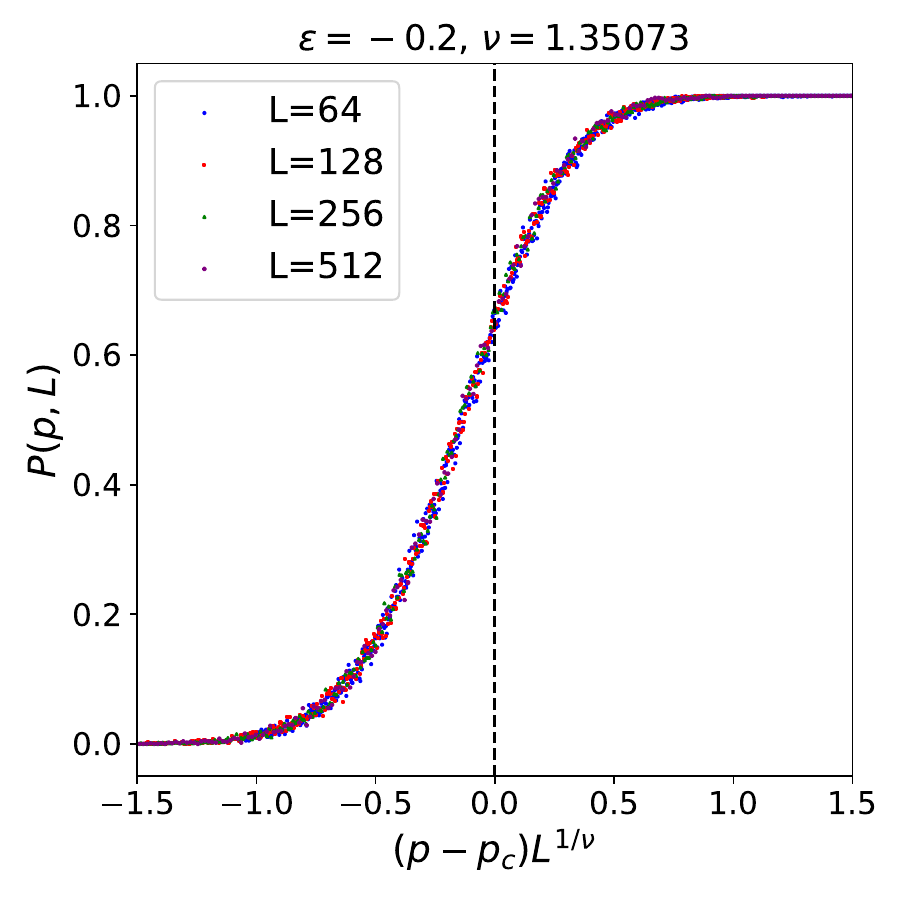} &
    \includegraphics[width=0.33\textwidth]{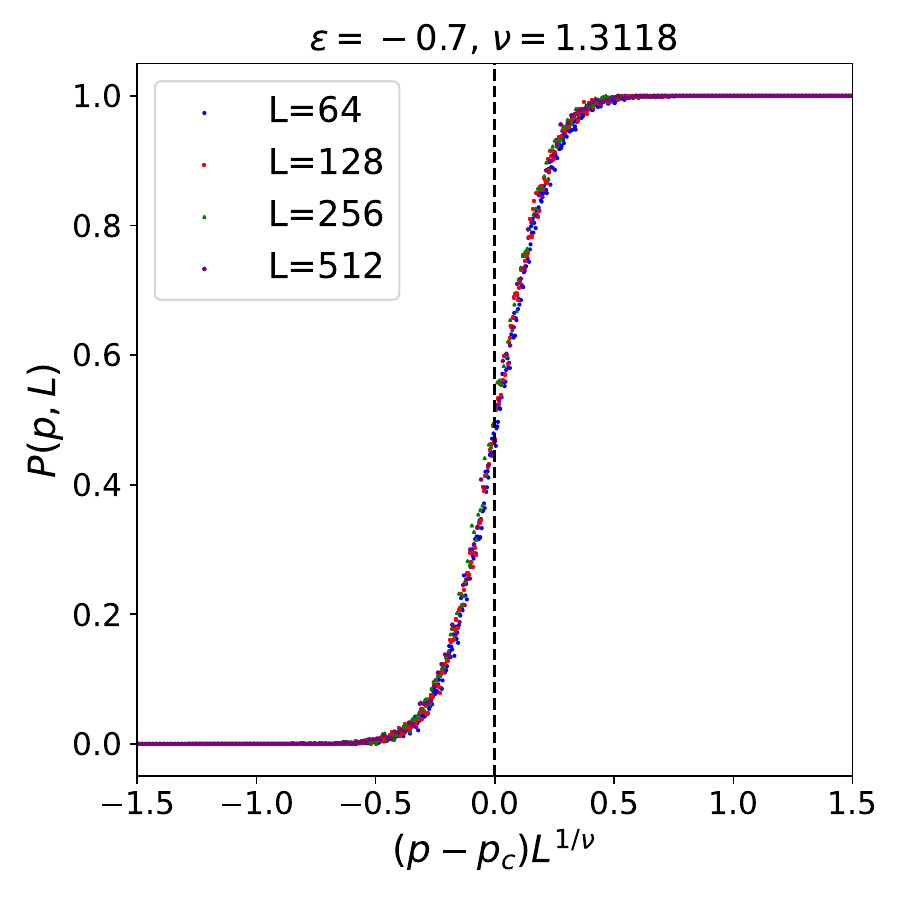} \\[-0.2cm]
    \textbf{(d) $\varepsilon =1.2$} & \textbf{(e) $\varepsilon =-0.2$} & \textbf{(f) $\varepsilon =-0.7$}
\end{tabular}
\caption{ The plot illustrates the scaling collapse of the wrapping probability function \( P(p, L) \) for different system sizes \( L \), computed from Metropolis Monte Carlo simulations. The critical exponent $\nu=\nu(\varepsilon)$ has been extracted for various cases of bond energy cost \( \varepsilon \) as follows:
\textbf{(a)} \( \varepsilon = 0.1 \): \( p_c \approx 0.65301 \),   
\textbf{(b)} \( \varepsilon = 0.3 \): \( p_c \approx 0.76348 \),   
\textbf{(c)} \( \varepsilon = 0.7 \): \( p_c \approx 0.85 \),  
\textbf{(d)} \( \varepsilon = 1.2 \): \( p_c \approx 0.9812 \),  
\textbf{(e)} \( \varepsilon = -0.2 \): \( p_c \approx 0.4683 \),  
\textbf{(f)} \( \varepsilon = -0.7 \): \( p_c \approx 0.2141 \).  
These plots illustrate how varying bond energies affect the critical occupation probabilities and exponents.}
\label{fig:corr_exp_sim_1}
\end{figure*}

\begin{figure*}[ht!]
\centering
\setlength{\tabcolsep}{4pt}
\begin{tabular}{ccc}
    \includegraphics[width=0.33\textwidth]{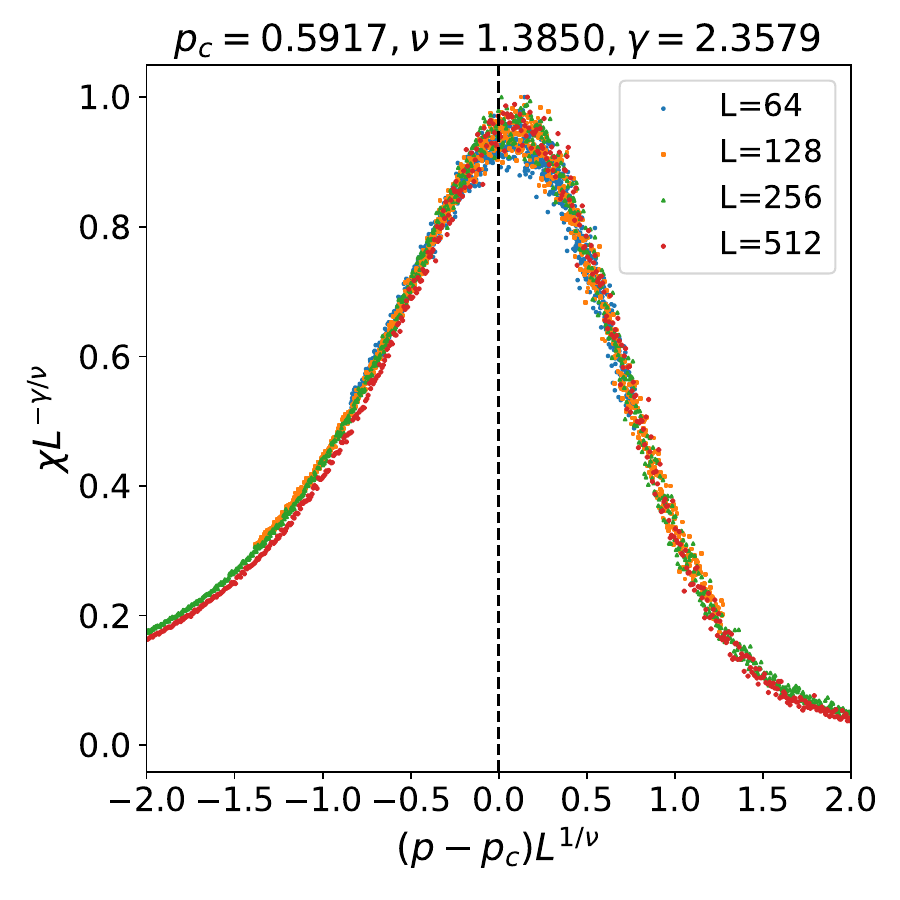} &
    \includegraphics[width=0.33\textwidth]{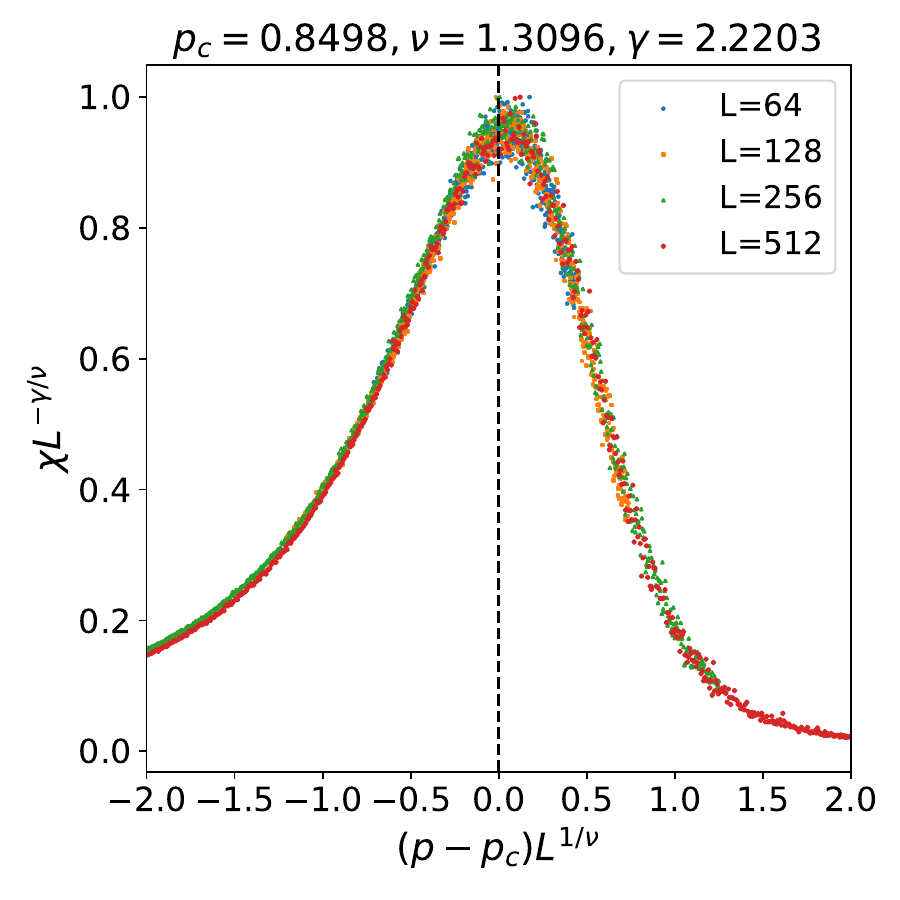} &
    \includegraphics[width=0.33\textwidth]{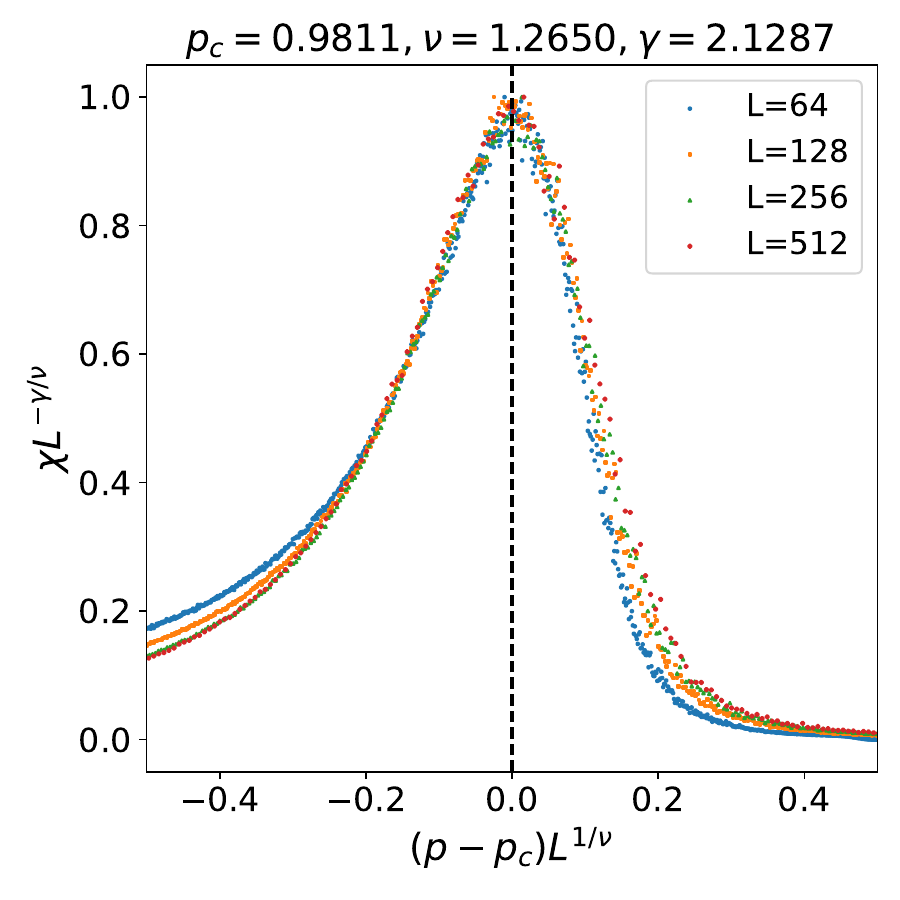} \\[-0.2cm]
    \textbf{(a) $\varepsilon=0$} & \textbf{(b) $\varepsilon=0.5$} & \textbf{(c) $\varepsilon=1.2$} \\[0.25cm]
    \includegraphics[width=0.33\textwidth]{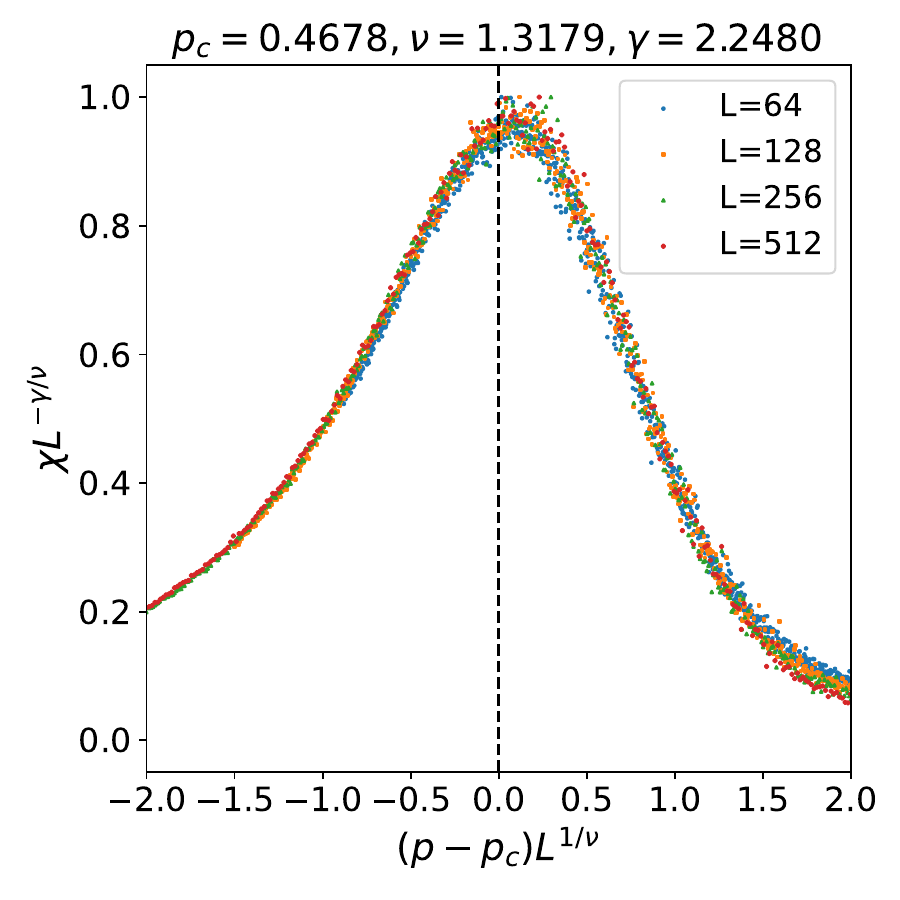} &
    \includegraphics[width=0.33\textwidth]{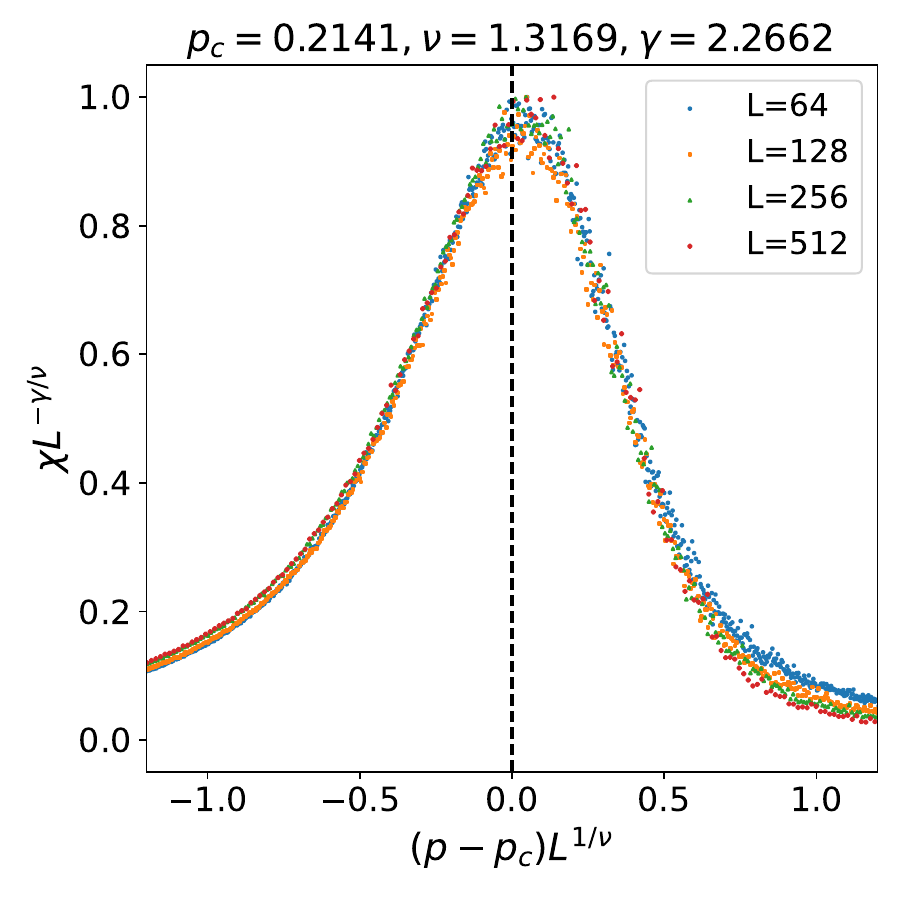} &
    \includegraphics[width=0.33\textwidth]{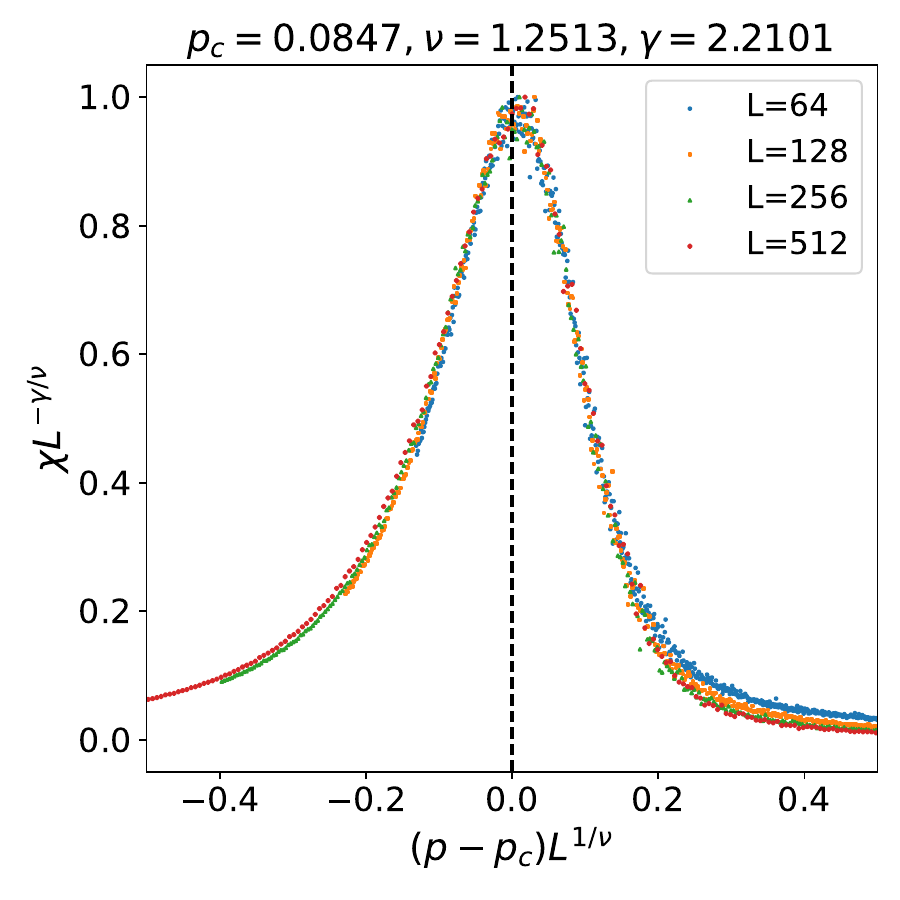} \\[-0.2cm]
    \textbf{(d) $\varepsilon=-0.2$} & \textbf{(e) $\varepsilon=-0.7$} & \textbf{(f) $\varepsilon=-1.2$}
\end{tabular}
\caption{The plot illustrates the scaling collapse of the susceptibility function \( \chi(p, L) \) for different system sizes \( L \), computed from Metropolis Monte Carlo simulations. The critical exponents $\nu=\nu(\varepsilon)$ and $\gamma=\gamma(\varepsilon)$ has been extracted for various cases of bond energy cost \( \varepsilon \) as follows:
\textbf{(a)} $\varepsilon=0$,
\textbf{(b)} $\varepsilon=0.5$,
\textbf{(c)} $\varepsilon=1.2$,
\textbf{(d)} $\varepsilon=-0.2$,
\textbf{(e)} $\varepsilon=-0.7$, 
\textbf{(f)} $\varepsilon=-1.2$.}
\label{fig:susc_exp_sim_1}
\end{figure*}

We simulate our model as an interacting lattice gas with nearest-neighbor interaction strength $\varepsilon$. To obtain the cluster configurations for various values of $\varepsilon$, we consider a two-dimensional square lattice of linear size $L$ with binary occupation variables $n_{ij}\in\{0,1\}$ defined on each lattice site $(i,j)$. For convenience, we map this occupation variable onto an Ising-like spin
variable \( S_{ij} \) according to
\begin{equation}
S_{ij} = 2n_{ij} - 1 =
\begin{cases}
+1, & \text{if } n_{ij} = 1, \\
-1, & \text{if } n_{ij} = 0 .
\end{cases}
\end{equation}
\begin{figure}[t!]
    \centering
    \includegraphics[keepaspectratio, width=1.02\linewidth]
    {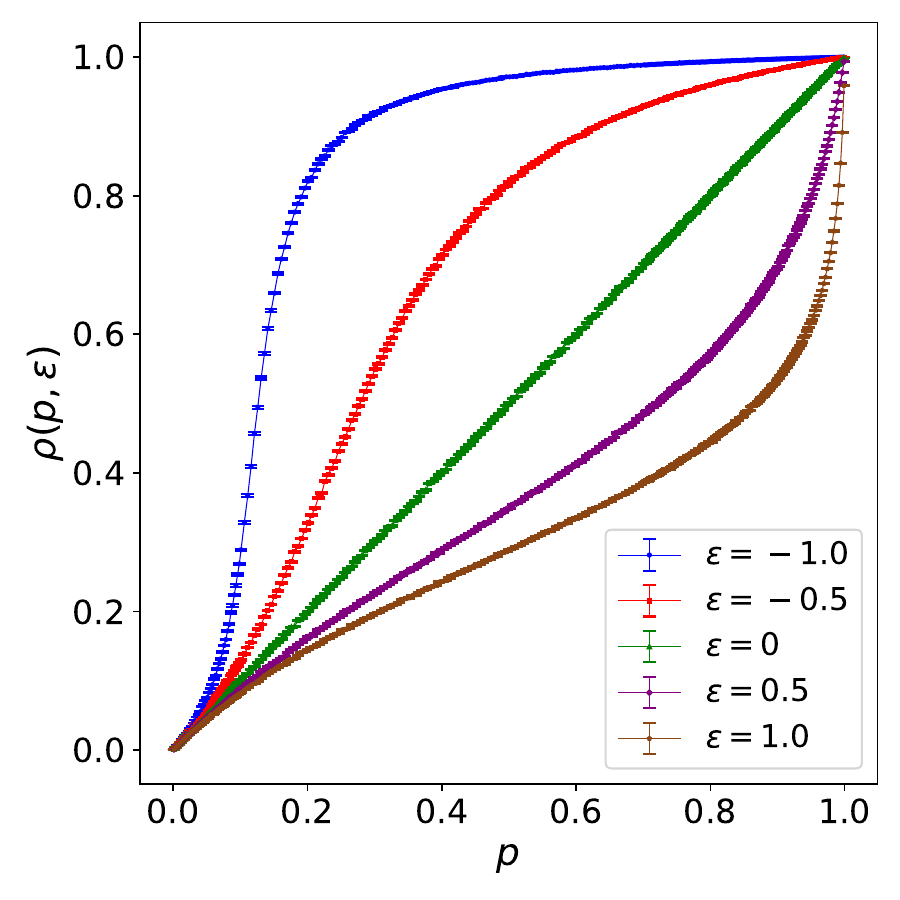}
     \caption{The plot illustrates the variation in the mean density of occupied sites, defined as \(\rho = \frac{N_{occ}}{L^{2}}\), where \(N_{occ}\) represents the number of occupied sites. This value is averaged over $500$ realizations of equilibrated samples while varying the site occupation probability \(p\) in the initial configuration, considering different values of bond energy cost \(\varepsilon\). The system size used in the study is \(N = 400\). The error bar represents the standard error of the mean density of occupied states. The equilibration is carried out using grand canonical Glauber-like Metropolis sweeps, as outlined in Sec.~\ref {Sec:moo9}.} 
    \label{fig:PD2a}
\end{figure}
Undirected bonds connect nearest-neighbor occupied sites, and connected components of occupied sites define clusters. For each cluster $\mathcal{C}$, we define its internal bond number $T(\mathcal{C})$ as the total number of bonds between nearest neighbor occupied sites within the cluster, counted once per bond, as defined previously in Sec.~\ref {sec:mo1}. The energy of a configuration is taken to be $E = \varepsilon \sum_{\mathcal{C}} T(\mathcal{C})$, where $T(\mathcal{C})$ denotes the energy of cluster $\mathcal{C}$. Configurations with a mean occupation probability $\rho \neq p$ are generated using a grand-canonical Monte Carlo using a Metropolis scheme \cite{metropolis1953equation}, where the chemical potential is defined as $\mu=\ln[p/(1-p)]$. Because we employ grand-canonical Glauber-like sweeps \cite{glauber1963timedependent,binder2019montecarlo} over occupation variables, the mean occupation probability of the equilibrated configuration will not equal the initial mean occupation probability $p$, but depends on both $p$ and bond energy cost $\varepsilon$. This equality holds only when $\varepsilon=0$. The site occupation probability $p$ of the initial non-equilibrated configuration sets the site fugacity $e^{\mu}$. Therefore, in all subsequent discussions, we will refer to $p$ instead of $\rho$. It is important to note this distinction. The variation in the density of occupied sites $\rho(p,\varepsilon)$, averaged over numerous realizations of equilibrated samples with a site occupation probability \( p \), is presented in Fig.~\ref{fig:PD2a}. Additionally, Fig.~\ref{fig:PD2aa1} illustrates the variation of the density of occupied sites \(\rho(p_c(\varepsilon))\) evaluated at the critical occupation probability \(p_c(\varepsilon)\) for different values of bond energy cost \(\varepsilon\). It is evident from computations conducted on finite system sizes that there is a minimum density of occupied sites, at a fixed bond energy cost $\varepsilon$, where a percolating phase can exist. This can be explored in greater detail, particularly in the thermodynamic limit, in future studies. In all the cases, the equilibration is performed using Glauber-like Metropolis sweeps \cite{glauber1963timedependent,metropolis1953equation,binder2019montecarlo}.
On the other hand, one can keep the density of occupied sites constant as the initial configurations by using Kawasaki swaps \cite{kawasaki1966diffusion1,binder2019montecarlo} for equilibration. However, since our objective is to examine the grand canonical partition function, we will not use Kawasaki swaps to compute the observables.
Starting from a random initial configuration with occupation probability $p$, we perform a sequence of Monte Carlo sweeps \cite{glauber1963timedependent,metropolis1953equation,binder2019montecarlo}; during each sweep, $L^{2}$ trial updates are attempted by selecting a site at random and proposing to flip its occupation state. The proposed move is accepted with probability $\min\{1,\exp[-\Delta H]\}$, where $\Delta H=\Delta E-\mu\,\Delta N$, $\Delta E$ is the change in interaction energy induced by the flip, and $\Delta N=\pm1$ is the corresponding change in occupation number \cite{FrenkelSmit2002}. After equilibration, clusters are identified by constructing the reduced adjacency matrix of occupied sites and computing its connected components. The cluster configurations are obtained from sampling equilibrated states. They are illustrated in Fig.~\ref{fig:cs1}(a) for the isotropic bond energy case and in Fig.~\ref{fig:cs1}(b) for the anisotropic bond energy case. As the value of $\varepsilon$ increases (with $\varepsilon > 0$) in the isotropic case, there is an energetic suppression of clusters, leading to sub-lattice ordered configurations. This ordering phenomenon is discussed in detail in the Supplementary Information~\cite{SI}.
The cluster size distribution \(n_{s}\) is calculated using this Monte Carlo sampling, as shown in Fig.~\ref{fig:csd_combined_2x1}(a) for isotropic bond energy cost and in Fig.~\ref{fig:csd_combined_2x1}(b) for anisotropic bond energy cost. From Fig.~\ref{fig:csd_combined_2x1}(b), it is evident that compared to the isotropic case, where $\varepsilon_x=\varepsilon_y$, the tails of the distribution are reduced and primarily consist of configurations not suppressed by the energetic cost of the bonds. Through this lattice gas simulation, we generate the phase diagram for percolation phases in the $p-\varepsilon$ plane, as illustrated in Fig.~\ref{fig:PD1}. One can also calculate the critical exponent for correlation length, denoted as $\nu(\varepsilon)$, along with the critical exponent for susceptibility or average cluster size, denoted as $\gamma(\varepsilon)$. These quantities diverge at the critical occupation probability $p_c(\varepsilon)$. This is done using finite-size scaling to collapse the wrapping probability $P(p, L,\varepsilon)$ and the susceptibility $\chi(p, L,\varepsilon)$ as follows:
\begin{equation}
    P(p,L, \varepsilon)\sim G(|p-p_c(\varepsilon)|L^{\frac{1}{\nu(\varepsilon)}}),
\end{equation}
\begin{equation}
    \chi(p,L, \varepsilon)\sim L^{-\frac{\gamma(\varepsilon)}{\nu(\varepsilon)}}H(|p-p_c(\varepsilon)|L^{\frac{1}{\nu(\varepsilon)}}),
\end{equation} where $G$ and $H$ are universal scaling functions, as shown in Fig.~\ref{fig:corr_exp_sim_1} and Fig.~\ref{fig:susc_exp_sim_1}.
\begin{figure*}[t!]
    \centering
    \includegraphics[keepaspectratio, width=0.46\textwidth]{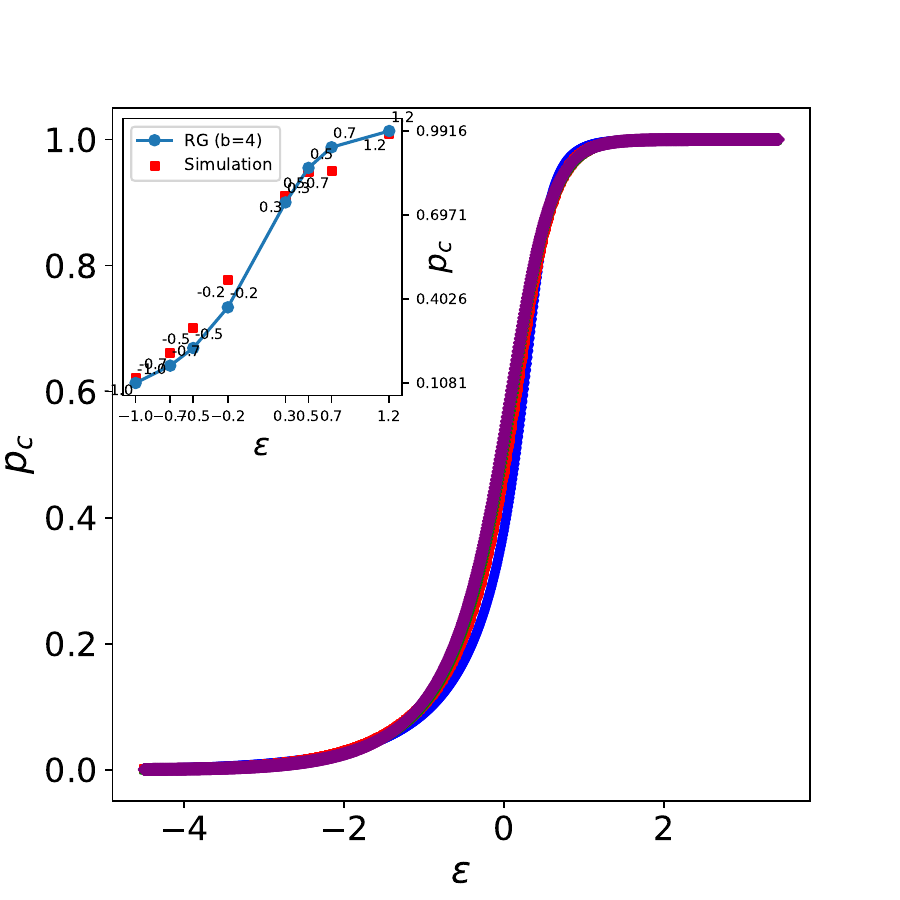}
    \includegraphics[keepaspectratio, width=0.46\textwidth]{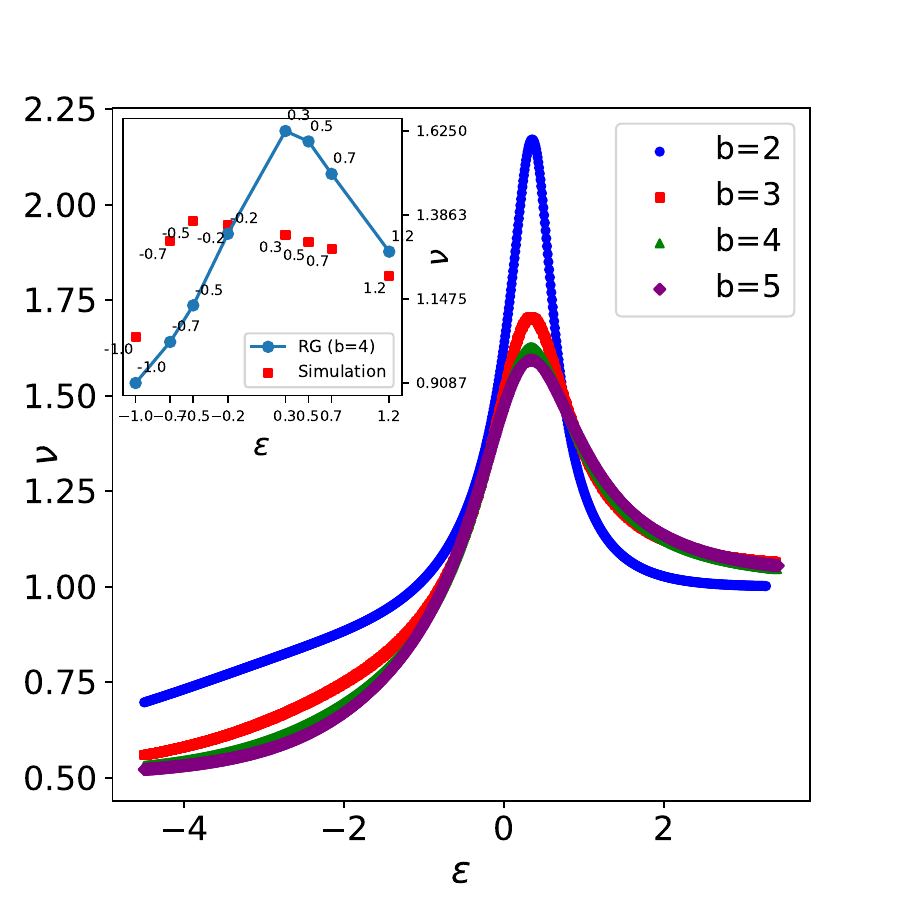}
    \put(-410,85){\textbf{(a)}}
    \put(-200,85){\textbf{(b)}}
    \\
    \includegraphics[keepaspectratio, width=0.46\textwidth]{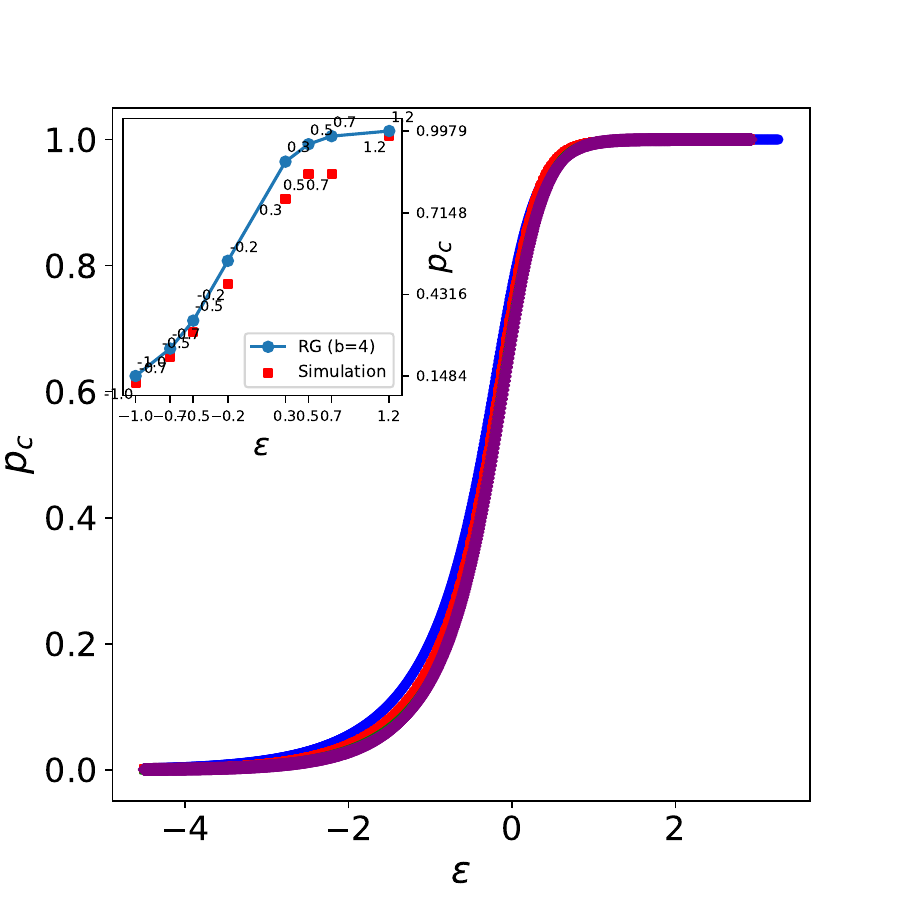}
    \includegraphics[keepaspectratio, width=0.46\textwidth]{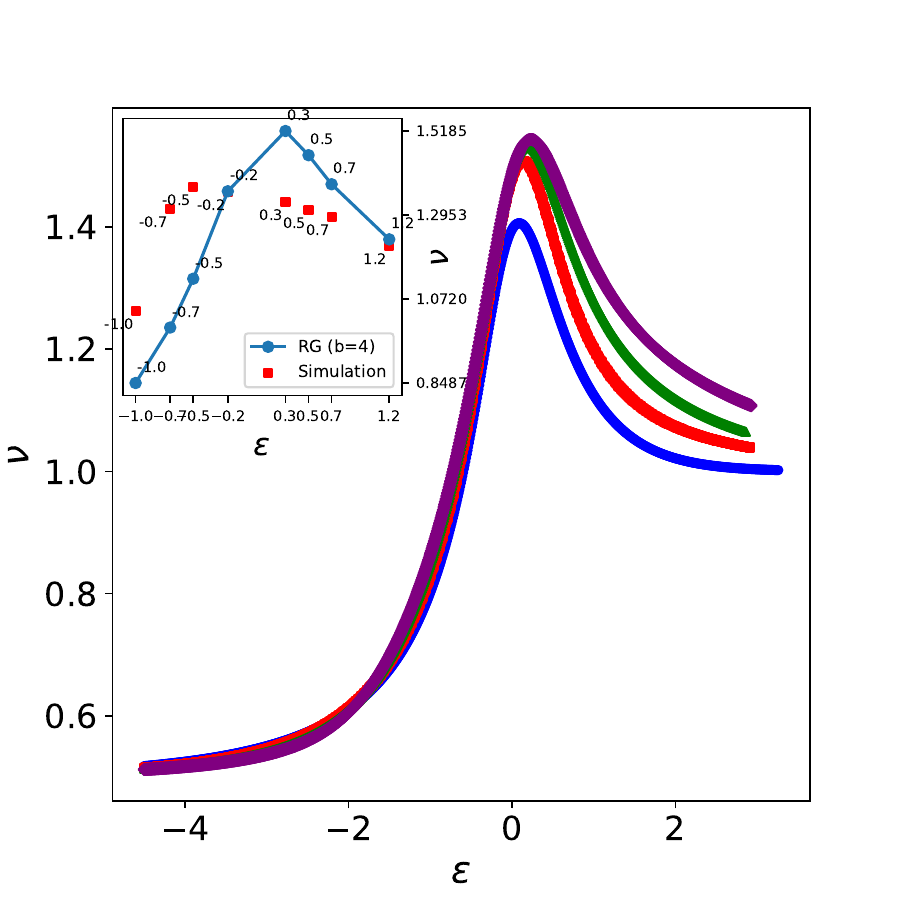}
    \put(-410,55){\textbf{(c)}}
    \put(-200,55){\textbf{(d)}}
     \caption{Panels \textbf{(a,b)}: $p_c$ and $\nu$ computed using the $R_0$ rule for block sizes
$b\in[2,5]$. Panels \textbf{(c,d)}: $p_c$ and $\nu$ computed using the $R_2$ rule for block sizes
$b\in[2,5]$.  
The insets display some values obtained from simulations and the spanning rule. All quantities are shown as functions of the energy cost per bond $\varepsilon$.}
    \label{fig:RG_2x3_combined}
\end{figure*}
\section{Real space Renormalization group analysis}\label{sec:RG1}
We consider the site-occupation model on the square lattice used in the numerics: each site is occupied independently with probability $p$, and a configuration with total $T$ bonds between nearest-neighbor occupied sites is weighted by
\begin{equation}
W(\mathcal{C}) =\frac{1}{Z_{\text{EWSP}}}p^{N_{\rm occ}} (1-p)^{N - N_{\rm occ}}\, e^{-\epsilon T},
\end{equation} where the Boltzmann weight for each cluster in the configuration is multiplied, and $Z_{\text{EWSP}}$ represents the partition function defined in Eq.~\eqref{eq:paf1}. The set of configurations is labeled by $\mathcal{C}$. In this section, we will rescale the original lattice to reduce the degrees of freedom in the system, allowing us to study its large-scale behavior using the real-space renormalization group approach, as presented in \cite{Kadanoff1966Scaling,PhysRevB.4.3174,Wilson1971RGCP2}. So the partition function for the rescaled block ($Z_{R}^{b}$) can be written as:
\begin{equation}\label{eq:yt1}
    Z_{R}^{b}=\sum_{k}\sum_{C}p^{k}(1-p)^{N-k}\prod_{i\in C}e^{-\varepsilon T_{i}},
\end{equation} where $b$ stands for the linear dimension of block, and $R$ specifies the rule one is considering for renormalization coarse-graining.
We rescale the lattice by a factor of $b$, the linear dimension of each block. We assign the occupation probability of each block $\tilde{p}$, according to a rule. The block occupation probability can be expanded using the usual procedure of real-space renormalization as:
\begin{equation}\label{eq:jh1}
    \tilde{p}=\frac{1}{Z_{R}^{b}}\sum_{k}\sum_{C^{'}}p^{k}(1-p)^{N-k}\prod_{i\in C^{'}}e^{-\varepsilon T_{i}},
\end{equation} where $C^{'}$ is the set of configurations satisfying the specific rule one is choosing for coarse-graining.
For rules like the ``majority rule" and other rules termed $R_{0}$, $R_{1}$, and $R_{2}$ as stated below, the numerator in the above expression in Eq.~\eqref{eq:jh1} simplifies to:
\begin{equation} \label{eq:Rgrec1}
    \tilde{p}= \frac{1}{Z_{R}^{b}}\sum_{k}M_{k}p^{k} q^{N-k}e^{-\varepsilon T_k},~~~~~q=1-p,
\end{equation} 
where $q$ is the probability that a site is not occupied, $T_k$ is the total number of bonds across all clusters of $k$ connected sites in each block, and the coefficient $M_k$ depends on the rule one chooses. We consider different such rules for analysis below. 
A rule can be such that $\tilde{p}=1$, where at least $k=[\frac{N}{2}]$ sites in a block are occupied, which corresponds to ``majority rule" in literature. The coefficient $M_k$ for this type of rule corresponds to the binomial coefficient. Other sets of rules are considered in \cite{PhysRevB.21.1223} termed as $R_0$, $R_1$, and $R_2$.  The rules state that a block is considered occupied if it contains a spanning cluster. For rule \( R_0 \), the cluster can span the block either horizontally or vertically. In the case of rule \( R_1 \), the cluster spans the block in a fixed direction (for example, horizontally), while rule \( R_2 \) requires the cluster to span the block both horizontally and vertically \cite{PhysRevB.21.1223}. 
These are the conditional statements for all three different rules. 
For example, below provides the polynomials for the renormalized site occupation probability $\tilde{p}=R_{0}^{b}(p)$ using rule $R_0$ (for $\varepsilon=0$) for block sizes $b\in[2,5]$ as:
For $b=2$:
\begin{equation}
    R_{0}^{2}(p,q,\varepsilon=0)=p^{4} + 4 p^{3} q + 4 p^{2} q^{2},
\end{equation}

for $b=3$:
\begin{equation}
\label{eq:Drok}
\begin{split}
R_{0}^{3}(p,q,\varepsilon=0) 
&= p^{9} + 9p^{8}q + 36p^{7}q^{2} + 82p^{6}q^{3} \\
&\quad + 93p^{5}q^{4} + 44p^{4}q^{5} + 6p^{3}q^{6}.
\end{split}
\end{equation}
for $b=4$:
\begin{equation}
\begin{split}
R_{0}^{4}(p, q, \varepsilon=0) =\;& p^{16} + 16 p^{15} q + 120 p^{14} q^{2} + 560 p^{13} q^{3} \\
&+ 1818 p^{12} q^{4}  + 4296 p^{11} q^{5} + 7196 p^{10} q^{6} \\
&+ 8136 p^{9} q^{7}+ 5988 p^{8} q^{8} + 2784 p^{7} q^{9} \\
&+ 780 p^{6} +q^{10} + 120 p^{5} q^{11} + 8 p^{4} q^{12},
\end{split}
\end{equation}
and for $b=5$:
\begin{equation}
\begin{split}
R_{0}^{5}(p,q, \varepsilon=0)=\;& p^{25} + 25 p^{24} q + 300 p^{23} q^{2} + 2300 p^{22} q^{3} \\
& + 12650 p^{21} q^{4} + 53128 p^{20} q^{5} + 176992 p^{19} q^{6} \\
& + 478316 p^{18} q^{7} + 1054923 p^{17} q^{8} \\
& + 2666712 p^{15} q^{10} + 2963364 p^{14} q^{11} \\
& + 2556058 p^{13} q^{12} 
 + 1699665 p^{12} q^{13} \\
 & + 865132 p^{11} q^{14} + 333630 p^{10} q^{15} \\
& + 95845 p^{9} q^{16} + 19916 p^{8} q^{17} + 2836 p^{7} q^{18} \\
& + 248 p^{6} q^{19} + 10 p^{5} q^{20}+ 1880864 p^{16} q^{9},
\end{split}
\end{equation}
where $q=1-p$, the probability that a site is not occupied. The expressions are also obtained in \cite{PhysRevB.21.1223}. For other rules like $R_1$ and $R_2$, similar expansions can be obtained by changing the conditional statement specified before. A full analytic derivation of this expression for a particular $b=3$ (Eq.~\eqref{eq:Drok}) using rule $R_{0}$ has been shown in Appendix~\ref {app:gh} using the inclusion-exclusion theorem. For $\varepsilon \neq 0$, the expansion using spanning rule $R_{0}$ can be computed as:
For $b=2$:
\begin{equation}
R_{0}^{2}(p,q,\varepsilon)=
\frac{4p^{2}q^{2}e^{-\varepsilon}
+
4p^{3}q\,e^{-2\varepsilon}
+
p^{4}e^{-4\varepsilon}}{Z_{0}^{2}},
\end{equation} with scaled block partition function as
\begin{equation*}
Z_{0}^{2}(p,q,\varepsilon)=
q^{4}
+
4pq^{3}
+
2p^{2}q^{2}
+
4p^{2}q^{2}e^{-\varepsilon}
+
4p^{3}q\,e^{-2\varepsilon}
+
p^{4}e^{-4\varepsilon},
\end{equation*}
for $b=3$:
\begin{equation}
\begin{split}
R_{0}^{3}(p,q,\varepsilon)=
\frac{1}{Z_{0}^{3}}
\Big(
6p^{3}q^{6}e^{-2\varepsilon}
+12p^{4}q^{5}e^{-2\varepsilon}
+32p^{4}q^{5}e^{-3\varepsilon} \\
+4p^{5}q^{4}e^{-2\varepsilon}
+40p^{5}q^{4}e^{-3\varepsilon}
+33p^{5}q^{4}e^{-4\varepsilon} \\
+16p^{5}q^{4}e^{-5\varepsilon}
+4p^{6}q^{3}e^{-3\varepsilon}
+22p^{6}q^{3}e^{-4\varepsilon} \\
+28p^{6}q^{3}e^{-5\varepsilon}
+24p^{6}q^{3}e^{-6\varepsilon}
+4p^{6}q^{3}e^{-7\varepsilon} \\
+14p^{7}q^{2}e^{-6\varepsilon}
+8p^{7}q^{2}e^{-7\varepsilon}
+14p^{7}q^{2}e^{-8\varepsilon} \\
+p^{8}q e^{-8\varepsilon}
+4p^{8}q e^{-9\varepsilon}
+4p^{8}q e^{-10\varepsilon} \\
+p^{9}e^{-12\varepsilon}
\Big),
\end{split}
\end{equation} with scaled block partition function given as:
\begin{equation}
\begin{aligned}
Z_{0}^{3}(p,q,\varepsilon)=&
q^{9}+p^{8}q\,e^{-8\varepsilon}
+4p^{8}q\,e^{-9\varepsilon}
+4p^{8}q\,e^{-10\varepsilon} 
+p^{9}e^{-12\varepsilon} \\
&9pq^{8}+24p^{2}q^{7}
+12p^{2}q^{7}e^{-\varepsilon} \\
&+22p^{3}q^{6}
+40p^{3}q^{6}e^{-\varepsilon}
+22p^{3}q^{6}e^{-2\varepsilon} \\
&+6p^{4}q^{5}
+28p^{4}q^{5}e^{-\varepsilon}
+56p^{4}q^{5}e^{-2\varepsilon} \\
&+32p^{4}q^{5}e^{-3\varepsilon}
+4p^{4}q^{5}e^{-4\varepsilon} \\
&+p^{5}q^{4}
+24p^{5}q^{4}e^{-2\varepsilon}
+48p^{5}q^{4}e^{-3\varepsilon} \\
&+37p^{5}q^{4}e^{-4\varepsilon}
+16p^{5}q^{4}e^{-5\varepsilon} \\
&+4p^{6}q^{3}e^{-3\varepsilon}
+24p^{6}q^{3}e^{-4\varepsilon}
+28p^{6}q^{3}e^{-5\varepsilon} \\
&+24p^{6}q^{3}e^{-6\varepsilon}
+4p^{6}q^{3}e^{-7\varepsilon} \\
&+14p^{7}q^{2}e^{-6\varepsilon}
+8p^{7}q^{2}e^{-7\varepsilon}
+14p^{7}q^{2}e^{-8\varepsilon} ,
\end{aligned}
\end{equation}
The expansions of the renormalized occupation probability for a block size of \( b = 4 \) are provided in the Supplementary Information~\cite{SI}.
The renormalized spanning probability calculated using the spanning rule $R_2$ is also detailed in Supplementary Information~\cite{SI}. 
We note that by setting $\varepsilon=0$, the polynomial expressions for the spanning probability simplify to the expression presented in \cite{PhysRevB.21.1223}. 
\par
The expression for critical exponent $\nu$ is given as:
\begin{equation}
    \nu(\varepsilon)=\frac{\ln(b)}{\ln\left(\frac{d\tilde{p}(p_c(\varepsilon))}{dp}\right)},
\end{equation} with $p_c(\varepsilon)$ are the fixed points of the RG flow for each value of bond energy cost $\varepsilon$. This expression is obtained by linearizing about each fixed point of the RG flow equation (Eq.~\eqref{eq:Rgrec1}) \cite{goldenfeld1992phase}. It is important to emphasize that while this real-space RG treatment is accurate for hierarchical lattices \cite{Hong1984RandomWalksHierarchical,PhysRevB.34.3403}, it is only approximately correct for regular lattices like the one we are considering.

The variation of the critical site occupation probability $p_c$ and the critical exponent $\nu$ using renormalization rules $R_{0}$ and $R_{2}$ are illustrated in Fig.~\ref{fig:RG_2x3_combined}(a,b) and Fig.~\ref{fig:RG_2x3_combined}(c,d) respectively. As the block size $b$ increases, the computed curves for the critical exponent $\nu$ using rules $R_0$ and $R_2$ become closer to each other, as can be inferred from Fig.~\ref{fig:RG_2x3_combined}(b,d).
In the general case, where $\varepsilon_x$ represents the energy cost for the bonds in the $x$ direction and $\varepsilon_{y}$ for the bonds in the $y$ direction, the variation of these quantities using rules $R_{0}$ and $R_{2}$ are shown in Fig.~\ref{fig:RGan_1b} and Fig.~\ref{fig:RGan_1}, respectively. All the fixed points \(p_c(\varepsilon)\) are unstable fixed points in the $p$ direction, with the only two basins of attraction or stable fixed points being at \(p_c(\varepsilon_{+})=1\) and \(p_c(\varepsilon_{-})=0\), respectively. In this context, \(\varepsilon_{+}\) and \(\varepsilon_{-}\) represent the values of \(\varepsilon\) when \(p_c=1\) and \(p_c=0\), with \(\varepsilon_{-}<0\) and \(\varepsilon_{+}>0\). \par
In the Monte Carlo simulations, we obtain equilibrated configurations for each value of the bond energy cost \(\varepsilon\). This parameter governs the effective interaction between neighboring sites and controls the connectivity properties of the system. As a result, variations in \(\varepsilon\) lead to systematic shifts in the critical site occupation probability \(p_c\). In this context, \(\varepsilon\) is treated as an externally tunable parameter with no assigned scale dependence. Similarly, in the initial RG schemes considered here, the bond energy is not renormalized. Instead, \(\varepsilon\) is again treated as a fixed, externally adjustable parameter, similar to its role in the Monte Carlo simulations. The RG transformations in this framework act solely on the configurational or connectivity degrees of freedom and do not incorporate feedback on the interaction strength. Consequently, the analysis captures how large-scale connectivity emerges for a given value of \(\varepsilon\), but it does not address how the interaction parameter itself evolves during coarse-graining. In the lattice gas RG framework discussed in Sec.~\ref{sec:mo6}, we consider both scenarios in a unified manner. First, we analyze the case in which the bond energy remains a non-renormalized, tunable parameter, thus maintaining a direct correspondence with the Monte Carlo simulations and the earlier RG treatment. Second, we extend the analysis to include the explicit renormalization of the bond energy, treating \(\varepsilon\) as a scale-dependent quantity that changes under the RG transformation. This latter approach provides a more complete and self-consistent description, as it accounts for the coupled evolution of both site occupation probability and interaction strength. The implications of these two treatments are examined and compared in detail in that section.
\section{Scaling exponents}\label{sec:CG1}
The scaling exponents of our model can be derived by relating it to a known loop model, as discussed in Sec.~\ref{sec:mo1}. We can expand the partition function in Eq.~\eqref{eq:yt1} in the vicinity of the limit as \(\varepsilon \to \infty\), focusing on the number of loops. In this limit, no loops are present as \(\varepsilon\) approaches infinity. We define a small expansion parameter as follows:
\begin{equation}
K \equiv e^{-\epsilon}, \qquad K \ll 1 \quad (\epsilon \to \infty).
\end{equation} The partition function in Eq.~\eqref{eq:yt1} can be re-written as:
\begin{equation}\label{eq:yt1aa}
Z = \sum_{s,\ell} \mathcal{N}(s,\ell)\, z^s K^{\ell},
\end{equation} 
where $\mathcal{N}(s,\ell)$ is the number of configurations with $s$ occupied sites and $l$ loops. Each loop contribute an extra factor of $e^{-\varepsilon}$ to the partition function in Eq.~\eqref{eq:yt1aa}, and $z= p e^{-\epsilon}$.
For $K \to 0$, only $\ell = 0$ contributes:
\begin{equation}
Z_0 = \sum_s \mathcal{N}(s,0)\, z^s.
\end{equation}
\begin{figure*}[ht!]
    \centering
    \includegraphics[keepaspectratio, width=0.95\textwidth]{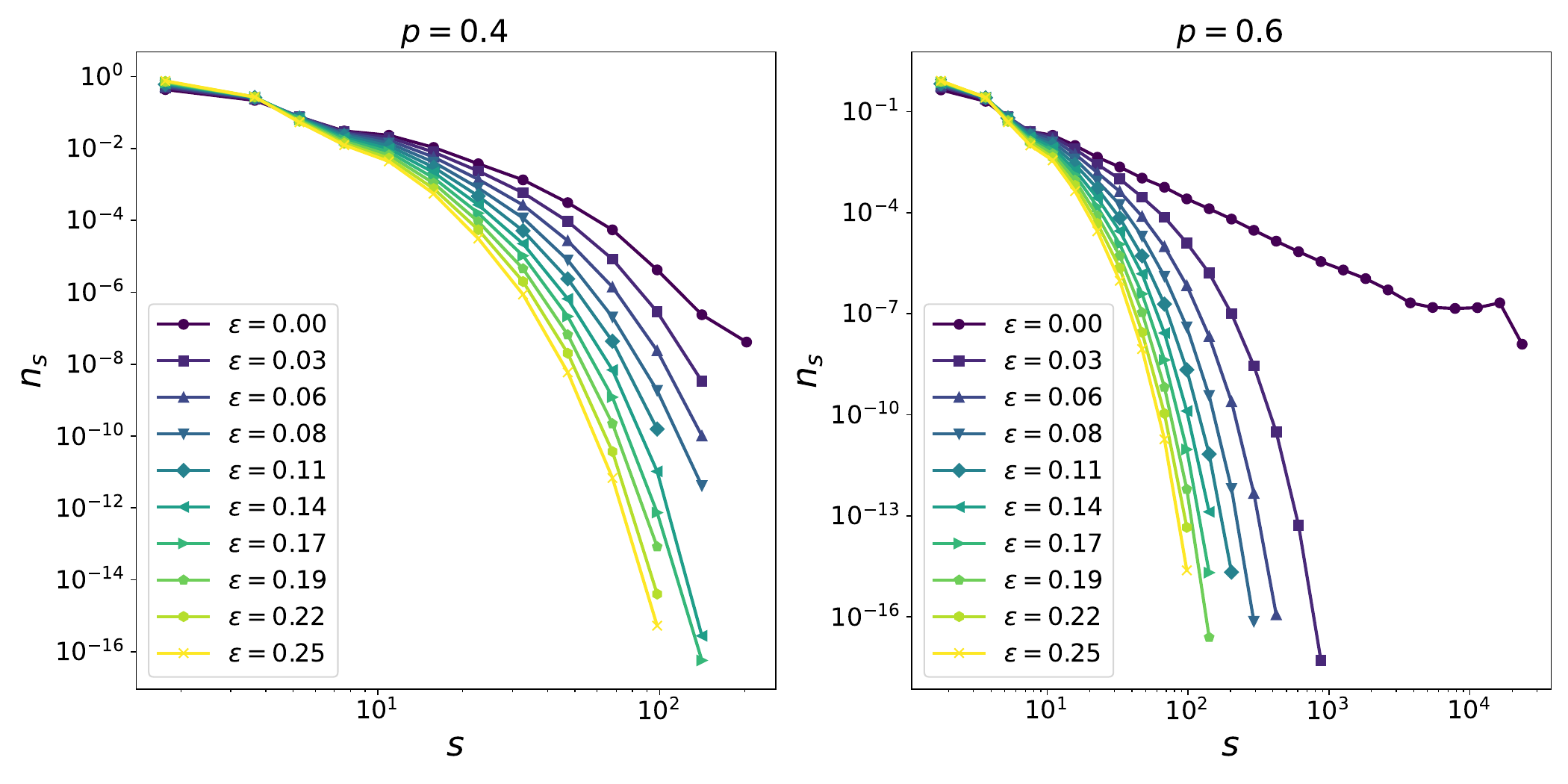}
     \caption{The plot illustrates the weighted cluster-size distribution for various values of the nearest-neighbor occupied-site bond energy cost, denoted by $\varepsilon$, while keeping the site-occupation probability fixed at $p$. The x-axis label $s$ is the cluster size. In both cases, it is evident that as $\varepsilon$ increases, the cluster size decreases.} 
    \label{fig:combined365iu1}
\end{figure*}
Expanding the partition function in terms of the number of loops $\ell$, one can obtain,
Expand:
\begin{equation}
Z = Z_0 + K Z_1 + K^2 Z_2+O(K^3),
\end{equation}
where:
\begin{equation}
Z_1 = \sum_s \mathcal{N}(s,1)\, z^s,
\end{equation}
and \begin{equation}
Z_2 = \sum_s \mathcal{N}(s,2)\, z^s,
\end{equation}
which are the one-loop and two-loop partition functions, respectively. The antiferromagnetic limit (as \(\varepsilon \to \infty\)) and the ferromagnetic limit (as \(\varepsilon \to -\infty\)) are particularly interesting. In the FK representation of the \(q\)-state Potts model, the partition function, as described in Eq.~\eqref{eq:yt1}, can be analyzed in the limits of \(\varepsilon \to 0\) and \(\varepsilon \to \pm \infty\). In these cases, it can be mapped to a loop model, whose critical behavior is characterized by a Coulomb gas (CG) or a conformal field theory with a dimensionless coupling constant in two dimensions (2D). This is because in the limit $\varepsilon \to \pm \infty$, the factor $\prod_{i\in C}e^{-\varepsilon T_{i}}$ in Eq.~\eqref{eq:yt1} becomes $e^{-\varepsilon C}$, where $C$ is the number of clusters in a particular configuration, since 
\begin{equation}
    \lim_{\varepsilon\to \infty}e^{-\varepsilon\sum_{i}T_{i}}=e^{-\varepsilon C}.
\end{equation}

\begin{figure}[t!]
    \centering
    \hspace{-1cm}
    \includegraphics[keepaspectratio, width=1.02\linewidth]{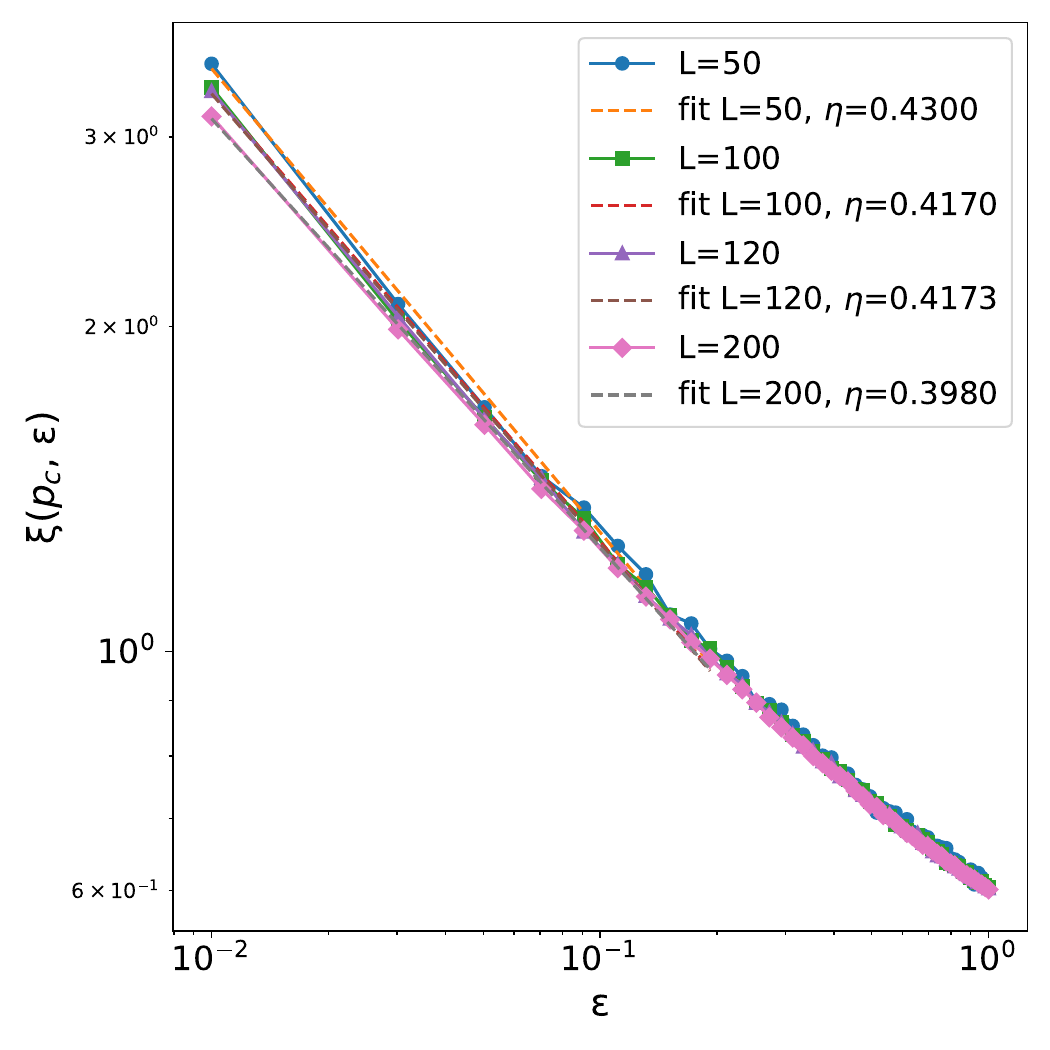}
     \caption{ Scaling of correlation length $\xi(\varepsilon)$ computed using Eq.~\eqref{eq:xi_energy_weighted} for site percolation on a square lattice, with open boundary condition (OBC) at critical percolation threshold (for $\varepsilon=0$) with varying energy cost $\varepsilon$. We fit the numerical values with $\varepsilon$ as $\xi(\varepsilon, p_c(\varepsilon=0))=A\varepsilon^{-\eta}$. The legend in the figure indicates the values of the exponent $\eta$ for different system sizes $L$.} 
    \label{fig:combined3658u1}
\end{figure}

The Coulomb gas is a massless free scalar field $\varphi(z,\bar z)$ with action \cite{Nienhuis1984}
\begin{equation}
    S = \frac{g}{4\pi} \int d^2r\, (\nabla \varphi)^2,
\end{equation}
where $g$ is the dimensionless coupling parameter, which acts like a stiffness constant controlling the strength of the logarithmic interaction between charges, in the Coulomb gas interpretation.  
The vertex operators are defined as \cite{DOTSENKO2001523,Kapec2021,PICCO2013719}:
\begin{equation}
    V_{\alpha}(z,\bar z) = e^{i\alpha \varphi(z,\bar z)}.
\end{equation}
The conformal weight $h(\alpha)$ of a vertex operator $V_{\alpha}$ in the presence of the background charge \cite{DiFrancesco1997CFT,DOTSENKO1984312} is
\begin{equation}
    h(\alpha) = \frac{\alpha(\alpha - 2\alpha_0)}{4g},
\end{equation}
and the full scaling dimension \cite{DiFrancesco1997CFT, JacobsenAIMEScomplete, DOTSENKO1984312} is:
\begin{equation}\label{eq:scd1}
    x(\alpha) = 2h(\alpha) = \frac{\alpha(\alpha - 2\alpha_0)}{2g}.
\end{equation}
The loop representation of the FK clusters assigns a weight $n$ to each closed loop.  
The Coulomb gas analysis gives the exact mapping \cite{Nienhuis1984}:
\begin{equation}
    n = -2 \cos(2\pi g),
\end{equation} where $g\in[0,1]$.
The total scaling dimension can be computed in terms of primary field indices $(r,s)$ as \cite{DOTSENKO1984312}:
\begin{equation}
    x_{r,s} = \frac{(r - s g)^2 - (1 - g)^2}{2g}.
    \label{eq:KacFormula}
\end{equation}
In two dimensions, the RG eigenvalue $y$ is related to the scaling dimension $x$ by
\begin{equation}
    y = 2 - x.
\end{equation}
The thermal operator corresponds to $(r,s) = (2,1)$ in the Kac table \cite{Kac1979Contravariant,PhysRevLett.52.1575,Belavin1984Infinite}.  
Substituting into Eq.~\eqref{eq:KacFormula},
\begin{equation}
\begin{aligned}
    x_{2,1}
    &= \frac{3}{2g} - 1.
\end{aligned}
\end{equation}
The RG eigenvalue \cite{PhysRevE.111.034108} is
\begin{equation}\label{eq:rgdst1}
    y_t = 2 - x_{2,1} = 3 - \frac{3}{2g}.
\end{equation}
The critical exponent $\nu$ is just the reciprocal of this RG eigenvalue, i.e., $\nu=\frac{1}{y_t}$.
This form reproduces the exponents of the Potts and percolation universality classes. 
For percolation ($\varepsilon = 0$), we have $n = 1$, which implies $g = \tfrac{2}{3}$, giving $y_t = \tfrac{3}{4}$ and therefore $\nu = \tfrac{4}{3}$. 
In dilute isolated cluster limit ($\varepsilon \to \infty$), we have $n = 0$, yielding $g = \tfrac{3}{4}$, which gives $y_t = 1$ and, consequently, $\nu = 1$. In the dense cluster limit, as \(\varepsilon\) approaches \(-\infty\), loops become abundant with a loop fugacity of \(n=2\). This results in the relation \(g=\frac{3}{2}\), which indicates a relevant eigenvalue of \(y_t=2\). As a consequence, the thermal exponent is computed as \(\nu=\frac{1}{2}\), which is consistent with the Ornstein-Zernike mean field result as well. Eq.~\eqref{eq:rgdst1} therefore interpolates naturally between dense ferromagnetic like cluster configuration ($\nu=\frac{1}{2}$), ordinary percolation ($\nu=4/3$) and isolated cluster configuration ($\nu=1$) as shown in Figs.~\ref{fig:RG_2x3_combined}(a), \ref{fig:RG_2x3_combined}(d), and \ref{fig:RG_2x3_combined}(f) corresponding to the isotropic case, where the parameter $\varepsilon$ is constant and identical in both the horizontal and vertical directions and Figs.~\ref{fig:RGan_1}(b) and \ref{fig:RGan_1b}(b) corresponding to the anisotropic case, where $\varepsilon$ is constant but takes different values in the horizontal and vertical directions and can also explain the finite behaviour of correlation length shown in Fig.~\ref{fig:combined3658u1}, obtained using the expression for correlation length written in Canonical ensemble, Eq.~\eqref{eq:xi_energy_weighted} and using lattice gas Monte Carlo simulations as shown in Fig.~\ref{fig:xi_eps_2x1}. 
A thorough description of the continuous transition between the two extremes as $\varepsilon$ varies is of interest; however we do not address this in this article. 
\section{Correlation Length and Cluster Size Statistics}\label{sec:mo2}
\subsection{Energy Weighted Correlation Length}
In ordinary site percolation on a square lattice, the critical point $p_c \approx 0.5927$ \cite{Mertens_2022} separates a phase with only finite clusters ($p<p_c$) from one in which an infinite connected cluster exists ($p>p_c$).  The value of $p_c$ in the case of bond percolation is analytically derived as $p_c=0.5$ \cite{bollobas2006sharp, PJReynolds_1977}.
The divergence of the correlation length near the critical point characterizes the critical behavior,
\[
\xi(p) \sim |p-p_c|^{-\nu}, \qquad \nu=\tfrac{4}{3} \;\; \text{in 2D}.
\]
In a walk-based approach, we define the disorder-averaged generating function as:
\begin{equation}\label{eq:mn1}
\overline{F(R;\alpha,p)} = \sum_{T=0}^{T_{\max}} \alpha^{-T} \overline{N_T(R,p)},
\end{equation}
where $\overline{N_T(R,p)}$ counts the number of walks of length $T$ that end at Euclidean distance $R$ from the origin, averaged over disorder realizations, and $\alpha=e^{\varepsilon}$. The same partition function for the case $p=1$ can be written as:
\begin{equation}\label{eq:cann1}
    F(R,\alpha)=\sum_{T}N_{T}(R)\alpha^{-T},
\end{equation}
The parameter $\alpha$ suppresses long walks: for $\alpha> \alpha_c$ (the critical threshold for energy cost) the series converges, whereas as $\alpha \to \alpha_c^+$ the suppression weakens and long walks proliferate. 
Thus, $\alpha$ plays the role of a second control parameter in addition to the percolation probability $p$. The partition function presented here is written in a grand canonical formalism, which allows for fluctuations in site occupancy across different realizations. This introduces the fugacity parameter in the average \(\overline{N_{T}(R,p)}\). In contrast, the partition function defined in Eq.~\eqref{eq:cann1} follows a canonical formalism, where no such fluctuations
occur across realizations. From the pair connectedness distribution, $G(R)$, obtained from the partition function, one can define the second-moment correlation length \cite{Pelissetto_2002} as:
\begin{equation}\label{eq:secondmoment1}
\xi_2^2 = \frac{1}{2d} \frac{\mu_2}{\mu_0}, 
\quad \mu_n = \sum_R R^n G(R),
\end{equation}
which coincides with the standard second-moment definition of correlation length used in statistical mechanics.  \par
In standard site percolation on a lattice, each site is independently occupied with
probability \(p\),
and the geometry of clusters is therefore purely random. Let \(k\) index the connected clusters of occupied sites in a given configuration.
Each cluster \(k\) contains \(s_k\) occupied sites and \(T_k\) bonds between nearest neighbor occupied sites . We now generalize the percolation problem by assigning an energy cost \(\varepsilon\) to each bond between occupied sites. A cluster with \(T_k\) bonds then carries a total energy:
\begin{equation}
E_{k} = \varepsilon \, T_k.
\end{equation}

\begin{figure}[ht!]
    \centering
    \hspace{-1cm}
    \includegraphics[keepaspectratio, width=1.02\linewidth]{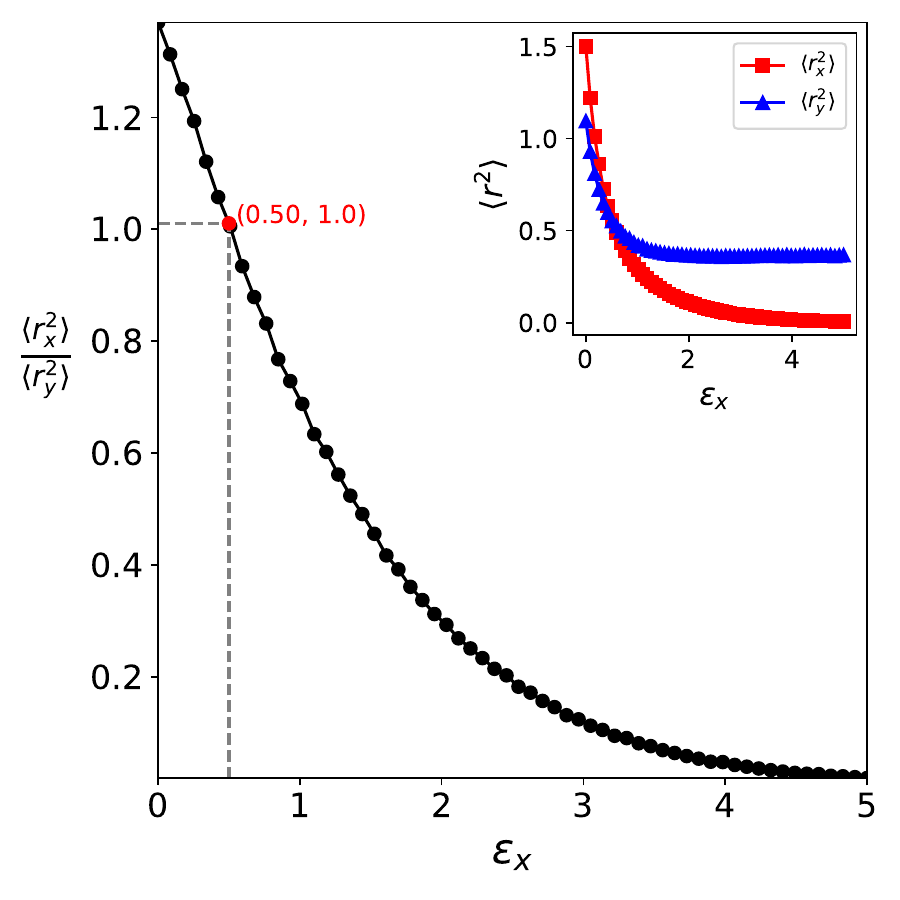}
     \caption{The plot shows the ratio of the mean-squared radii in the \( x \) and \( y \) directions for the energy-weighted cluster ensemble. This ratio is shown relative to the anisotropic energy costs, with \(\varepsilon_x\) denoting the cost in the \( x \) direction and \(\varepsilon_y = 0.5\) in the \( y \) direction. The inset illustrates the mean square radius in the \( x \) and \( y \) directions. The site-occupation probability is fixed at \( p = 0.6 \), and the system size is \(N = 1600 \). The mean square radius in both directions is computed using Eq.~\eqref{eq:xi_standard} and the equilibration is reached by performing a Glauber-like Metropolis sweep. At $\varepsilon_x=0.5$, the two inset curves of mean-squared radii in the \( x \) and \( y \) directions cross each other. This indicates that at $\varepsilon_x=\varepsilon_y=\varepsilon=0.5$, the anisotropic ratio is equal to $1$, as can be observed in the main plot.} 
    \label{fig:Aniso1}
\end{figure}

Clusters with many bonds (large $T_k$) are thus energetically penalized. In statistical mechanical language, for each cluster labelled as $k$, there is a total of $s_{k}$ number of microstate degeneracies contributing to the same energy $E_{k} = \varepsilon \, T_k$, so the probability for a randomly chosen site in a system to belong to cluster 
$k$ is:
\begin{equation}\label{eq:Bolz_cs1}
    P_{\varepsilon}^{k}=\frac{s_ke^{-E_{k}}}{\sum_{k}s_ke^{-E_{k}}},
\end{equation}
where $s_k$ is the count of microstates or degeneracy associated with energy state $k$, and $e^{-E_k}$ is the Boltzmann weight associated with the canonical ensemble of clusters or energy states. 
Here $E_k$ is the total energy of the energy state $k$ or $E_{\mathrm{cluster},k}$ for cluster $k$.
Here \(\beta\), the inverse temperature parameter is set as $1$.
This construction introduces an additional control parameter~\(\varepsilon\)
that suppresses large connected clusters.
In the absence of an energy penalty ($\varepsilon = 0$), the system reduces to standard percolation, whereas for finite $\varepsilon > 0$ the formation of large clusters is exponentially suppressed; in the strongly penalized regime $\varepsilon \gg \varepsilon_c$, the system is therefore dominated by small, isolated clusters.
Above a critical value \(\varepsilon_c\), the correlation length becomes finite even if \(p \approx p_c\). Here $p_c$ denotes the critical percolation threshold for $\varepsilon=0$. In classical percolation, the correlation length is obtained from
cluster statistics as:
\begin{equation}
\xi^2
= \frac{\sum_k s_k \, \langle r^2 \rangle_k}{\sum_k s_k},
\label{eq:xi_standard}
\end{equation}
where, \(\langle r^2 \rangle_k\), the mean-square distance of sites
within cluster~\(k\) from its center of mass is defined as:
\begin{equation}
\langle r^2 \rangle_k
= \frac{1}{s_k} \sum_{i=1}^{s_k}
|\mathbf{r}_{i} - \mathbf{r}_{\mathrm{cm},k}|^2 ,
\qquad
\mathbf{r}_{\mathrm{cm},k} = \frac{1}{s_k}\sum_{i=1}^{s_k} \mathbf{r}_i,
\end{equation} and the weighting is done in terms of cluster size or number of sites in the cluster. This results from the probability
that the chosen site belongs to cluster \(k\) is
\begin{equation}
\mathbb P(\text{site}\in k) = \frac{s_k}{\sum_j s_j},
\end{equation} and the correlation length is just the mean of the mean-square distance of sites within a cluster. 
\par
The correlation length result presented in Eq.~\eqref{eq:xi_standard} can be readily reduced to the standard definition found in the literature \cite{Stauffer1979Percolation}, which involves the number of clusters of size \( s \) per site, denoted as \( n_s(p) \) or the cluster size strength. We can reorganize the summation over clusters into a summation over cluster sizes as:
\begin{equation}
\sum_k s_k
=
\sum_s \sum_{k=1}^{N_s} s
=
\sum_s s N_s,
\end{equation} where, $N_s$ is the total of clusters of size $s$.
Using $N_s =Nn_s$, where $N$ is the total number of clusters of all sizes, this can be written as:
\begin{equation}\label{eq:tg1}
\sum_k s_k
=
N \sum_s s n_s.
\end{equation}
Hence, the radius of gyration squared summed over each cluster can be written as:
\begin{equation}
\sum_k s_k \langle r^2 \rangle_k
=
\sum_s \sum_{k=1}^{N_s}
s \langle r^2 \rangle_k.
\end{equation}
We next define the average radius of gyration squared for clusters of size $s$:
\begin{equation}
\langle R^2 \rangle_s
=
\frac{1}{N_s}
\sum_{k=1}^{N_s}
\langle r^2 \rangle_k.
\end{equation}
Then,
\begin{equation}
\sum_{k=1}^{N_s}
s \langle r^2 \rangle_k
=
s N_s \langle R^2 \rangle_s.
\end{equation}
Thus, the radius of gyration squared summed over each cluster can be written as:
\begin{equation}\label{eq:tg2}
\sum_k s_k \langle r^2 \rangle_k
=
\sum_s s N_s \langle R^2 \rangle_s
=
N \sum_s s n_s \langle R^2 \rangle_s.
\end{equation}
Substituting the expressions in Eq.~\eqref{eq:tg1} and Eq.~\eqref{eq:tg2} into Eq.~\eqref{eq:xi_standard}, we can obtain the standard expression for correlation length as:
\begin{equation}
\xi^2
=
\frac{
N \sum_s s n_s \langle R^2 \rangle_s
}{
N \sum_s s n_s
}=\frac{
\sum_s s\, n_s\, \langle R^2 \rangle_s
}{
\sum_s s\, n_s
},
\end{equation}
where the sum runs over finite clusters only. \par
\begin{figure}[t!]
    \centering
    \hspace{-1cm}
    \includegraphics[keepaspectratio, width=0.52\textwidth]
    {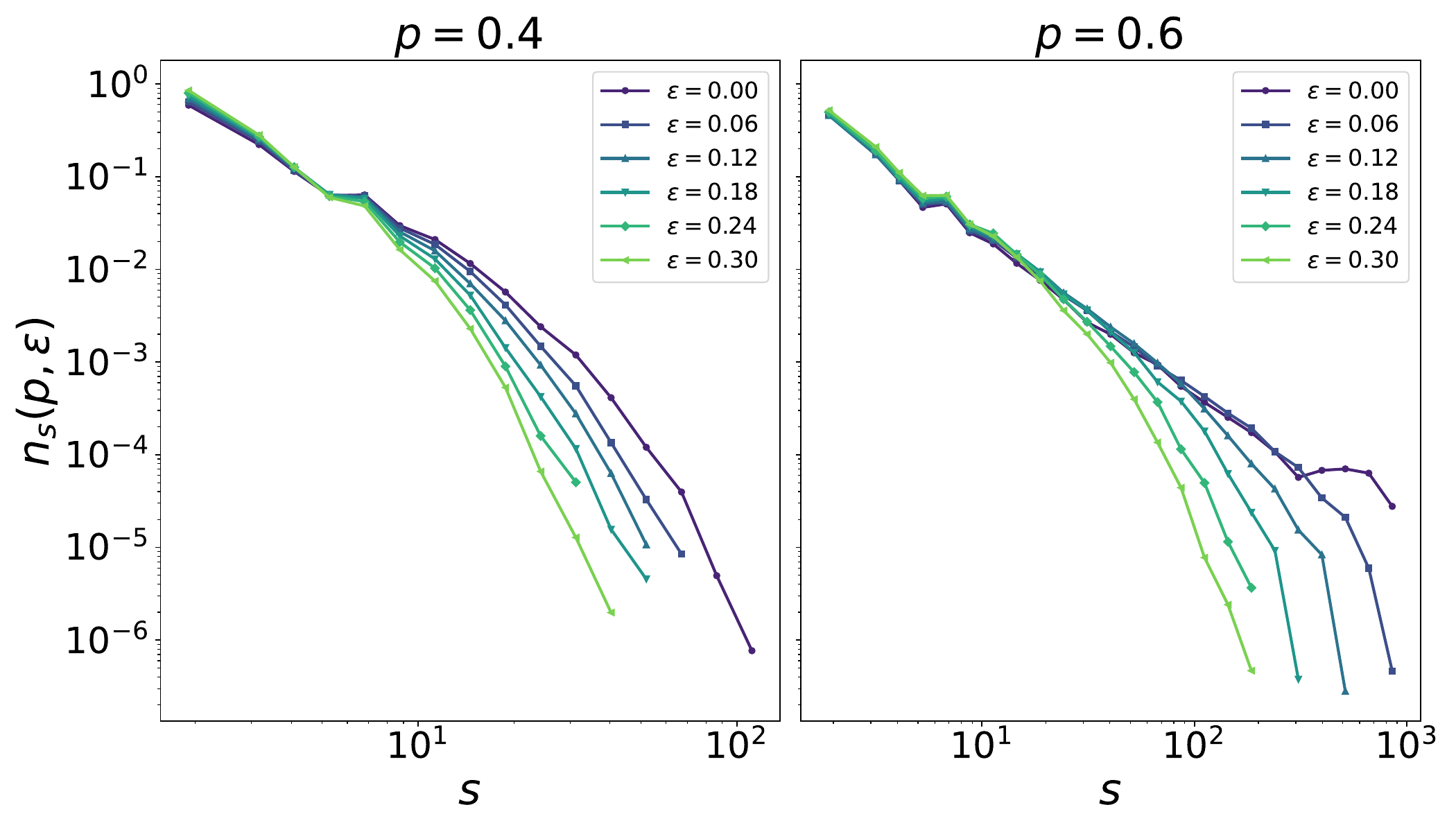}
    \put(-260,150){\textbf{(a) }}
    \\
    \centering
    \hspace{-1cm}
    \includegraphics[keepaspectratio, width=0.52\textwidth]{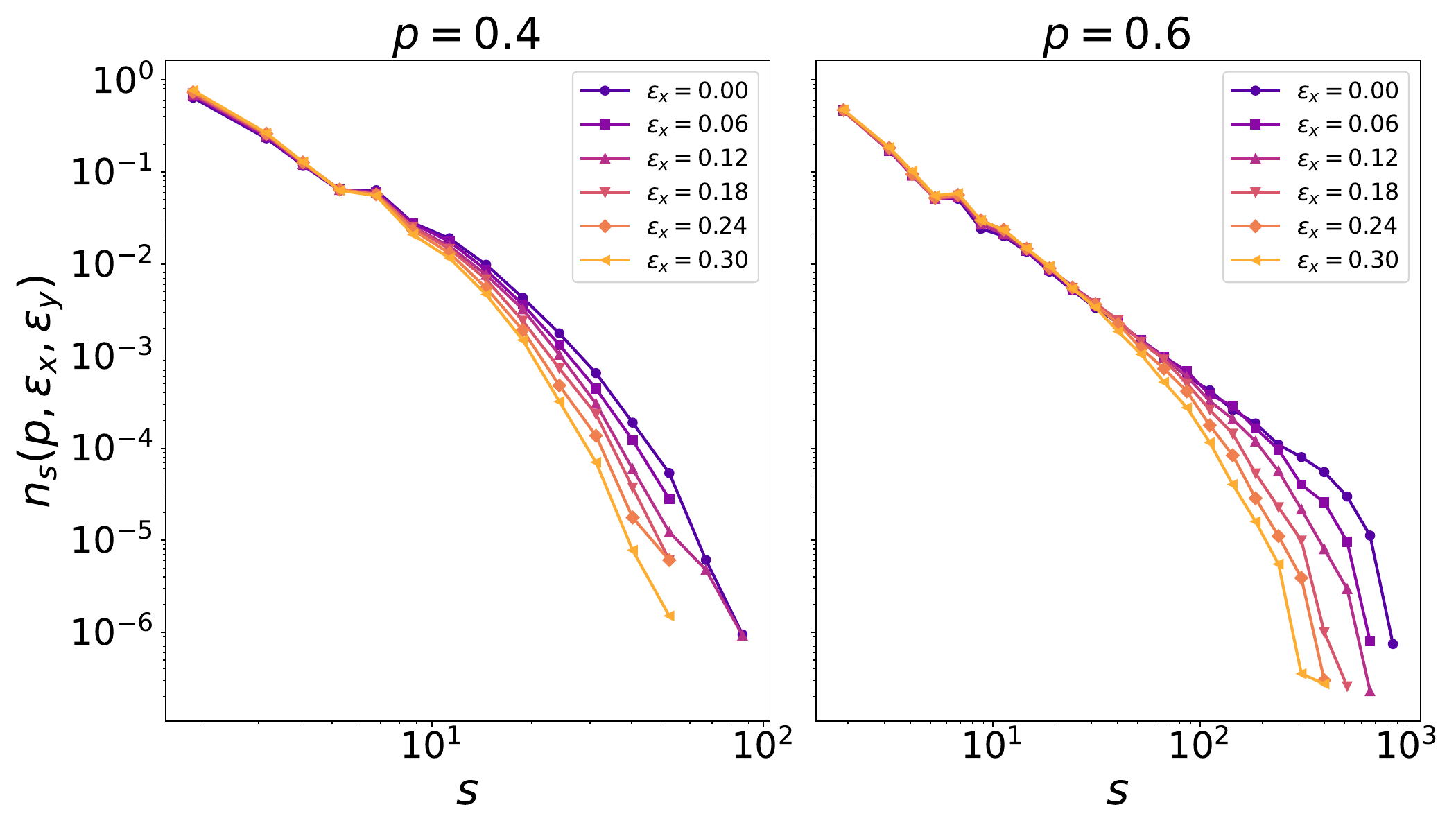}
     \put(-260,150){\textbf{(b)}}
     \\
\caption{Cluster-size distributions obtained from Glauber-like Metropolis sweep as described in Sec.~\ref{Sec:moo9} for a square lattice of size $N=1600$.
The top figure \textbf{(a)} shows the distributions for isotropic bond energy costs $\varepsilon$
for different site occupation probabilities $p=0.4$ and $p=0.6$.
The bottom figure \textbf{(b)} shows the corresponding distributions for anisotropic bond energy
costs, with $\varepsilon_x$ and $\varepsilon_y$ associated with bonds along
the $x$ and $y$ directions, respectively. The bond energy cost along the 
$y$-direction is taken as $\varepsilon_y=0.1$. The bond energy values along the $x$ direction are indicated in the legends.}
\label{fig:csd_combined_2x1}
\end{figure}

The correlation length can also be expressed in terms of the pair-connectedness function. The connectedness function \(G(\mathbf r)\) is defined as the probability that two sites
separated by vector \(\mathbf r\) lie in the same cluster.

The second-moment correlation
length is often defined by as Eq.~\eqref{eq:secondmoment1}:
\begin{equation}
\xi^2 \;\; =\frac{\sum_{\mathbf r} |\mathbf r|^2 G(\mathbf r)}
{\sum_{\mathbf r} G(\mathbf r)},
\label{eq:xi_pair}
\end{equation}
where \(d\) is the spatial dimension.
We now relate this to cluster sums for a single configuration. Summing over all sites
pairs in a given configuration, one can have:
\begin{equation}
\sum_{i,j} \mathbf 1_{ij}
= \sum_k \sum_{i\in k}\sum_{j\in k} 1
= \sum_k s_k^2,
\end{equation}
and
\begin{equation}\label{eq:corrsum11}
\sum_{i,j} |\mathbf r_j-\mathbf r_i|^2\,\mathbf 1_{ij}
= \sum_k \sum_{i\in k}\sum_{j\in k} |\mathbf r_j-\mathbf r_i|^2,
\end{equation}
and performing the summation in Eq.~\eqref{eq:corrsum11} over \(i\in k\) and \(j\in k\), \cite{Stauffer1979Percolation}:
\begin{equation}
\sum_{i\in k}\sum_{j\in k} |\mathbf r_j-\mathbf r_i|^2
= 2\,s_k \sum_{i\in k} |\mathbf r_i-\mathbf r_{\mathrm{cm},k}|^2
= 2\,s_k\cdot s_k\cdot \langle r^2\rangle_k.
\end{equation}

Therefore, inserting these into Eq.~\eqref{eq:xi_pair}, yields a cluster representation with \(s_k^2\) weights:
\begin{equation}
\xi^2 (\varepsilon=0)=
\frac{\sum_k s_k^2\,\langle r^2\rangle_k}{\sum_k s_k^2}.
\end{equation}
When the energy cost per link is introduced,
each cluster is weighted by its Boltzmann factor \(e^{-\varepsilon T_k}\) following Eq.~\eqref{eq:Bolz_cs1},
yielding the generalized, energy-weighted correlation length per realization of occupation probability ($p$):
\begin{equation}
\xi^2(\varepsilon)
= \frac{\displaystyle \sum_k s_k \, e^{-\varepsilon T_k} \, \langle r^2 \rangle_k}
        {\displaystyle \sum_k s_k \, e^{-\varepsilon T_k}} .
\label{eq:xi_energy_weighted}
\end{equation}
This definition reduces to the usual percolation result~\eqref{eq:xi_standard} when
\(\varepsilon = 0\). The above result Eq.~\eqref{eq:xi_energy_weighted} is the energy-weighted correlation length for a particular realization, which can be treated in the canonical formalism. 
For an infinite system size, near the critical occupation probability $p_c(\varepsilon)$, the correlation length $\xi(\varepsilon)$ diverges as \cite{Stauffer1979Percolation}:
\begin{equation}
    \xi(\varepsilon)\sim |p-p_c(\varepsilon)|^{-\nu(\varepsilon)}.
\end{equation}
For finite system size $L$,
\begin{equation}
    \xi(p,\varepsilon)\sim L \sim |p-p_c(\varepsilon)|^{-\nu(\varepsilon)},
\end{equation} which implies 
\begin{equation}
    |p-p_c(\varepsilon)|L^{\frac{1}{\nu(\varepsilon)}}\sim 1.
\end{equation} 
\begin{figure}[ht!]
    \centering
    \hspace{-1cm}
    \includegraphics[keepaspectratio, width=1.02\linewidth]{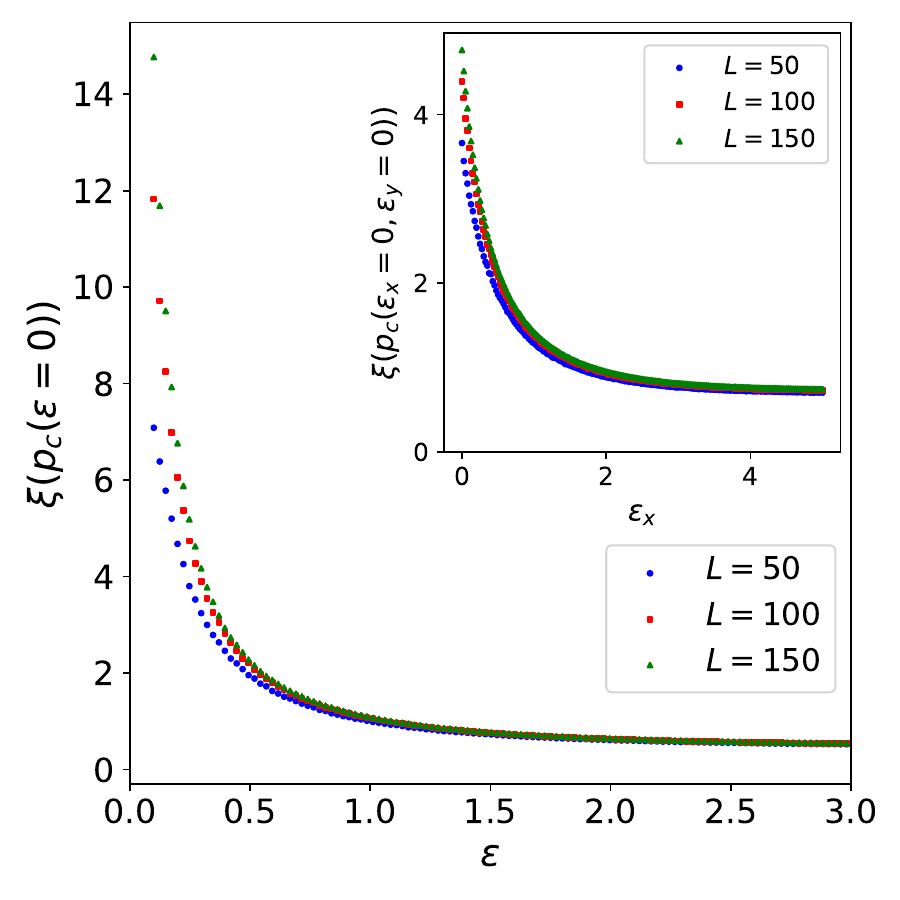}
     \caption{Correlation length for site percolation on a square lattice with open boundary conditions at the critical percolation threshold $p_c(\varepsilon=0)$, computed using Eq.~\eqref{eq:xi_standard}.
The main plot shows the variation of $\xi(p_c(\varepsilon=0))$ with the energy cost
$\varepsilon$ for different system sizes $L$.
The inset plot shows the variation of $\xi(p_c(\varepsilon_x=0,\varepsilon_y=0))$ with anisotropic energy cost
$\varepsilon_x$ for fixed $\varepsilon_y=0.5$ and different $L$.
The data are obtained from Monte Carlo simulations in the grand canonical
ensemble as described in Sec.~\ref {Sec:moo9}. The system size used is $N=1600$.} 
    \label{fig:xi_eps_2x1}
\end{figure}
The plot of the ratio $\frac{\langle r_{x}^{2}\rangle}{\langle r_{x}^{2}\rangle}$, computed using Eq.~\eqref{eq:xi_energy_weighted}, as shown in Fig.~\ref{fig:Aniso1}, indicates that as the anisotropy in bond energy increases, the clusters effectively become one-dimensional. Summing Eq.~\eqref{eq:xi_energy_weighted} over the fluctuations of site occupation across each realization will yield the complete energy-weighted correlation length in a grand canonical ensemble, similar in fashion to that which can be obtainable from the partition function expressed in Eq.~\eqref{eq:mn1}. 
\subsection{Energy Weighted Cluster Size Statistics}
For \(\varepsilon=0\), large clusters contribute significantly,
leading to a diverging correlation length as \(p \to p_c(\varepsilon=0)\).
As \(\varepsilon\) increases, the contribution of large clusters is exponentially
suppressed, and the effective correlation length \(\xi(\varepsilon)\) decreases sharply.
For \(\varepsilon > \varepsilon_c(p)\),
the system is dominated by small clusters and
\(\xi(\varepsilon)\) becomes finite even at \(p = p_c(\varepsilon=0)\), as shown in Fig.~\ref{fig:xi_eps_2x1}.
Near the percolation threshold \( p_c \), the cluster size distribution follows \cite{Stauffer1979Percolation}:
\begin{equation}
n_s(p) \sim s^{-\tau} e^{-s/s^*(p)} ,
\end{equation}
where the  size $s^*(p)$ scales as
\begin{equation}\label{eq:cc1}
s^*(p) \sim |p - p_c|^{-1/\sigma} ,
\end{equation} 
where the critical exponent is calculated as $\sigma=\frac{1}{\nu D}$, with $D$ as the fractal dimension of the lattice, and $\nu$ as defined previously, the critical exponent for correlation length. 
Now, suppose each bond between nearest-neighbor occupied sites carries an additional energy cost 
\(\varepsilon\).
Then a cluster of size \( s \) acquires a Boltzmann weight 
\( e^{-\varepsilon s} \).
The modified cluster size distribution takes the form
\begin{equation}\label{eq:Ewccs}
n_s(p, \varepsilon) 
\sim s^{-\tau} 
e^{-s/s^*(p, \varepsilon)} ,
\end{equation}
where the effect of bond energy cost impacts only the cut-off cluster size; thus, the overall cluster size distribution retains the same form.
From Fig.~\ref{fig:combined365iu1} one can verify that the energy-dependent cut-off cluster size \( s^*(p, \varepsilon) \) decreases with increasing bond energy cost $\varepsilon$. In the full grand canonical ensemble, as shown in Fig.~\ref{fig:csd_combined_2x1}, the cluster size cut-off diverges near the critical occupation threshold $p_c(\varepsilon)$ as:
\begin{equation}
    s^*(p, \varepsilon)\sim |p-p_c(\varepsilon)|^{-\frac{1}{\sigma(\varepsilon)}},
\end{equation} where $\sigma(\varepsilon)$ is the cut-off cluster size critical exponent as mentioned before. In this article, we have not included the observed values for \(\sigma(\varepsilon)\) at varying bond energy costs \(\varepsilon\). This is because we have already presented the relationship between the thermal exponent \(\nu(\varepsilon)\) and the bond energy cost \(\varepsilon\). Additionally, the cluster cut-off size critical exponent is not an independent critical exponent; it inversely depends on the thermal critical exponent by a constant proportionality factor, which is a function of the fractal dimension \(D\), as specified earlier.
\section{Renormalization Group Analysis for the Lattice gas}\label{sec:mo6}
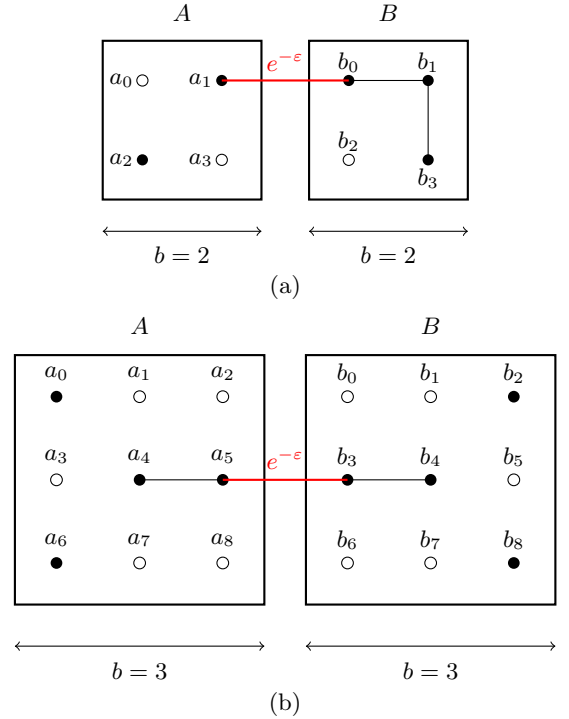
\begin{figure}[ht!]
\centering
\begin{tikzpicture}[scale=1.05]
\def\b{2}
\def\gap{0.6}
\draw[thick] (0,0) rectangle (\b,\b);
\node at (\b/2,\b+0.35) {$A$};
\draw[thick] (\b+\gap,0) rectangle (2*\b+\gap,\b);
\node at (\b+\gap+\b/2,\b+0.35) {$B$};
\draw (0.5,1.5) circle (2pt);
\node[left] at (0.5,1.5) {$a_0$};
\fill (1.5,1.5) circle (2pt);
\node[left] at (1.5,1.5) {$a_1$};
\fill (0.5,0.5) circle (2pt);
\node[left] at (0.5,0.5) {$a_2$};
\draw (1.5,0.5) circle (2pt);
\node[left] at (1.5,0.5) {$a_3$};
\fill (\b+\gap+0.5,1.5) circle (2pt);
\node[above] at (\b+\gap+0.5,1.5) {$b_0$};
\fill (\b+\gap+1.5,1.5) circle (2pt);
\node[above] at (\b+\gap+1.5,1.5) {$b_1$};
\draw (\b+\gap+0.5,0.5) circle (2pt);
\node[above] at (\b+\gap+0.5,0.5) {$b_2$};
\fill (\b+\gap+1.5,0.5) circle (2pt);
\node[below] at (\b+\gap+1.5,0.5) {$b_3$};
\draw (\b+\gap+0.5,1.5) -- (\b+\gap+1.5,1.5);
\draw (\b+\gap+1.5,1.5) -- (\b+\gap+1.5,0.5);
\draw[red, thick] (1.5,1.5) -- (\b+\gap+0.5,1.5);
\node[red] at (2.30,1.75) {$e^{-\varepsilon}$};
\draw[<->] (0,-0.4) -- (\b,-0.4);
\node at (\b/2,-0.7) {$b=2$};
\draw[<->] (\b+\gap,-0.4) -- (2*\b+\gap,-0.4);
\node at (\b+\gap+\b/2,-0.7) {$b=2$};
\node at (\b+\gap/2,-1.1) {(a)};
\end{tikzpicture}
\vspace{0.01cm}
\begin{tikzpicture}[scale=1.1]
\def\b{3}
\def\gap{0.5}
\draw[thick] (0,0) rectangle (\b,\b);
\node at (\b/2,\b+0.35) {$A$};
\draw[thick] (\b+\gap,0) rectangle (2*\b+\gap,\b);
\node at (\b+\gap+\b/2,\b+0.35) {$B$};
\foreach \x/\y/\label in {
0.5/2.5/a_0,
1.5/2.5/a_1,
2.5/2.5/a_2,
0.5/1.5/a_3,
1.5/1.5/a_4,
2.5/1.5/a_5,
0.5/0.5/a_6,
1.5/0.5/a_7,
2.5/0.5/a_8}
{
\node at (\x,\y+0.28) {$\label$};
}
\fill (0.5,0.5) circle (2pt);
\fill (1.5,1.5) circle (2pt);
\fill (2.5,1.5) circle (2pt);
\fill (0.5,2.5) circle (2pt);
\draw (1.5,0.5) circle (2pt);
\draw (2.5,0.5) circle (2pt);
\draw (0.5,1.5) circle (2pt);
\draw (1.5,2.5) circle (2pt);
\draw (2.5,2.5) circle (2pt);
\draw (1.5,1.5) -- (2.5,1.5);
\foreach \x/\y/\label in {
4.0/2.5/b_0,
5.0/2.5/b_1,
6.0/2.5/b_2,
4.0/1.5/b_3,
5.0/1.5/b_4,
6.0/1.5/b_5,
4.0/0.5/b_6,
5.0/0.5/b_7,
6.0/0.5/b_8}
{
\node at (\x,\y+0.28) {$\label$};
}
\fill (4.0,1.5) circle (2pt);
\fill (5.0,1.5) circle (2pt);
\fill (6.0,2.5) circle (2pt);
\draw (5.0,0.5) circle (2pt);
\draw (4.0,0.5) circle (2pt);
\fill (6.0,0.5) circle (2pt);
\draw (4.0,2.5) circle (2pt);
\draw (5.0,2.5) circle (2pt);
\draw (6.0,1.5) circle (2pt);
\draw (4.0,1.5) -- (5.0,1.5);
\draw[red, thick] (2.5,1.5) -- (4.0,1.5);
\node[red] at (3.25,1.75) {$e^{-\varepsilon}$};
\draw[<->] (0,-0.5) -- (\b,-0.5);
\node at (\b/2,-0.8) {$b=3$};
\draw[<->] (\b+\gap,-0.5) -- (2*\b+\gap,-0.5);
\node at (\b+\gap+\b/2,-0.8) {$b=3$};
\node at (3.25,-1.2) {(b)};
\end{tikzpicture}
\caption{Schematic diagram showing two neighboring Kadanoff blocks, labeled \(A\) and \(B\). 
Filled circles denote occupied sites (weight \(e^\mu\)), open circles denote empty sites, and bonds carry weight \(e^J=e^{-\epsilon}\). 
The red bond illustrates an inter-block connection contributing to coarse-grained bond energy. 
(a) Block size \(b=2\). (b) Block size \(b=3\).}
\label{fig:lattgas_rg_blocks}
\end{figure}
We have a site percolation on a square lattice where each bond between two nearest-neighbor occupied sites has an energy cost $\varepsilon$. A cluster of $T$ bonds contributes weight
\begin{equation}
    W_T = e^{-\varepsilon T}.
\end{equation}

Now we introduce fugacity $e^{\mu} = p/(1-p)$ for sites and $e^J = e^{-\varepsilon}$ for bonds. Consider two nearest neighbor Kadanoff blocks $A$ and $B$ (after rescaling the original lattice into blocks of linear dimensions $b$) \cite{Kadanoff1966Scaling,PhysRevB.4.3174,Wilson1971RGCP2}, the weight of a block $A$ is given as:
\begin{equation}
W_A = e^{\mu k_A + J b_A}, \quad k_A = \sum_{i\in A}a_i,\ b_A = \sum_{\langle ij\rangle\in A}a_i a_j.
\end{equation}
The sites of the $A$ block (with $b=2$) are numbered as ${0,1,2,3}$ as shown in Fig.~\ref{fig:lattgas_rg_blocks}, and the internal bonds are given by ${(0,1), (0,2), (1,3), (2,3)}$ for a fully occupied block.
Fixing the right-column boundary sites $(a_1,a_3)$ and sum over the rest sites $(a_0,a_2)$ gives the weights corresponding to boundary sites:
\begin{equation}
\begin{aligned}
Q(a_1,a_3) 
&= \sum_{a_0,a_2\in\{0,1\}} e^{\mu k_A + J b_A}, \\
k_A &= a_1 + a_3 + a_0 + a_2, \\
b_A &= a_1 a_3 + a_0 a_1 + a_0 a_2 + a_2 a_3,
\end{aligned}
\end{equation}
Here $a_0$, $a_1$, $a_2$ and $a_3$ label the occupancy variables for sites $0$, $1$, $2$ and $3$ for $A$ block respectively as shown in Fig.~\ref{fig:lattgas_rg_blocks}.
Explicitly evaluating the weights gives:
\begin{equation}
\begin{aligned}
Q(0,0) &= 2 e^{\mu} + e^{2\mu} e^{J}, \\
Q(1,0) &= Q(0,1) = e^{\mu} + e^{2\mu}(1+e^J) + e^{3\mu} e^{2J}, \\
Q(1,1) &= e^{2\mu} e^J + 2 e^{3\mu} e^{2J} + e^{4\mu} e^{4J}.
\end{aligned}
\end{equation}
We denote $Q(0,0)=Q_{00}$, $Q(1,0)=Q(0,1)=Q_1$, and $Q(1,1)=Q_{11}$ for the sake of convenience.
The combined weights of two nearest neighbor blocks $A$ and $B$, denoted as $W_{N_{A}, N_{B}}$, where $N_{A}$ and $N_{B}$ denotes the occupancy variables of blocks $A$ and $B$ respectively. The weights can be computed as:
\begin{equation}
W_{00} = 1,
\end{equation}
\begin{equation}\label{eq:wei1}
\begin{aligned}
W_{10} &= Q_{00}+2Q_1 + Q_{11} =4e^{\mu}+2e^{2\mu}+4e^{2\mu}e^{J} \\
&\quad +4e^{3\mu}e^{2J}+e^{4\mu}e^{4J}.
\end{aligned}
\end{equation}
\begin{equation}\label{eq:weii1}
W_{11} = \sum_{a_1,a_3,b_0,b_2} Q(a_1,a_3) Q(b_0,b_2) e^{J(a_1 b_0 + a_3 b_2)},
\end{equation}
Grouping the terms in the above expression by boundary configurations, the expression simplifies to:
\begin{equation}
    \begin{aligned}
W_{11} &= Q_{00} (Q_{00}+2Q_1+Q_{11}) \\
&\quad + 2Q_1(Q_{00} + Q_1(1+e^J) + Q_{11}e^J) \\
&\quad + Q_{11}(Q_{00} + 2Q_1 e^J + Q_{11} e^{2J})
\end{aligned}.
\end{equation}
Substituting the $Q$ values and collecting terms for each power of $e^{\mu}$ and $e^J$ yields a finite polynomial expansion as:
\begin{equation}\label{eq:weib1}
\begin{aligned}
W_{11} ={}&
14 e^{2\mu}
+ 2 e^{2\mu} e^{J}
+ 12 e^{3\mu}
+ 28 e^{3\mu} e^{J}
+ 8 e^{3\mu} e^{2J} \\
&+ 2 e^{4\mu}
+ 10 e^{4\mu} e^{J}
+ 37 e^{4\mu} e^{2J}
+ 18 e^{4\mu} e^{3J}
+ e^{4\mu} e^{4J} \\
&+ 4 e^{5\mu} e^{2J}
+ 24 e^{5\mu} e^{3J}
+ 20 e^{5\mu} e^{4J}
+ 8 e^{5\mu} e^{5J} \\
&+ 2 e^{6\mu} e^{4J}
+ 16 e^{6\mu} e^{5J}
+ 8 e^{6\mu} e^{6J}
+ 2 e^{6\mu} e^{7J} \\
&+ 4 e^{7\mu} e^{7J}
+ 4 e^{7\mu} e^{8J}
+ e^{8\mu} e^{10J}.
\end{aligned}
\end{equation}
\begin{figure}[ht!]
    \centering
    \includegraphics[keepaspectratio, width=0.49\textwidth]{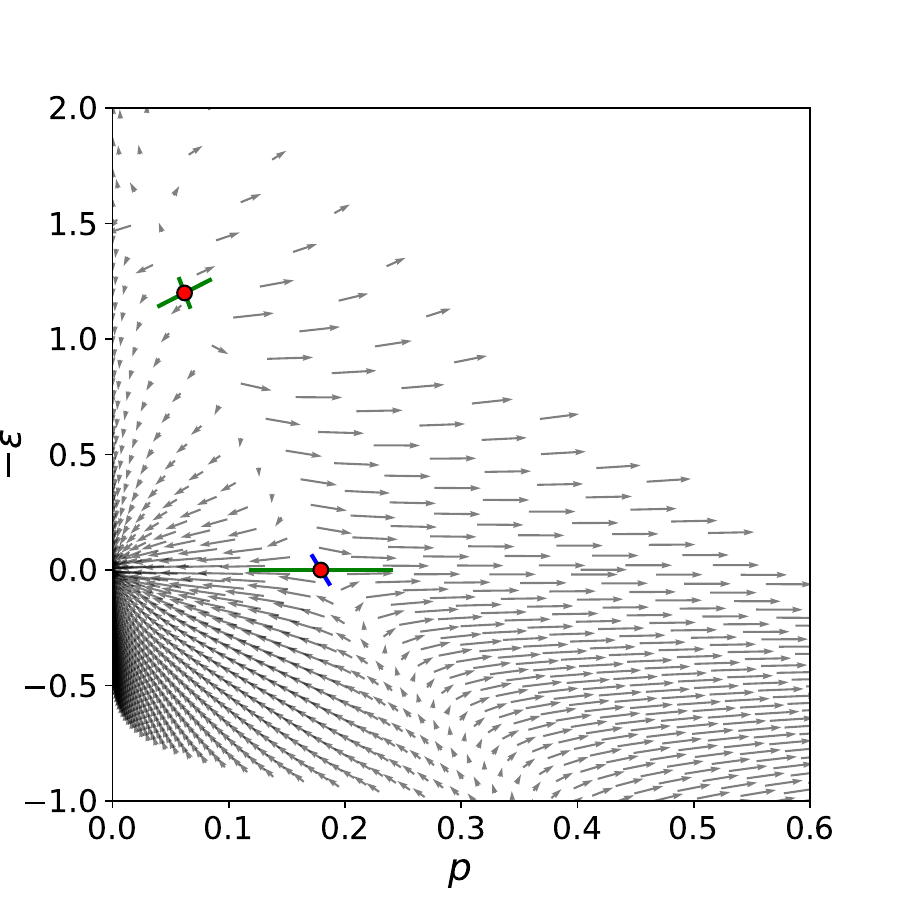}
    \put(-130,200){\textbf{(a) $b=3$ }}
    \\
    \includegraphics[keepaspectratio, width=0.49\textwidth]{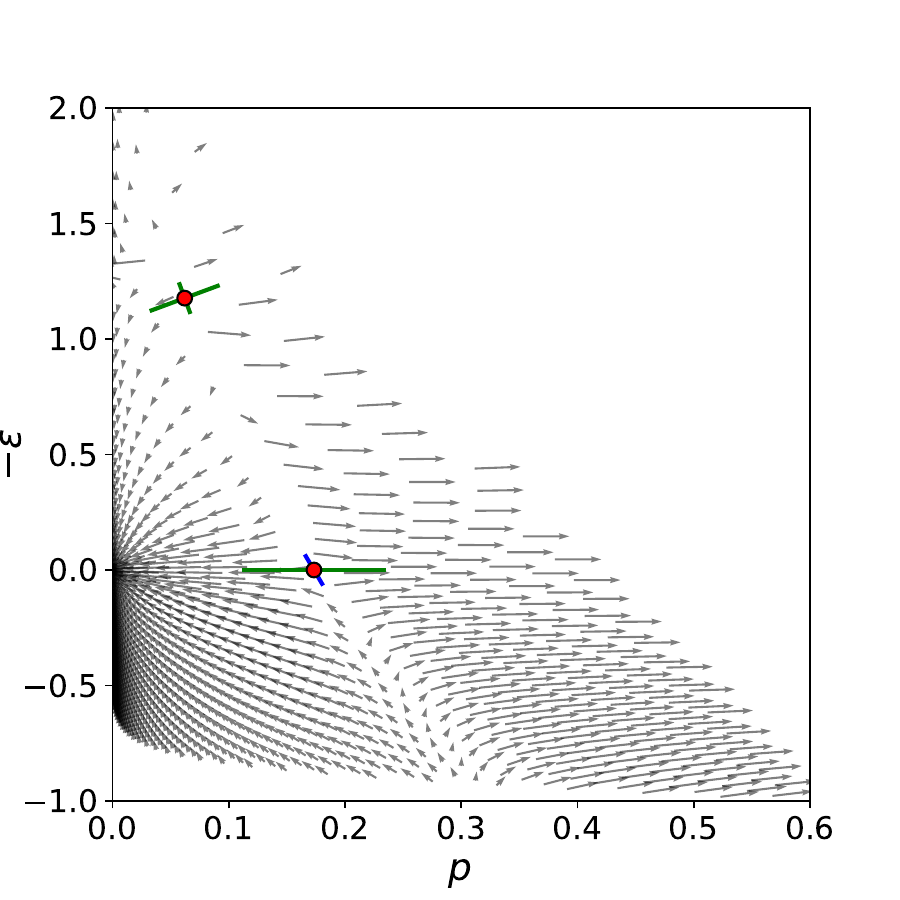}
     \put(-130,200){\textbf{(b) $b=4$}}
     \\
\caption{ RG flows around the fixed points, where both the site occupation probability $p$ and the energy cost $\varepsilon$ are renormalized for block sizes $b=3,4$, respectively. The y-axis represents the energy cost \( -\varepsilon \). The linearized flow around each fixed point is depicted by blue and green lines. The green lines indicate the relevant direction, while the blue lines represent the irrelevant direction. The overall flow is illustrated with black arrows.}
\label{fig:lgrg2}
\end{figure}
After rescaling, the weight for the two nearest neighbor blocks $A$ and $B$ can be written as:
\begin{equation}\label{eq:LG1}
    W_{N_A,N_B}=e^{\tilde{\mu}(N_A+N_B)+\tilde{J}N_AN_B},
\end{equation} where $\tilde{\mu}$ and $\tilde{J}$ are the renormalized site fugacity and bond energy respectively.
The RG flow equation for the renormalized site occupation probability and the renormalized energy cost can be derived by comparing the derived weights with those given in Eq.~\eqref{eq:LG1}:
\begin{equation}\label{eq:LGp1}
p'= \frac{W_{10}(p,y)}{1+W_{10}(p,y)}
\end{equation}
\begin{equation}\label{eq:LGp2}
    J'= \ln \left( \frac{W_{11}(p,y)}{W_{10}^2(p,y)} \right),
\end{equation}
where $W_{10}, W_{11}$ are explicit polynomials in $e^{\mu}=p/(1-p)$ and $y=e^J$ as derived above. 
In a real-space RG, we coarse-grain the lattice by a factor $b$ (here $b=2$ for $2\times 2$ blocks).  
At the critical fixed point $(\mu^*,J^*)$, the system is scale-invariant \cite{PhysRevB.4.3174}. Small deviations from the fixed point obey the linearized RG map as:
\begin{equation}
\begin{pmatrix} \delta \mu' \\ \delta J' \end{pmatrix} =
\mathcal{J} \begin{pmatrix} \delta \mu \\ \delta J \end{pmatrix}, \quad
\mathcal{J} = \left. \frac{\partial (\mu',J')}{\partial (\mu,J)} \right|_{(\mu^*,J^*)},
\end{equation}
where $\delta \mu = \mu-\mu^*$, $\delta J = J-J^*$. Let $\lambda_{\rm rel}$ be the eigenvalue corresponding to the relevant direction (growing under RG), so that
\begin{equation}
\delta' = \lambda_{\rm r} \, \delta.
\end{equation}
The correlation length $\xi$ transforms under RG as:
\begin{equation}
\xi(\delta') = \frac{\xi(\delta)}{b}.
\end{equation}
Near criticality, $\xi \sim \delta^{-\nu}$. Using $\delta' = \lambda_{\rm r} \, \delta$, we have
\begin{equation}
\xi(\delta') = (\delta')^{-\nu} = (\lambda_{\rm r} \delta)^{-\nu} = \lambda_{\rm r}^{-\nu} \, \delta^{-\nu}.
\end{equation}
Comparing with $\xi(\delta') = \xi(\delta)/b = \delta^{-\nu}/b$ gives:
\begin{equation}
\lambda_{\rm r}^{-\nu} = \frac{1}{b} \quad \Rightarrow \quad
\nu = \frac{\ln b}{\ln \lambda_{\rm r}}.
\end{equation} \par
\begin{figure}[t!]
    \centering
    \includegraphics[keepaspectratio, width=1.02\linewidth]{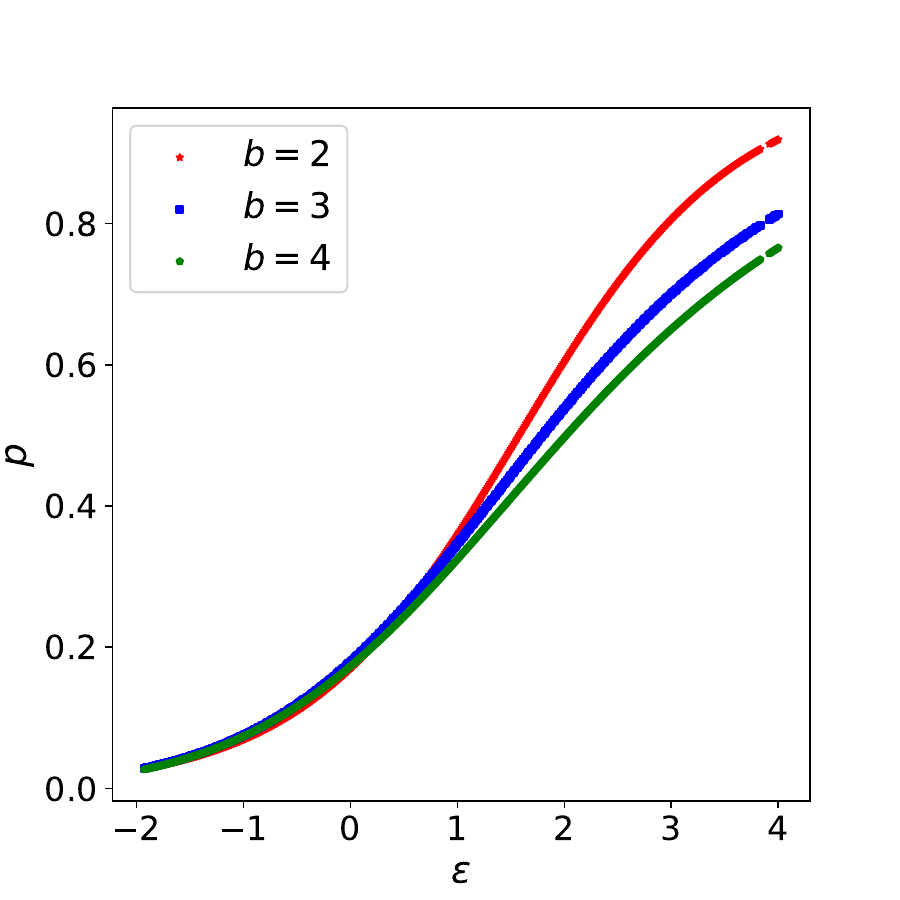}
     \caption{Variation of the critical site percolation threshold, calculated from renormalization group analysis in real space using Kadanoff blocks with a lattice gas analogy, as described in Eq.~\eqref{eq:LGp1}, considering an energy cost per link $\epsilon$ for different block sizes, $b$ as $b=2$, $b=3$ and $b=4$ as derived in Eq.~\eqref{eq:wei1a}, Eq.~\eqref{eq:wei2a} and Eq.~S25 given in the Supplementary Information~\cite{SI}. We can infer from the plot that as $\varepsilon$ increases, the critical occupation probability $p_c$ tends to unity.} 
    \label{fig:lgrg1}
\end{figure}

Let $X \equiv W_{10}$, $Y \equiv W_{11}$. Then
\[
\mu' = \ln X, \quad J' = \ln Y - 2\ln X.
\]
The Jacobian entries at any point $(\mu,J)$ can be computed as:
\begin{equation}\label{eq:jac1}
\begin{split}
\frac{\partial \mu'}{\partial \mu} &= \frac{\tilde X_\mu}{X}, \\
\frac{\partial \mu'}{\partial J}   &= \frac{\tilde X_J}{X}, \\
\frac{\partial J'}{\partial \mu}   &= \frac{\tilde Y_\mu}{Y} - 2\frac{\tilde X_\mu}{X}, \\
\frac{\partial J'}{\partial J}     &= \frac{\tilde Y_J}{Y} - 2\frac{\tilde X_J}{X},
\end{split}
\end{equation}
where 
\begin{equation}
\begin{split}
\tilde X_\mu &= e^\mu \frac{\partial X}{\partial e^\mu}, \\
\tilde X_J   &= e^J \frac{\partial X}{\partial e^J}, \\
\tilde Y_\mu &= e^\mu \frac{\partial Y}{\partial e^\mu}, \\
\tilde Y_J   &= e^J \frac{\partial Y}{\partial e^J}.
\end{split}
\end{equation}
The RG eigenvalue $\{\lambda_i\}$ can be evaluated from  Eq.~\eqref{eq:jac1} at fixed point $(\mu^*,J^*)$. The relevant eigenvalue $\lambda_{\mathrm{rel}}$ gives the correlation-length exponent similar to that done in Sec.~\ref{sec:RG1}, with
\begin{equation}
\nu = \frac{\ln b}{\ln \lambda_{\mathrm{r}}}.
\end{equation}
In our original model, \(\varepsilon\) represents a constant energy cost at all length scales, which we will later relax to consider the renormalization of the energy cost as well.
Therefore, we focus on the flow equation for the site-occupation probability, as derived in Eq.~\eqref{eq:LGp1}. For \(b=2\), the way we derived Eq.~\eqref{eq:wei1} indicates that a block is considered occupied if at least one site within it is occupied. When we solve the flow equation for the fixed point, we find only the trivial fixed point \(p_c=1\) and do not obtain any non-trivial fixed points for values of \(\varepsilon\). For that reason, while deriving Eq.~\eqref{eq:wei1}, we now only consider configurations that span the block along at least one direction. 
To compute the modified weight for a block size of \( b = 2 \), we can use Eq.~\eqref{eq:weii1} along with the modified boundary weights computed as follows:
\begin{equation}\label{eq:bw1}
\begin{aligned}
\tilde{Q}_{00} &= e^{2\mu} e^{J}, \\
\tilde{Q}_{10} &= \tilde{Q}_{01} = e^{2\mu} e^{J} + e^{3\mu} e^{2J}, \\
\tilde{Q}_{11} &= e^{2\mu} e^{J} + 2 e^{3\mu} e^{2J} + e^{4\mu} e^{4J}.
\end{aligned}
\end{equation}
Taking this into account, the modified weights become:
\begin{equation}\label{eq:wei1a}
\tilde{W}^{2}_{10} =4e^{2\mu}e^{J}+4e^{3\mu}e^{2J}+e^{4\mu}e^{4J},
\end{equation} where $\tilde{W}^{2}$ indicates the modified weights for block size $b=2$.
For block size $b=3$, the modified boundary weights can be computed similarly as:
\begin{equation}\label{eq:bw2}
\begin{aligned}
\tilde{Q}(0,0,0)
= {} & 2 y^2 z^3
+ 8 y^3 z^4
+ 2 y^4 z^5
+ 4 y^5 z^5
+ y^7 z^6, \\[6pt]
\tilde{Q}(1,0,0)
= {} & y^2 z^3
+ 3 y^2 z^4
+ 4 y^3 z^4
+ 6 y^3 z^5
+ 5 y^4 z^5
+ y^5 z^5 \\
& + 3 y^5 z^6
+ 3 y^6 z^6
+ y^8 z^7, \\[6pt]
\tilde{Q}(0,1,0)
= {} & y^2 z^3
+ y^2 z^4
+ 7 y^3 z^4
+ 4 y^3 z^5
+ 6 y^4 z^5
+ y^4 z^6 \\
& + 2 y^5 z^5
+ y^5 z^6
+ 4 y^6 z^6
+ y^8 z^7, \\[6pt]
\tilde{Q}(0,0,1)
= {} & y^2 z^3
+ 3 y^2 z^4
+ 4 y^3 z^4
+ 6 y^3 z^5
+ 5 y^4 z^5
+ y^5 z^5 \\
& + 3 y^5 z^6
+ 3 y^6 z^6
+ y^8 z^7, \\[6pt]
\tilde{Q}(0,1,1)
= {} & 3 y^3 z^4
+ 4 y^3 z^5
+ 5 y^4 z^5
+ 4 y^4 z^6
+ 3 y^5 z^5 \\
& + 4 y^5 z^6 
+ 5 y^6 z^6
+ y^6 z^7
+ y^7 z^6
+ 2 y^7 z^7\\
& + 3 y^8 z^7
+ y^{10} z^8, \\[6pt]
\tilde{Q}(1,1,0)
= {} & 3 y^3 z^4
+ 4 y^3 z^5
+ 5 y^4 z^5
+ 4 y^4 z^6
+ 3 y^5 z^5\\
& + 4 y^5 z^6 
+ 5 y^6 z^6
+ y^6 z^7
+ y^7 z^6
+ 2 y^7 z^7\\
& + 3 y^8 z^7
+ y^{10} z^8, \\[6pt]
\tilde{Q}(1,0,1)
= {} & 2 y^2 z^4
+ 3 y^2 z^5
+ 8 y^3 z^5
+ 3 y^3 z^6
+ y^4 z^5
+ 7 y^4 z^6 \\
& + 5 y^5 z^6
+ 4 y^6 z^7
+ 2 y^7 z^7
+ y^9 z^8, \\[6pt]
\tilde{Q}(1,1,1)
= {} & y^2 z^3
+ 3 y^2 z^4
+ y^2 z^5
+ 3 y^3 z^4
+ 8 y^3 z^5
+ y^3 z^6 \\
& + 4 y^4 z^5
+ 6 y^4 z^6
+ 2 y^5 z^5
+ 8 y^5 z^6
+ 4 y^6 z^6 \\
& + 8 y^6 z^7
+ y^7 z^6
+ 2 y^7 z^7
+ 5 y^8 z^7
+ y^8 z^8 \\
& + 3 y^9 z^8
+ 2 y^{10} z^8
+ y^{12} z^9,
\end{aligned}
\end{equation}
with permutation symmetry $\tilde{Q}(1,0,0)=\tilde{Q}(0,0,1)$, and $\tilde{Q}(1,1,0)=\tilde{Q}(0,1,1)$ respectively.
A similar calculation for block size $b=3$ yields the following expansion for modified weight:
\begin{equation}\label{eq:wei2a}
\begin{aligned}
\tilde{W}^{3}_{10}(z,y) &= 6z^3 y^2 + 12z^4 y^2 + 32z^4 y^3 \\
&\quad + 4z^5 y^2 + 40z^5 y^3 + 33z^5 y^4 + 16z^5 y^5 \\
&\quad + 4z^6 y^3 + 22z^6 y^4 + 28z^6 y^5 + 24z^6 y^6 + 4z^6 y^7 \\
&\quad + 14z^7 y^6 + 8z^7 y^7 + 14z^7 y^8 \\
&\quad + z^8 y^8 + 4z^8 y^9 + 4z^8 y^{10} + z^9 y^{12},
\end{aligned}
\end{equation}
where $z=e^{\mu}$ and $y=e^{-\varepsilon}$ respectively. Setting $y=1$, i.e $\varepsilon=0$, the coefficients in the above expansion, Eq.~\eqref{eq:wei2a} reduces to that in Eq.~\eqref{eq:Drok}, and that given in \cite{PhysRevB.21.1223}. Thus, one can reproduce the same RG flow equations directly from a hard-core lattice gas analogy relevant to our model and Monte Carlo simulation. The critical threshold for site occupation probability is calculated using the recursion relation outlined in Eq.~\eqref{eq:LGp1}. 
This calculation is performed for various bond energy costs, denoted as $\varepsilon$, and for block sizes of $b=2$ and $b=3$. The corresponding weights used in this analysis are given in Eq.~\eqref{eq:wei1a} and Eq.~\eqref{eq:wei2a}. The results are illustrated in Fig.~\ref{fig:lgrg1}. \par
One significant advantage of this technique arises when the energy cost \(\varepsilon\) is not constant but renormalizable, similar to the probability of site occupation. In this case, it is possible to express the weights \(W_{11}\) as derived in Eq.~\eqref{eq:weib1} and the modified weight \(\tilde{W}_{11}\). By solving the coupled polynomial equations in Eq.~\eqref{eq:LGp1} and Eq.~\eqref{eq:LGp2}, we can determine the fixed point for \(p\), and \(\varepsilon\) flows. \par 
Using this modified boundary weights derived in Eq.~\eqref{eq:bw1}, we can determine the modified weight $\tilde{W}_{11}^{2}$ for block size $b=2$ similar to that in Eq.~\eqref{eq:wei2a} following Eq.~\eqref{eq:weii1} as:
\begin{equation}
\begin{aligned}
\tilde{W}_{11}^{2}(z,y)
= {} & 9 y^2 z^4
+ 6 y^3 z^4
+ 12 y^3 z^5 
+ y^4 z^4
+ 16 y^4 z^5
+ 2 y^4 z^6 \\
& + 4 y^5 z^5
+ 12 y^5 z^6 
+ 8 y^6 z^6
+ 2 y^7 z^6 \\
& + 4 y^7 z^7
+ 4 y^8 z^7 
+ y^{10} z^8.
\end{aligned}
\end{equation}
where $z=e^{\mu}$, and $y=e^{-\varepsilon}$ as defined previously. 
Similarly, the modified block weight $\tilde{W}_{11}^{3}$ for block size $b=3$ can be computed as:
\begin{equation}
\tilde{W}_{11}^{3}
= \sum_{\substack{a_2,a_5,a_8\in [0,1] \\ b_0,b_3,b_6\in [0,1]}}
Q(a_2,a_5,a_8)\, Q(b_0,b_3,b_6)\,
e^{J(a_2 b_0 + a_5 b_3 + a_8 b_6)}.
\end{equation}
By explicitly computing the summation using the modified boundary weights derived in Eq.~\eqref{eq:bw2}, one can arrive at,
\begin{equation}\label{eq:weif1}
\begin{aligned}
\tilde{W}_{11}^{3}(z,y) =\;&
z^{18}y^{27}
+ 4z^{17}y^{25}
+ 10z^{17}y^{24}
+ 4z^{17}y^{23} \\
&+ 14z^{16}y^{23}
+ 38z^{16}y^{22}
+ 65z^{16}y^{21}
+ 33z^{16}y^{20} \\
&+ 3z^{16}y^{19}
+ 2z^{15}y^{22}
+ 32z^{15}y^{21}
+ 130z^{15}y^{20} \\
&+ 232z^{15}y^{19}
+ 268z^{15}y^{18}
+ 132z^{15}y^{17} \\
&+ 8z^{14}y^{20}
+ 83z^{14}y^{19}
+ 296z^{14}y^{18}
+ 693z^{14}y^{17} \\
&+ 852z^{14}y^{16}
+ 716z^{14}y^{15}
+ 282z^{14}y^{14} \\
&+ 24z^{13}y^{18}
+ 178z^{13}y^{17}
+ 628z^{13}y^{16} \\
&+ 2088z^{13}y^{14}
+ 1928z^{13}y^{13}
+ 1054z^{13}y^{12} \\
&+ 20z^{13}y^{10}
+ 3z^{12}y^{17}
+ 46z^{12}y^{16}
+ 342z^{12}y^{15} \\
&+ 1135z^{12}y^{14}
+ 2638z^{12}y^{13}
+ 3745z^{12}y^{12}\\
&+ 2033z^{12}y^{10}
+ 556z^{12}y^{9}
+ 85z^{12}y^{8}
+ 6z^{12}y^{7} \\
&+ 8z^{11}y^{15}
+ 96z^{11}y^{14}
+ 516z^{11}y^{13}
+ 1794z^{11}y^{12} \\
&+ 3796z^{11}y^{11}
+ 5186z^{11}y^{10}
+ 4562z^{11}y^{9} \\
&+ 482z^{11}y^{7}
+ 30z^{11}y^{6}
+ 20z^{10}y^{13}
+ 140z^{10}y^{12} \\
&+ 712z^{10}y^{11}
+ 2281z^{10}y^{10}
+ 4302z^{10}y^{9} \\
&+ 3061z^{10}y^{7}
+ 781z^{10}y^{6}
+ 30z^{10}y^{5}
+ 2z^{9}y^{12} \\
&+ 24z^{9}y^{11}
+ 182z^{9}y^{10}
+ 792z^{9}y^{9}
+ 2248z^{9}y^{8} \\
&+ 3194z^{9}y^{7}
+ 2318z^{9}y^{6}
+ 402z^{9}y^{5}
+ 6z^{9}y^{4} \\
&+ 4z^{8}y^{10}
+ 29z^{8}y^{9}
+ 162z^{8}y^{8}
+ 668z^{8}y^{7} \\
&+ 1370z^{8}y^{6}
+ 763z^{8}y^{5}
+ 56z^{8}y^{4}
+ 6z^{7}y^{8} \\
&+ 18z^{7}y^{7}
+ 106z^{7}y^{6}
+ 318z^{7}y^{5}
+ 80z^{7}y^{4} \\
&+ z^{6}y^{7}
+ 9z^{6}y^{5}
+ 26z^{6}y^{4}
+ 266z^{13}y^{11}\\
& +16z^{15}y^{16}+ 28z^{14}y^{13}+ 1480z^{13}y^{15}\\
& + 3635z^{12}y^{11}+ 2058z^{11}y^{8}+ 4970z^{10}y^{8}.
\end{aligned}
\end{equation}
Solving the coupled equations in Eq.~\eqref{eq:LGp1} and Eq.~\eqref{eq:LGp2} using the modified block weights derived in Eq.~\eqref{eq:wei2a}, Eq.~\eqref{eq:weif1} and in Supplementary Information~\cite{SI}, we obtain the fixed points $(p^{*},\varepsilon^{*})$ for block sizes $b=3,4$ as listed below, with the RG flows around the fixed points shown in Fig.~\ref{fig:lgrg2}.
For a block size \( b = 3 \), there are three fixed points:
(a) \( p = 0 \), \( \varepsilon = 0 \): This is a stable fixed point, remaining stable in both directions.
(b) \( p \approx 0.179202 \), \( \varepsilon = 0 \): This is a critical fixed point, stable in one direction and unstable in the other. The scaling dimension associated with the relevant direction is \( y_t \approx 1.2662 \), which implies a thermal exponent of \( \nu \approx 0.78975 \).
(c) \( p \approx 0.062002 \), \( \varepsilon \approx -1.199417 \): This is an unstable fixed point, unstable in both directions. The scaling dimension associated with the maximum relevant direction is \( y_t \approx 1.407 \), leading to a thermal exponent of \( \nu \approx 0.710751 \). Additionally, the crossover exponent corresponding to the second relevant direction is \( \phi \approx 0.295249 \).
For a block size of \( b = 4 \), the unstable fixed point is at \( p \approx 0.0620712 \) with \( \varepsilon \approx -1.177 \). The thermal exponent associated with the maximum relevant direction is \( \nu \approx 0.69934 \), and the crossover exponent due to the second relevant direction is \( \phi \approx 0.290565 \).  The critical fixed point for this block size is at \( p \approx 0.17328 \) with \( \varepsilon = 0 \), and the thermal exponent associated with the relevant direction is \( \nu \approx 0.760657 \). We do not want to comment on the universality class, since to do so would require performing similar calculations for very large block sizes. The full expressions for the modified boundary weights and modified block weights for block size \( b = 4 \) are given in the Supplementary Information~\cite{SI}.
The critical fixed point is the classical percolation phase transition point $(p_c,\varepsilon=0)$, indicating a phase transition from non-percolating to percolating phase as shown in Fig.~\ref{fig:lgrg1} with tunable bond energy cost. At the unstable fixed point with two relevant directions, the crossover exponent \(\phi \approx 0.3\) indicates that the second relevant direction affects the system much more slowly than the primary one. As a result, the system behaves according to the first scaling regime over a wide range before eventually transitioning to a different behavior. The second relevant direction becomes useful at large length scales and guides the flow towards $\varepsilon=0$ and $\varepsilon=-\infty$, respectively.
\section{Conclusions and Discussion}
In this work, we generalized site percolation on the two-dimensional square lattice by assigning an energy cost $\varepsilon$ to each occupied nearest-neighbor bond. This model interpolates between a dense-cluster phase ($\varepsilon\to -\infty$) and a dilute phase ($\varepsilon\to\infty$) dominated by minimally connected clusters. Numerically, energy weighting preserved the universal scaling form of the cluster-size distribution $n_s(p,\varepsilon)$ but lowered its cutoff $s^\ast(p,\varepsilon)$ as $\varepsilon$ increased, suppressing large, highly connected clusters. The correlation length $\xi(\varepsilon)$ similarly decreased with $\varepsilon$ and became finite at the classical percolation threshold, showing that energetic constraints could destroy geometric criticality even at fixed $p=p_c(\varepsilon=0)$. Monte Carlo simulations with varying $p$ and bond energy $\varepsilon$ revealed antiferromagnetic site ordering at large $\varepsilon$ as $p$ increases. Using Kadanoff block scaling, we constructed a real-space RG that tracks the renormalization of effective occupation and bond-energy parameters under coarse-graining. For both spanning rules ($R_{0}$, $R_{2}$), the RG recursions showed that increasing $\varepsilon$ continuously drove the correlation-length exponent from the classical $\nu=4/3$ to the isolated-cluster limit $\nu=1$, in quantitative agreement with Coulomb-gas predictions that identified the bond energy cost $\varepsilon$ as tuning the effective loop fugacity. A complementary lattice-gas RG with site and bond fugacities gave exact recursion relations for block-renormalized parameters and reproduced the full flow, including the shift of $p_c(\varepsilon)$ toward unity as $\varepsilon$ grows. This framework can therefore provide a unified description of the crossover behavior between percolation, self-avoiding walks, and broader random-cluster universality classes.

Energy-Weighted Site Percolation could capture the behavior relevant to polymer networks \cite{PhysRevLett.91.108102, RevModPhys.86.995}, random media with activation energies \cite{Storm2005}, force-balanced soft materials \cite{Storm2005,Flory1941, Stockmayer1944_GelFormation2,ramola2016disordered}, and networks with resource-dependent connectivity costs \cite{PhysRevE.75.045103, PhysRevE.96.062302, Kim_2025}. This framework can also be extended to analyze lattice walks, where the asymptotic behavior of the generating function in Eq.~\eqref{eq:mn1} becomes directly connected to the emergence of the critical percolation threshold and the growth of connected trajectories across the lattice. 
Generalizing the framework to non-equilibrium directed percolation (DP)~\cite{PhysRevLett.81.1646} further allows the incorporation of non-Hermitian transition rates and irreversible stochastic dynamics that violate detailed balance. Applying large anisotropic bond energy penalties ($\varepsilon_{x} \gg \varepsilon_{y}$ or $\varepsilon_{y} \gg \varepsilon_{x}$) biases cluster formation preferentially along selected directions, producing anisotropic connected structures that map onto directed rigidity percolation \cite{PhysRevE.60.5699}. This is particularly relevant in gels, fiber networks, and polymers \cite{Storm2005, PhysRevLett.91.108102, RevModPhys.86.995}, where microscopic deformation energies associated with bending, stretching, and cross-linking \cite{Sahimi, StaufferAharony, PhysRevLett.95.178102, PhysRevE.76.031906, PhysRevLett.91.108102} constrain connectivity and alter gelation thresholds \cite{PhysRevA.43.5412,Flory1941, Flory1941_Tetrafunctional, Stockmayer1944_GelFormation2}. It can also connect to modern elasticity-based descriptions of polymer networks \cite{DoiEdwards1986,deGennes1979, Treloar1975, Edwards1985} by integrating additional structural features—such as loop formation and spatial correlations. In the context of transport in disordered media, $\varepsilon$ acts as a thermal activation factor representing microscopic energy barriers for hopping or tunneling \cite{PhysRevB.66.075417,Kirkpatrick, PhysRevB.72.125121}, exponentially suppressing energetically costly connections and producing heterogeneous transport pathways. In neural networks, the same framework captures how resource constraints shift robustness thresholds and modify cluster statistics \cite{PhysRevE.75.045103, PhysRevE.96.062302, Kim_2025}. In disordered solids and jamming transitions \cite{PhysRevLett.100.028001,Sahimi}, penalizing internal connectivity can suppress highly coordinated bond configurations, providing a connection between rigidity percolation and mechanically marginal elastic states. Finally, extensions to higher dimensions and higher-order neighbor interactions could provide a statistical framework for slow relaxation dynamics in supercooled liquids \cite{PhysRevLett.132.067101} and lattice glass models \cite{merrigan2020arrested,PhysRevLett.102.015702,PhysRevLett.132.067101} near the solid-to-liquid transition.

\section{Acknowledgments}
We thank Mustansir Barma and Sreenath K. Manikandan for useful discussions. This project was funded by intramural funds at TIFR Hyderabad
from the Department of Atomic Energy (DAE), Government
of India, under Project Identification No. RTI4007.
\bibliographystyle{apsrev4-2}
\bibliography{References}
\appendix
\section{COMBINATORIAL DERIVATION OF THE COEFFICIENTS $M_k$ FOR RULE $R_0$, $b=3$}\label{app:gh}
We consider a \(3 \times 3\) block of sites, which contains a total of \(N = 9\) sites. Each site has an occupation probability of \(p\), while the probability of being unoccupied is \(q = 1 - p\). A configuration is said to span horizontally if there exists a connected path of occupied sites, using nearest-neighbor adjacency, that connects the left edge (at \(x = 1\)) to the right edge (at \(x = 3\)). Similarly, a configuration spans vertically if a path connects the top edge (at \(y = 1\)) to the bottom edge (at \(y = 3\)).
The renormalized occupation probability according to Rule $R_0$ is then
\begin{equation}
p' \;=\; \sum_{k=0}^{9} M_k\, p^{k} q^{9-k},
\end{equation}
where $M_k$ counts the number of distinct $k$--site configurations that span either horizontally or vertically.
We index the $3\times3$ sites in row-major order:
\[
\begin{array}{ccc}
0 & 1 & 2\\
3 & 4 & 5\\
6 & 7 & 8
\end{array}
\]
A subset \( A \subset \{0, \dots, 8\} \) is considered a horizontal minimal spanning set if it contains a connected path that connects the left edge to the right edge, and no proper subset of \( A \) possesses this property. The minimal spanning sets that span the block from left to right for \( b = 3 \) can be listed as follows:
\begin{align}
A_1 &= \{0,1,2\}, &
A_2 &= \{3,4,5\}, &
A_3 &= \{6,7,8\}, \nonumber\\
A_4 &= \{0,1,4,5\}, &
A_5 &= \{1,2,3,4\}, &
A_6 &= \{3,4,7,8\}, \nonumber\\
A_7 &= \{4,5,6,7\}, &
A_8 &= \{0,1,4,7,8\}, &
A_9 &= \{1,2,4,6,7\}. &
\end{align}
Thus, there are $9$ minimal sets in total:
three of size $3$, four of size $4$, and two of size $5$.
\begin{figure*}[t!]
    \centering
    \includegraphics[keepaspectratio, width=0.45\textwidth]{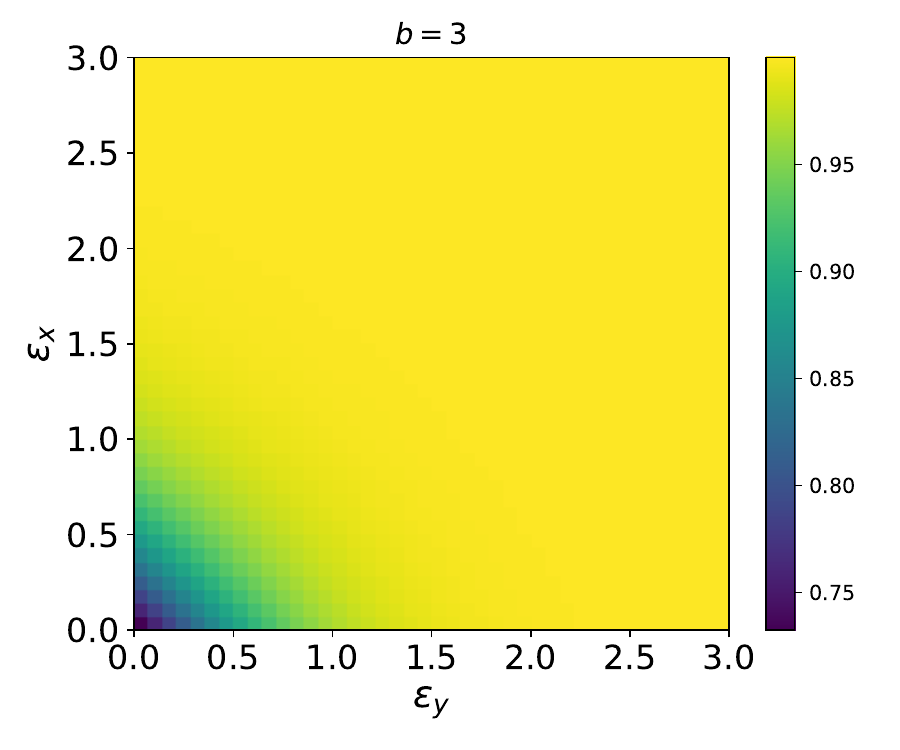}
    \includegraphics[keepaspectratio, width=0.45\textwidth]{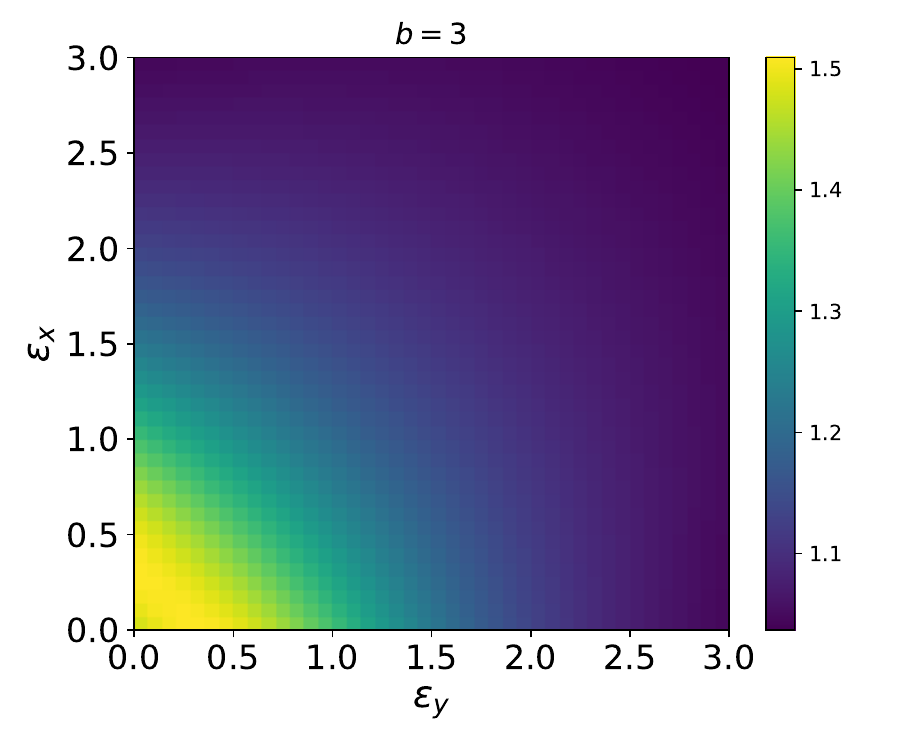}
    \put(-350,85){\textbf{(a)}}
    \put(-150,85){\textbf{(b)}}
    \\
     \caption{\textbf{(a)} The variation of the critical site percolation threshold was computed from RG analysis in real space using Kadanoff blocks. This analysis utilized the \( R_{2} \) rule, incorporating an anisotropic energy cost per link in the \( x \) direction, denoted as \( \varepsilon_x \), and an energy cost per link in the \( y \) direction, denoted as \( \varepsilon_y \), for a block size of \( b = 3 \), \textbf{(b)} variation of the correlation length exponent \( \nu \), also computed from RG analysis in real space using Kadanoff blocks, following the same \( R_{2} \) rule and considering the anisotropic energy costs \( \varepsilon_x \) and \( \varepsilon_y \) for the block size \( b = 3 \).}
    \label{fig:RGan_1}
\end{figure*}
\begin{figure*}[t!]
    \centering
    \includegraphics[keepaspectratio, width=0.45\textwidth]{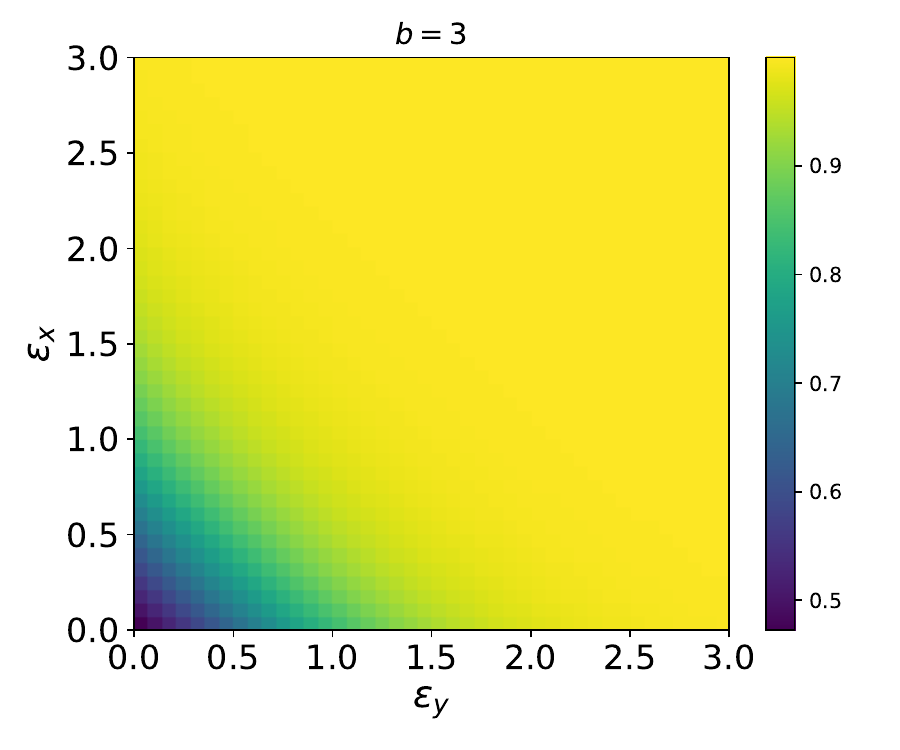}
    \includegraphics[keepaspectratio, width=0.45\textwidth]{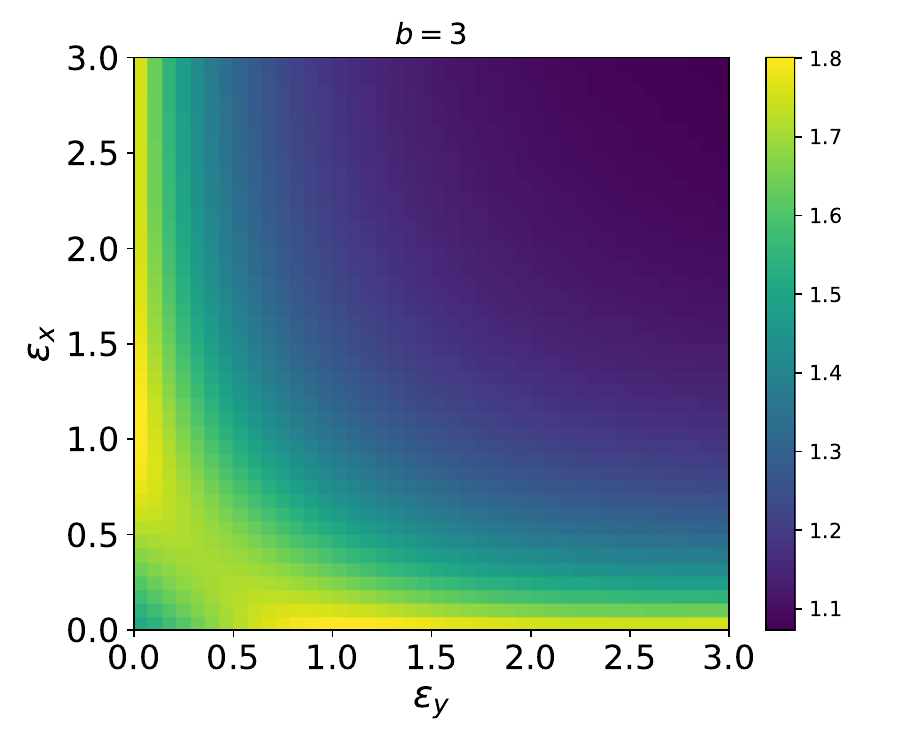}
    \put(-350,85){\textbf{(a)}}
    \put(-150,85){\textbf{(b)}}
    \\
     \caption{\textbf{(a)} The variation of the critical site percolation threshold was computed from RG analysis in real space using Kadanoff blocks. This analysis utilized the \( R_{0} \) rule, incorporating an anisotropic energy cost per link in the \( x \) direction, denoted as \( \varepsilon_x \), and an energy cost per link in the \( y \) direction, denoted as \( \varepsilon_y \), for a block size of \( b = 3 \), \textbf{(b)} variation of the correlation length exponent \( \nu \), also computed from RG analysis in real space using Kadanoff blocks, following the same \( R_{0} \) rule and considering the anisotropic energy costs \( \varepsilon_x \) and \( \varepsilon_y \) for the block size \( b = 3 \).}
    \label{fig:RGan_1b}
\end{figure*}
\subsection*{Inclusion--Exclusion Expansion}
According to the principle of inclusion-exclusion, for a finite collection of finite sets $X_1, \cdots, X_n$,
\begin{equation}
\left|\bigcup_{i=1}^{n} X_i \right|
=
\sum_{k=1}^{n} (-1)^{k+1}
\left(
\sum_{1 \le i_1 < \cdots < i_k \le n}
\left| X_{i_1} \cap \cdots \cap X_{i_k} \right|
\right),
\end{equation} which can be written compactly as:
\begin{equation}
\left|\bigcup_{i=1}^{n} X_i \right|
=
\sum_{\emptyset \neq I \subseteq \{1,\ldots,n\}}
(-1)^{|I|+1}
\left|\bigcap_{i \in I} X_i \right|.
\end{equation}
Let $\mathcal{A} = \{A_1, \dots, A_9\}$ be the set of all minimal sets. For any subset $I \subseteq \{1, \dots, 9\}$, let $U_I = \bigcup_{i \in I} A_i$ represent its union. By inclusion-exclusion, the number of configurations with exactly $k$ occupied sites that span from left to right is:
\begin{equation}
H_k \;=\;
\sum_{\varnothing\neq I\subseteq\mathcal{A}}
(-1)^{|I|+1}
\binom{9-|U_I|}{k-|U_I|},
\end{equation}
with the convention that $\binom{n}{r}=0$ if $r<0$ or $r>n$. \par
Let \(\mathcal{A}\) represent the set of all minimal spanning subsets of sites in a \(b \times b\) block that extend from the left side to the right side of the block. From the principle of inclusion-exclusion, the number of configurations in which exactly \(k\) sites are occupied, and span from left to right, is given by:
\begin{equation}
H_k 
\;=\;
\sum_{\varnothing \neq I \subseteq \mathcal{A}}
(-1)^{|I|+1}
\binom{b^2 - |U_I|}{k - |U_I|},
\qquad
U_I = \bigcup_{A \in I} A,
\end{equation}
where we adopt the convention that $\binom{n}{r} = 0$ if $r < 0$ or $r > n$.
\begin{proof}
A configuration with exactly \( k \) occupied sites corresponds to a subset \( S \subseteq [b^2] \) where \( |S| = k \). This configuration is said to span left to right if it includes at least one of the minimal spanning subsets \( A \in \mathcal{A} \). Therefore, the collection of configurations of size \( k \) that span left to right can be expressed as:
\begin{equation}
\mathcal{H}_k 
= \bigcup_{A \in \mathcal{A}} 
\mathcal{X}_A,
\qquad 
\text{where } 
\mathcal{X}_A = \{\, S \subseteq [b^2]: |S| = k,\, S \supseteq A \,\}.
\end{equation}
Therefore,
\begin{equation}
H_k = |\mathcal{H}_k|
= \left| \bigcup_{A \in \mathcal{A}} \mathcal{X}_A \right|.
\end{equation}
Applying the inclusion--exclusion principle gives:
\begin{equation}
H_k 
= 
\sum_{\varnothing \neq I \subseteq \mathcal{A}}
(-1)^{|I|+1}
\left| 
\bigcap_{A \in I} \mathcal{X}_A
\right|.
\end{equation}
For any nonempty subset $I \subseteq \mathcal{A}$,
the intersection $\bigcap_{A \in I} \mathcal{X}_A$ consists of all 
$k$-site configurations that contain every $A \in I$, i.e.
\begin{equation}
\bigcap_{A \in I} \mathcal{X}_A
=
\{\, S \subseteq [b^2]: |S| = k,\, S \supseteq U_I \,\},
\qquad
U_I = \bigcup_{A \in I} A.
\end{equation}
Given $U_I$, the number of $k$-site super sets $S \supseteq U_I$
is the number of ways to choose the remaining 
$k - |U_I|$ sites from the $b^2 - |U_I|$ sites that are not in $U_I$:
\begin{equation}
\left| \bigcap_{A \in I} \mathcal{X}_A \right|
= \binom{b^2 - |U_I|}{k - |U_I|},
\end{equation}
where $\binom{n}{r} = 0$ if $r < 0$ or $r > n$.
Substituting this expression into the inclusion--exclusion formula yields
\begin{equation}
H_k 
= 
\sum_{\varnothing \neq I \subseteq \mathcal{A}}
(-1)^{|I|+1}
\binom{b^2 - |U_I|}{k - |U_I|},
\end{equation}
which proves the stated result.
\end{proof} \par
We group all such non-empty subsets $I$ by their union size $u=|U_I|$ and define
\begin{equation}
S_u \;=\;
\sum_{\substack{\varnothing\neq I\subseteq\mathcal{A}\\ |U_I|=u}}
(-1)^{|I|+1}.
\end{equation}
Then the inclusion--exclusion formula for $b=3$ becomes the compact expression
\begin{equation}
H_k \;=\;
\sum_{u=1}^{9} S_u
\binom{9-u}{k-u}.
\end{equation}
Enumerating all such subsets of $\mathcal{A}$ and summing signs grouped by $u=|U_I|$
gives the following signed multiplicities:
\begin{align}
S_1 &= 0, &
S_2 &= 0, &
S_3 &= 3, &
S_4 &= 4, \nonumber\\
S_5 &= -6, &
S_6 &= -9, &
S_7 &= 14, &
S_8 &= -6, &
S_9 &= 1.
\end{align}
Evaluating the values of $H_k$ for all possible values of $k$ gives,
for $k=3$:
\begin{align}
H_3 &= S_3\binom{6}{0} = 3\cdot1 = 3.
\end{align}
\clearpage
\newpage
\onecolumngrid
Since vertical symmetry gives $V_3=3$ and no 3-site configuration spans both directions,
\begin{equation}
M_3 = H_3 + V_3 = 6.
\end{equation}
For $k=4$:
\begin{align}
H_4 &= S_3\binom{6}{1} + S_4\binom{5}{0}
      = 3\cdot6 + 4\cdot1 = 22,\\
M_4 &= 2H_4 = 44.
\end{align}
Similarly for $k=5$:
\begin{align}
H_5 &= S_3\binom{6}{2} + S_4\binom{5}{1} + S_5\binom{4}{0}\\
      =& 3\cdot15 + 4\cdot5 - 6\cdot1 = 59, \notag\\[4pt]
(H\cap V)_5 &= 25, \notag\\[4pt]
M_5 &= 2H_5 - (H\cap V)_5
     = 2\cdot59 - 25 = 93.
\end{align}
Among the $\binom{9}{5}=126$ configurations with 5 occupied sites, exactly $25$
span both horizontally and vertically.
Enumerating them in row-major site indices (0--8):
\begin{equation}
\begin{aligned}
[H\cap V]_5=\Big\{&
(0,1,2,3,6),\; (0,1,2,4,7),\; (0,1,2,5,8), \\[4pt]
&(0,1,4,5,7),\; (0,1,4,5,8),\; (0,1,4,7,8), \\[4pt]
&(0,3,4,5,6),\; (0,3,4,5,7),\; (0,3,4,5,8), \\[4pt]
&(0,3,4,7,8),\; (0,4,5,6,7),\; (0,4,5,7,8), \\[4pt]
&(1,2,3,4,6),\; (1,2,3,4,7),\; (1,2,3,4,8), \\[4pt]
&(1,2,4,5,7),\; (1,2,4,6,7),\; (1,4,5,6,7), \\[4pt]
&(1,4,5,7,8),\; (2,3,4,5,6),\; (2,3,4,6,7), \\[4pt]
&(2,3,4,6,8),\; (2,4,5,6,7),\; (3,4,5,6,7), \\[4pt]
&(3,4,5,7,8)
\Big\}.
\end{aligned}
\end{equation}
From the same procedure, one obtains
\begin{align}
H_k &= [0,\,0,\,0,\,3,\,22,\,59,\,67,\,36,\,9,\,1],\\
(H\cap V)_k &= [0,\,0,\,0,\,0,\,0,\,25,\,52,\,36,\,9,\,1],\\
M_k &= [0,\,0,\,0,\,6,\,44,\,93,\,82,\,36,\,9,\,1].
\end{align}
The polynomial for rule $R_0$, $b=3$ is therefore:
\begin{align}
p' &= \sum_{k=0}^{9} M_k\, p^k q^{9-k} \notag \\
   &= p^9 + 9p^8 q + 36p^7 q^2 + 82p^6 q^3 + 93p^5 q^4 \notag \\
   &\quad + 44p^4 q^5 + 6p^3 q^6.
\end{align}
where $q=1-p$. 
The polynomial expansion for other $b$ cases and other rules like $R_1$ and $R_2$ as described in \cite{PhysRevB.21.1223} can be similarly derived using the Inclusion-Exclusion principle. Nevertheless, they can be easily computed by writing a computer program, up to $b=5$. For $b \geq 6$, Monte Carlo techniques need to be implemented.
\end{document}